\documentclass[amsmath,amssymb,amsfonts,aps,pre,superscriptaddress,bibnotes,showpacs,showkeys,longbibliography,10pt]{revtex4-1}
\usepackage{epsfig}
\usepackage{ulem}
\usepackage{cancel}
\usepackage[english]{babel}
\usepackage{subcaption}
\usepackage{color}
\usepackage{hyperref}

\begin{document}

\title{Statistical Physics of Active Matter, Cell Division and Cell Aggregation}

\author{Jean-Fran\c cois Joanny}
\affiliation{Physicochimie Curie (CNRS-UMR168 and Universit\'e Pierre et Marie Curie), Institut Curie Section de Recherche, 26 rue d’Ulm, 75248 Paris Cedex, 05 France}
\author{Joseph O. Indekeu \footnote{Given his role as Editor of this journal, Joseph O. Indekeu had no involvement in the peer review of articles for which he was an author and had no access to information regarding their peer review. Full responsibility for the peer-review process for this article was delegated to another Editor.}}
\affiliation{Institute for  Theoretical Physics, KU Leuven, BE-3001 Leuven, Belgium}
\date{\today}

\begin{abstract}
In these Lecture Notes we aim at clarifying how soft matter physics, and herein notably statistical mechanics and fluid mechanics, can be engaged to understand and manipulate non-equilibrium systems consisting of numerous (microscopic) constituents that convert (chemical) energy to mechanical energy, or vice versa, and that are known as active matter. 
Hydrodynamic theory, vitally extended to include (anisotropic) active stress, provides an astonishingly successful scaffold for tackling the problem of spontaneous flow in active nematics, all the way to active turbulence. The laws of physics, nonchalantly tresspassing the border crossing between inanimate particle and living cell, are seen to perform cum laude in describing the bi-directional coupling between division and apoptosis on the one hand and mechanical stress on the other. Fluidization of cellular tissue by cell division is a conceptual leap in this arena. 
The active behavior of nematic tissues (cell extrusion, multilayer formation, ...) turns out to be controlled by topological defects in the orientational order. Playgrounds by excellence for exhibiting stress-growth coupling are multicellular spheroids serving as model tumors, and cysts used as stem cell factories for cell therapy. Finally, our study of villi and crypts in the intestine furnishes a synthesis of various concepts explored. Cell mechanical pressure and cell layer geometrical curvature turn out to provide the dynamical ingredients which, when coupled to the cell division rate, allow one to develop a physical theory of tissue morphology which hopefully will have practical impact on cancer research. 
\end{abstract}

\maketitle
\tableofcontents

\section{Prelude}
In these lectures we will be dealing with various aspects of the 
physics of both active and biological matter. When we say physics 
we mean macroscopic considerations on the one hand, as in 
thermodynamics and hydrodynamics away from equilibrium, and on the 
other hand we have in mind more mesoscopic and microscopic 
descriptions and explanations, as in statistical mechanics, which 
in recent years has been more and more successful in dealing with complex and even living systems. 

\begin{figure}[h!]
\centering
\includegraphics[width=0.4\linewidth]{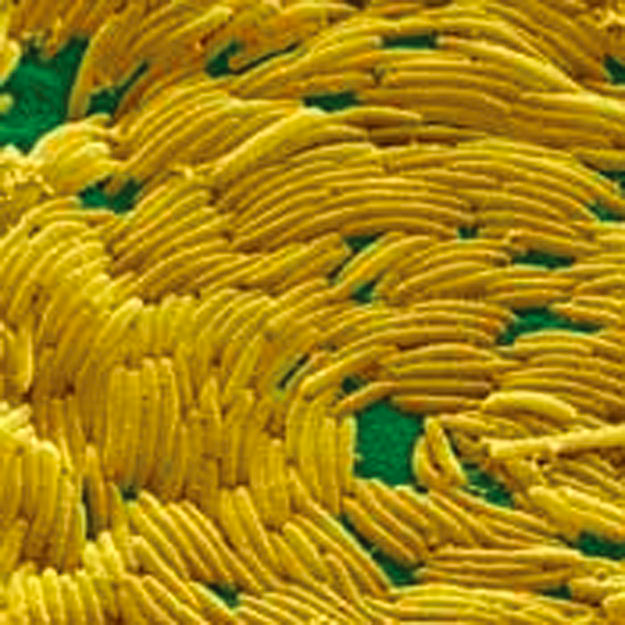}
\caption{Liquid-crystalline order in a myxobacterial flock.
Picture from Gregory Velicer (Indiana University Bloomington) and J\"urgen Bergen (Max-Planck Institute for Developmental Biology).}
\label{teaser}
\end{figure}

We distinguish three parts in these lectures. The first is devoted 
to active matter, an example of which is shown in Fig.\ref{teaser}. 
We define active systems, discuss the hydrodynamic theory of 
active matter and devote time to deal with the conceptually and 
experimentally challenging domain of active turbulence. In the 
second part we turn to the phenomenon of cell division, elucidate 
the concept of homeostatic pressure and treat the fluidization of 
tissues by cell division. In the third and final part we develop 
an exploration of the active behavior of tissues around four 
themes. We touch upon topological defects in nematic tissues, we 
discuss multicellular spheroids, we describe cyst growth and dwell 
upon villi and crypts in the intestine.

In closing this prelude, we  provide a list of general references which the reader may find useful to get acquainted with the background, the main lines of thought, the experimental observations and theoretical results in this arena \cite{MarRMP,ProstNP,JulicherRPP,Ramas,Needle,AlertCJ,JBlesHouches,ABAJ,BNL,GLScience,HBCell,DArcyT,Wolpert,Weinberg}

\section{Active Matter}
What is active matter \cite{Das, ProstNP, MarRMP}? Firstly, let us emphasize the 
material aspect. From a statistical physics point of view, we 
have in mind a system with sufficiently many constituents (or 
``particles") in order for a statistical description to be 
necessary and useful to understand its behavior at the meso- or 
macroscopic level. The key property of  an active system  is then that detailed balance is broken by supplying energy (of any form but more specifically chemical energy in many biological systems) at the microscopic/particle/agent level which couples in a nontrivial way to particle motion or interactions \cite{ramaswamy}. The numerous constituents in active 
matter, therefore, differ from molecules or atoms in ordinary matter 
in that they possess motor-like active character. They convert 
non-mechanical forms of available energy, chemical or electromagnetic 
(light, ...), into mechanical energy in a way that modifies their 
motion (so they appear ``self-propelled") and/or modifies the way they interact. If this modification emerges as time-dependent patterns observable on mesoscopic or macroscopic length 
scales, we call this form of matter active. Generic collective 
properties of active matter include spontaneous flows, 
orientational order and non-equilibrium phase separation. 

Let us test this fairly heuristic definition in two directions. Is 
a turbulent fluid active? Not really, because the patterns result 
from conversion of mechanical energy into mechanical energy, at a 
hierarchy of length scales, from macroscopic to molecular. The drive is applied by stirring at the largest scale. 

A 
similar conversion takes place in the Brownian motion of a passive particle in a fluid, which is driven by thermal fluctuations (mechanical energy of microscopic origin called heat). Thermal Brownian motion is not ``active matter" because the drive (heat) and dissipation (drag) are balanced, enforcing detailed balance. Any departure from detailed balance would make the particles active.
It is interesting to note that the original work of R. Brown \cite{brown} detected the random motions of particles in pollen grains both in living plants and in dead plants.

Is a living organism active matter? A bacterium? A human? 
From the point of view of activity, we can safely say yes. From 
the material point of view, the criterion of sufficiently many 
constituents prompts us to be cautious. Probably it is correct to 
consider that a bacterium, or a human, contain active matter. It 
may also be appropriate to describe a large collection of 
interacting bacteria, or to attempt to explain features of human 
collective behavior, using the concepts and methods that apply to active 
matter. 

From a statistical mechanics (and thermodynamics) point of view, 
active matter is intrinsically non-equilibrium and its non-equilibrium nature can be characterized by its 
entropy production. Established thermodynamic 
variables such as pressure must be reexamined \cite{TK}. From a 
mechanical point of view, in active matter even  Newton's law of reciprocity of mechanical action and reaction 
among constituents may not be valid. 

\subsection{Active systems}
Let us mention some examples of non-equilibrium and active matter of particular interest for these 
lectures. Bartolo {\it et al.} combined model experiments and 
theory for studying the motion of an artificial crowd of motorized (self-propelled) spherical beads \cite{Bartolo}. Polar liquids 
assembled from motile micrometer-sized colloids undergo a motility-induced solidification, a dynamical phase transition from a 
spontaneously flowing liquid to an amorphous active solid. This 
type of transition is expected for any flocking group \cite{Vicsek}
in which the individuals slow down as the distance to their 
neighbors decreases. 

\begin{figure}[h!]
\centering
\includegraphics[width=0.5\linewidth]{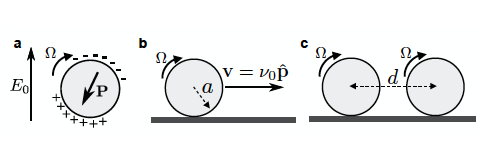}
\caption{Quincke rollers. a. Effect of an applied DC electric
field larger than the Quincke threshold.  b. Rotation to translation conversion. c. Hindering action of the lubrication torque on two neighboring colloids rolling in the same direction.}
\label{1}
\end{figure}

What makes the crowd of beads qualify for active matter? First of 
all their large number, permitting cooperative behavior in a 
statistical physics sense. Secondly, the important energy 
consumption and conversion at the level of an individual bead is 
realized making use of the Quincke electro-rotation instability. 
The insulating colloidal beads are immersed in a conducting fluid 
and develop a surface-charge dipole when a DC electric field is 
applied. 
Increasing the electric field strength beyond  a threshold 
destabilizes the dipole orientation, causing the bead to spin 
driven by an electric torque. This fulfills our criterion of 
``activity" since electric energy is converted to rotational 
energy at the level of an individual constituent. Part of this 
energy is dissipated in the surrounding fluid due to friction. A 
viscous frictional torque balances the electric torque and a 
constant limiting angular velocity results. Next, the rotational 
energy is converted to translational energy when beads sediment on 
a groove and start to roll. The direction of these ``Quincke 
rollers" is random in a dilute phase, but in a denser state the 
electric and hydrodynamic inter-bead interactions promote velocity 
alignment, but also entail hindrance to the motion (due to a 
lubrication torque). Fig.$\ref{1}$ depicts this setting.

Our second example is an out-of-equilibrium ``driven" system of 
manufactured rod-like particles confined between closely spaced 
horizontal parallel plates that vibrate vertically and so transfer 
mechanical energy to the particles which affects their horizontal 
motion. This system was studied experimentally by Narayan {\it et 
al.} \cite{NRM}. The particles are agitated but do not convert or 
``consume" energy. They constitute 
 a ``simple" driven liquid crystal with 
spontaneous orientational order. Still, this non-equilibrium 
system qualifies for ``active matter", as the vertical (mechanical) drive is dissipatively rectified by individual particles to generate biased horizontal motion. The system is interesting for us because it shares a number of properties 
with active matter exhibiting spontaneous orientational symmetry 
breaking such as flocks of birds or bacteria.

\begin{figure}[h!]
\centering
\includegraphics[width=0.2\linewidth]{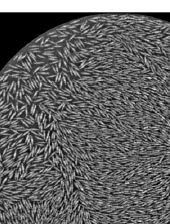}
\caption{Giant number fluctuation in nematically ordered agitated rods in the horizontal plane. The confining plates vibrate vertically. The dilute region seen in the top left corner is an extraordinary density fluctuation. It takes a long time to relax and form elsewhere. Picture extracted from \cite{NRM}. }
\label{numberfluc}
\end{figure}

One of those properties is giant, and long-lived, number 
fluctuations. In contrast with equilibrium systems, we encounter 
here density fluctuations that do not scale down with system size 
as the central limit theorem would predict. As a consequence, the 
well-established concept of a sharply defined density (for a 
sufficiently large system) must be applied with more care in this
quasi-two-dimensional non-equilibrium system of agitated rods. Fig. 
$\ref{numberfluc}$ illustrates this. 

Our third example is concerned with the phenomenon of propagating 
density waves in, e.g., bird flocks, which are complex patterns 
emerging in response to an imminent attack by a predator. We are 
dealing here with living matter with thousands of constituent 
individuals. Such waves originate at (very) high density spots, 
where the individuals are almost in direct collisional contact due 
to a ``panic" event. Quantitative characterization of this 
phenomenon by Procaccini {\it et al.} \cite{Procaccini} has 
confirmed that these wave-like density variations are effective in 
confusing predators and reducing the success of an attack. 

\subsection{Hydrodynamic theory of active matter}
In a hydrodynamic theory of soft active matter \cite{MarRMP} we 
begin with identifying the slow variables and studying cooperative 
effects at long length scales and long time scales. Our system 
under consideration is a liquid consisting of elongated particles, 
each with local (unit vector) direction ${\bf n}$, capable of 
nematic order. The director is the continuum field $\bf p$, a 
coarse-grained version of ${\bf n}$,  and if, in addition, the 
particles are polar (head and tail are distinct), $\bf p$ is the 
(dimensionless) polarization field. Unless the system is near a 
critical point, it suffices to consider unit vectors, $||{\bf 
p}||=1$. We also note that systems with a scalar order parameter such as a concentration or a volume fraction can become active, the local vector is in this case the gradient of the order parameter. 

Let us discuss the slow variables. Typically, these are the 
densities associated with the conserved quantities and the density 
associated with the order parameter (either polarization vector or 
nematic orientational order parameter tensor). In a 
one-component 
active system,  a 
slow variable therefore is the number density $\rho$ of the 
particles (of mass $m$), satisfying the continuity equation
\begin{equation}
    \partial_t \rho + \nabla (\rho {\bf v})=0,
\end{equation}
with ${\bf v}$ the local velocity.
The momentum density $\rho m {\bf v}$ of the active fluid, may or 
may not be conserved. The momentum of the active matter is not 
conserved when momentum is transfered to another phase such as a 
solid substrate in two dimensions or a porous medium in three 
dimensions. The dynamics of the active particles that transfer
momentum  is then overdamped. However, for a bulk active fluid, a 
description taking into account conservation of 
total momentum is appropriate. In this case we deal with an 
``active gel" and have the additional continuity equation for the 
momentum,
\begin{equation}
    \partial_t (\rho m v_{\alpha})+\partial_{\beta} (\rho m v_{\alpha} v_{\beta} - {\bar\sigma}_{\alpha \beta})=0,
\end{equation}
where $\bar\sigma$ is the tensor of the total stress, which is the 
negative of the momentum flux. Finally, the energy of active 
matter is of course not conserved as its constituents ``consume" 
energy.

\begin{figure}[h!]
\centering
\includegraphics[width=0.4\linewidth]{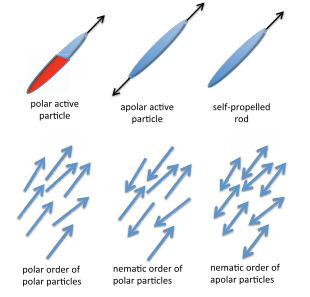}
\caption{Active particles and possible orientational order. Polar particles (top left; e.g., bacteria, fish, birds, ...), with distinct heads and tails, are generally 
self-propelled along their axis. They can display polar (bottom left) or nematic order (bottom center). 
Apolar particles (top center) are head/tail symmetric
and can order nematically (bottom right). Active rods
(top right) are head/tail symmetric, but each rod is self-propelled
in a specific direction along its axis. The self-propulsion
renders them polar, however, for exclusively apolar
interactions (e.g., steric hindrance), the rods can order
only nematically (bottom center). }
\label{cartoon1}
\end{figure}

If we attempt to classify active matter of elongated self-propelled 
particles possibly into universality classes, another 
criterion, besides the momentum (non-)conservation, is the nature 
of the broken (continuous) rotational symmetry in the ordered 
phase, polar or nematic. Polar order (oriented parallel with net 
direction) and nematic order (parallel with no net direction) are possible 
for  elongated self-propelled particles with head/tail distinction 
(e.g., bacteria or fish).  For head/tail symmetric particles 
(``apolar") nematic order is possible (e.g., melanocytes). The 
distinction between polar and nematic order is illustrated in the 
following cartoon (Fig.$\ref{cartoon1}$).

The order parameter for nematic ordering is the traceless nematic tensor, with $\alpha, \beta = x,y,z$ (for $d=3$),
\begin{equation}
\label{nematic tensor}
    q_{\alpha\beta} = \langle (n_{\alpha}n_{\beta} - \frac{\delta_{\alpha\beta}}{3})\rangle  = S (p_{\alpha}p_{\beta} - \frac{\delta_{\alpha\beta}}{3}),
\end{equation}
where $\langle . \rangle$ denotes the average over a bunch of 
particles about some point in space and moment in time. $S$ is the 
local magnitude of the nematic order (for our purposes positive 
and $0 < S \leq 1$).
 
In the vicinity of a continuous order-disorder transition (as, 
e.g., predicted in a mean-field theory) the amplitude of the 
polarization is a critical variable, and critical slowing down of 
fluctuations of the order parameter is expected. However, the intrinsic 
relaxation time of the amplitude of the polarization is long but does not diverge with system size, and strictly speaking this amplitude is not a hydrodynamic variable. In the following, we mostly consider a system deep in the nematic phase for which the modulus of the director field is equal to one.

The basic variables in the hydrodynamic theory are thus the local 
velocity field, the local density field, the local orientational 
field and the energy consumption, e.g., an amount of chemical 
energy $\Delta \mu$ per particle due to ATP consumption. This 
energy consumption entails that the entropy production density 
contains a contribution $\Delta \mu$ times the number of ATP 
molecules consumed per unit time and unit volume. 

Using a linear generalized hydrodynamic theory, or ``Onsager 
theory" close to equilibrium, one can derive, for example, the 
entropy production rate. The constitutive equations of active 
matter are obtained by i) identifying the fluxes and forces, and 
ii) writing down the most general linear relation between fluxes 
and forces that respects the conservation laws and symmetries of 
the problem. These include translational and rotational 
symmetries, and time-reversal symmetry. For example, in a nematic 
active gel, the relevant fluxes (forces) are stress (velocity 
gradient), total time derivative of the polarization (orientational 
field conjugate to the polarization) and energy consumption rate 
(energy consumption). 

The fluxes must be separated into a reactive component and a dissipative
component. For example, for the stress the reactive component is the
elastic stress and the dissipative component is the viscous
stress. Only the dissipative component of a flux contributes to 
the entropy production. Using this decomposition one arrives at 
the reactive (antisymmetric) and dissipative (symmetric) Onsager 
matrices of kinetic coefficients.

Away from bulk criticality, the polarization free energy of the active gel
is the classical Frank free energy of a nematic liquid crystal, 
\begin{equation}
 {F_p}=\int d{\bf x}\Big[ \frac{K_1}{2} ({\bf\nabla} \cdot {\bf p})^2 + 
 \frac{K_2}{2} ({\bf p}\cdot ({\bf\nabla}\times {\bf p}))^2 
 + \frac{K_3}{2}
({\bf p}\times ({\bf\nabla}\times {\bf p}))^2 + k_1  {\bf\nabla}\cdot {\bf p}- 
\frac{1}{2}h^0_{\parallel} {\bf p}^2 \Big],
\label{Frank}
\end{equation}
where the $K_i >0$ are the Frank elastic constants (for splay, 
twist and bend), the term $k_1  {\bf\nabla}\cdot {\bf p}$ has been 
added to allow for polar systems, and a Lagrange multiplier 
$h^0_{\parallel}$ has been added to ensure that the polarization 
is a unit vector. Note that in {\it two dimensions} (active gel on a 
substrate) the twist term vanishes, and if we make the additional 
``one-elastic constant" approximation that splay and bend feature 
the same elastic constant, which we denote by $\kappa$ (now with dimension of energy, not force), then 
standard vector identities applied to the unit vector ${\bf p} = 
\cos \phi \,\hat {\bf e}_x + \sin \phi \,\hat {\bf e}_y $ entail 
that the Frank free energy, for an apolar system, reduces to 
\begin{equation}
 {F_p}=\int d{\bf x}\Big[ \frac{\kappa}{2} ({\bf\nabla}  {\bf \phi})^2  
  \Big], \; \mbox{with} \; {\bf x} = (x,y).
\label{Frank2D}
\end{equation}

In general dimensionality, the orientation field ${\bf h}$ conjugate to ${\bf p}$ is obtained by functional differentiation,
\begin{equation}
    {\bf h}=-\frac{\delta {F_p}}{\delta {\bf p} }
    \label{orienfield}
\end{equation}
It is useful to decompose it into a component
parallel to the polarization, $h_{\parallel}$ and a component perpendicular
to it, $h_{\perp}$. For $d=2$ and ``one-elastic",
\begin{equation}
    h_{\perp}= \kappa \,\nabla^2 \phi
\end{equation} 

In an anisotropic medium there is an antisymmetric component of the stress, $\sigma^a$, associated with torques due to rotational symmetry breaking. Its tensor elements are given by 
\begin{equation}
    \sigma^a_{\alpha \beta}= \frac {1}{2}  (h_{\alpha}p_{\beta}-h_{\beta}p_{\alpha})
\end{equation}
Note that in $d=2$ ($d=3$) the dimension of stress is force per unit length (area).

We can now provide an explicit expression, in three dimensions, for the entropy production rate in this nonequilibrium system,
\begin{equation}
    T \dot{\cal S}=\int d{\bf r}\,  \{ \sigma_{\alpha \beta} 
  v_{\alpha \beta}+\dot{P}_{\alpha}h_{\alpha}+ r\Delta \mu     \}\; , 
\end{equation}
where the dot denotes time derivative. In this expression we identify the following constituents. To the ``fluxes" contribute the symmetric deviatoric tensor $\sigma_{\alpha \beta}$, given by 
\begin{equation}
    \sigma_{\alpha \beta} = \bar{\sigma}_{\alpha \beta}+
 P \delta_{\alpha \beta}-  \sigma^a_{\alpha \beta}, 
\end{equation}
where $P$ is the pressure, as well the (co-moving and co-rotational) rate of change of the polarization,
\begin{equation}
\label{dynaP}
    \dot{P}_{\alpha} \equiv  \frac{D {\bf p}}{Dt} = \frac{\partial {\bf p}}{\partial t} +({\bf v}\cdot {\bf \nabla}){\bf p} +{\bf \omega} \times  {\bf p}, 
\end{equation}
with $({\bf \omega} \times  {\bf p})_{\alpha} \equiv \omega_{\alpha\beta}p_{\beta}$, and the ATP consumption rate $r$.
To the ``forces" contribute the strain rate tensor $v_{\alpha \beta}$, which features velocity gradient components in a symmetric manner, whereas the vorticity tensor is the antisymmetric part,
\begin{eqnarray}
\label{vgrad}
v_{\alpha \beta}  =\frac 12(\partial_{\beta} v_{\alpha}+ \partial_{\alpha} v_{\beta}), \\
\label{vorticity}
\omega_{\alpha \beta} = \frac 12(\partial_{\beta} v_{\alpha}- \partial_{\alpha} v_{\beta}).
\end{eqnarray}
To the ``forces" also contribute the orientation field ${\bf h}$ and the chemical potential difference
\begin{equation}
    \Delta \mu=\mu_{ \rm ATP}-\mu_{\rm ADP}- \mu_{\rm P_i}
\end{equation}
between fuel (ATP) and its reaction
products ADP and inorganic phosphate P$_i$. Each of these forces has signature $1$ or $-1$ under time reversal. 

We now present the constitutive equations for an incompressible (${\nabla \cdot {\bf v}}=0$) one-component system,
\begin{eqnarray}
    {\tilde \sigma}_{\alpha
\beta}&=& 2\eta \,v_{\alpha \beta} - 
 \zeta \,\Delta  \mu \, q_{\alpha \beta} + \frac{\nu}{2}
 (p_{\alpha}h_{\beta}+p_{\beta}h_{\alpha})\;,\\ 
 \frac{D p_{\alpha}}{Dt} &=& \frac{h_{\alpha}}{ \gamma} -\nu \,v_{\alpha \beta}p_{\beta}\;,\\
  r &=& r_0 \Delta \mu + \zeta \,v_{\alpha \beta} q_{\alpha \beta}\;,
 \end{eqnarray}
 where the tilde in $\tilde \sigma$ denotes the traceless part. Note that the tensor $v_{\alpha \beta}$ is already traceless in view of ${\nabla \cdot {\bf v}}=0$. Further, $\eta$ is the shear viscosity, $\zeta$ is a transport coefficient associated with the activity of the system, $\nu$ is a flow-coupling coefficient and $\gamma $ is the rotational viscosity. The tensor $q$ is the nematic tensor introduced in \eqref{nematic tensor}.

\begin{figure}[h!]
\centering
\includegraphics[width=0.4\linewidth]{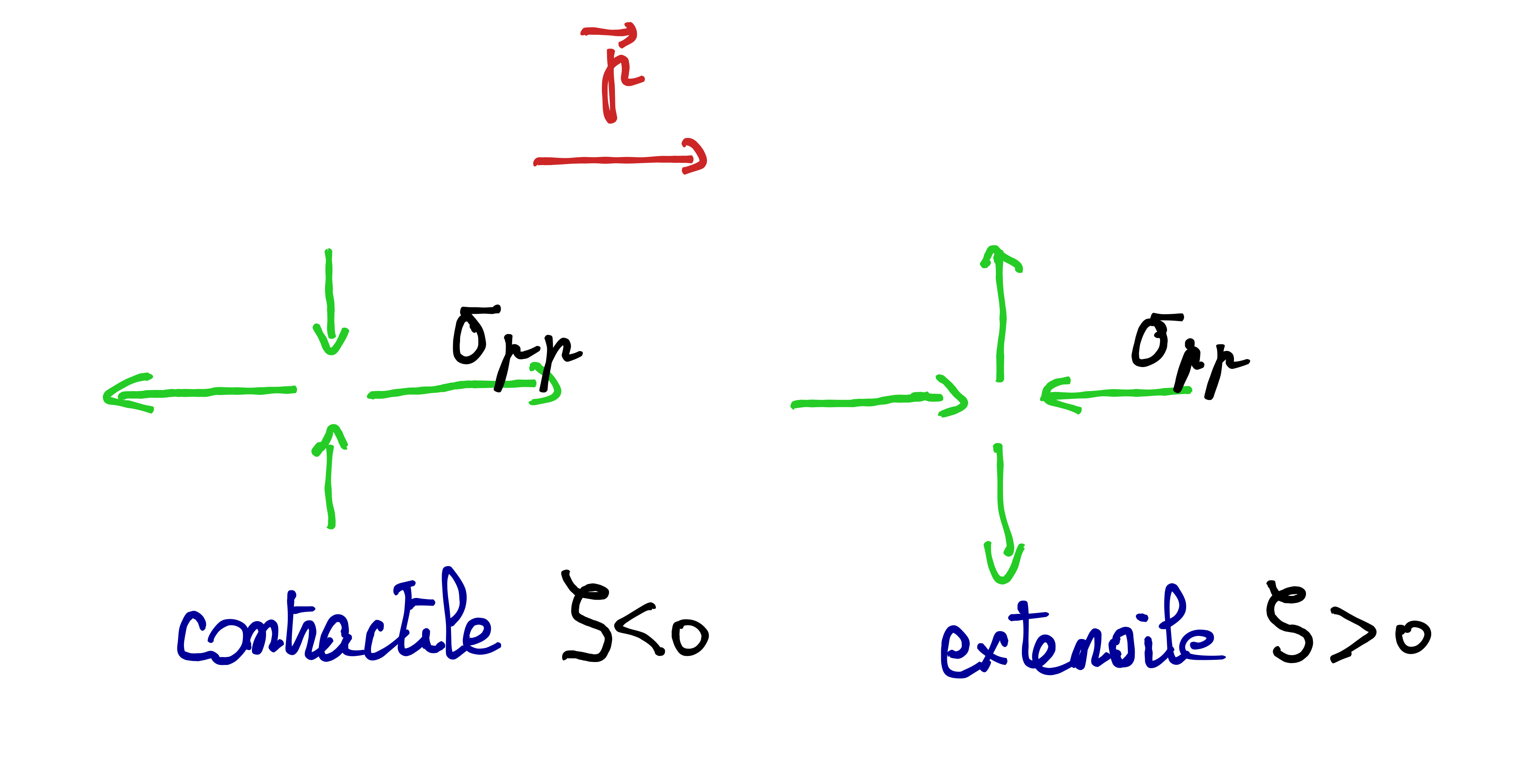}
\caption{Active stress: contractile versus extensile stress. The stress is extensile if the local fluid element pushes on its environment in the direction of the director. The environment then pulls on the local volume element and the stress is negative. The signs are reversed for a contractile stress}
\label{contractile}
\end{figure}

The hydrodynamic description of active polar or nematic
gels strongly resembles that of nematic liquid crystals.
However, the presence of a non-equilibrium contribution entails
nonintuitive and spectacular phenomena. The non-equilibrium term here is the active stress of the system, defined through
 \begin{equation}
 \sigma_{\alpha\beta}^{\rm active}  = - \zeta \,\Delta  \mu \, q_{\alpha \beta}
 \label{actstress}
 \end{equation}
A negative transport coefficient, $\zeta <0$,
corresponds to a contractile stress along {\bf p} and extensile in the perpendicular direction, and vice versa for $\zeta >0$ (for an example, see the sketch in Fig.\ref{contractile}). The active stress is reactive, not dissipative, and is due to the transformation of chemical energy into
mechanical energy. It includes all forms of consumption of ATP. Its introduction does not require any microscopic model of the cell behavior and energy consumption. The non-equilibrium character is due to the fact that the chemical reaction does not relax and $\Delta  \mu$ remains constant. It has zero trace and is, in general, non-diagonal. An important phenomenon to note is that a distortion of the polarization induces a gradient of active stress that must be balanced by a viscous stress. This causes a spontaneous flow.

Next, we turn to how the hydrodynamic theory is extended for describing more complex active systems. For polar systems, there is a polar splay contribution to the free energy \cite{Voituriez}. It is, for example, the term with coefficient $k_1$ included in \eqref{Frank}. This spontaneous splay
term is a surface term if $k_1$ is a constant. Note that for a 
one-component system there is no polar term in the constitutive equation. In general, nonlinear terms can be added to the dynamical equation for the polarization \eqref{dynaP}. For example,
\begin{equation}
    \frac{D{\bf p}}{Dt} = \lambda {\bf p}\cdot \nabla {\bf p}
\end{equation}
 is a polar nonlinear contribution. Furthermore, when the theory is extended to chiral systems, care must be taken that the local rotation is no longer described by the vorticity as we have defined it in \eqref{vorticity} \cite{Furt}. Also, at larger scales one can
extend the active gel theory to construct a hydrodynamic theory of (polarized) tissues. 

Most active systems are of multicomponent nature. They contain
a solvent and active constituents.
In many cases they can be treated by an effective one-component theory.
However, when viscoelastic effects are important
a one-component theory ignores the permeation of the
solvent through the active gel. A two-component
theory that properly takes into account permeation
effects was developed in \cite{Callan,Adar}.  What is the essence of an extension to two components? It is that each component can display a current. The composition variable comes into play and the process of diffusion takes place in response to concentration gradients. There occurs an interesting coupling between diffusion and activity. For example, in the presence of the polar order parameter ${\bf p}$, there is an additional flux, in the form of an active contribution to the current,
\begin{equation}
    {\bf j}= j_0 \,{\bf p} \,\Delta \mu, 
\end{equation}
with $j_0$ a kinetic Onsager coefficient, which entails a spontaneous velocity. This is relevant for mesenchymal tissues (which give organs shape and strength), in which spontaneous cell motion takes place, which can be described in terms of self-propulsion of particles. 

Let us comment further on the complication of visco-elasticity. An active gel is often not a simple liquid but a
more complex medium with a finite viscoelastic relaxation time,
which is liquid only at long time scales. In a first step, a
viscoelastic medium can be described by the Maxwell model, which features a single relaxation time $\tau$. The constitutive equation among the stress and velocity gradient tensors is   
\begin{equation}
     \frac{D \tilde \sigma_{\alpha\beta}}{Dt}+ \frac{\tilde \sigma_{\alpha\beta}}{\tau}= 2 G \, v_{\alpha\beta},
\end{equation}
where the shear modulus $G$ is related to the viscosity $\eta$ through  $\eta =G \tau$. As we already alluded to, there is an interesting coupling between viscoelasticity and activity, which is expressed, e.g., in permeation effects in multicomponent systems \cite{Callan, Adar}.

Let us now describe some remarkable hydrodynamic
phenomena induced by activity in active gels,
such as thin liquid film instabilities and rheological properties.
One generic property of active orientable liquids, whether
polar or apolar, is the instability of any homogeneous nonflowing
steady state towards an inhomogeneous spontaneously
flowing state (with no pressure gradient applied) \cite{SimhaRamas}. We illustrate this instability in the simple geometry of a thin active nematic liquid film \cite{Voituriez}.

The instability we will describe is akin to the fascinating Fredericks transition, which a nematic liquid crystal slab of finite thickness can undergo in an external magnetic or electric field \cite{deGP}. So let us recall that briefly. The director field throughout the sample may be distorted, away from a homogeneous configuration dictated by anchoring boundary conditions, by a sufficiently strong field. The director alignment remains unaffected as long as the field strength is less than the critical or ``Fredericks" threshold, but will start to become non-uniform in space when the magnitude of the field exceeds the critical field strength. Of course, this is the important property of liquid crystals that underlies many of its applications as displays.

\begin{figure}[h!]
\centering
\includegraphics[width=0.4\linewidth]{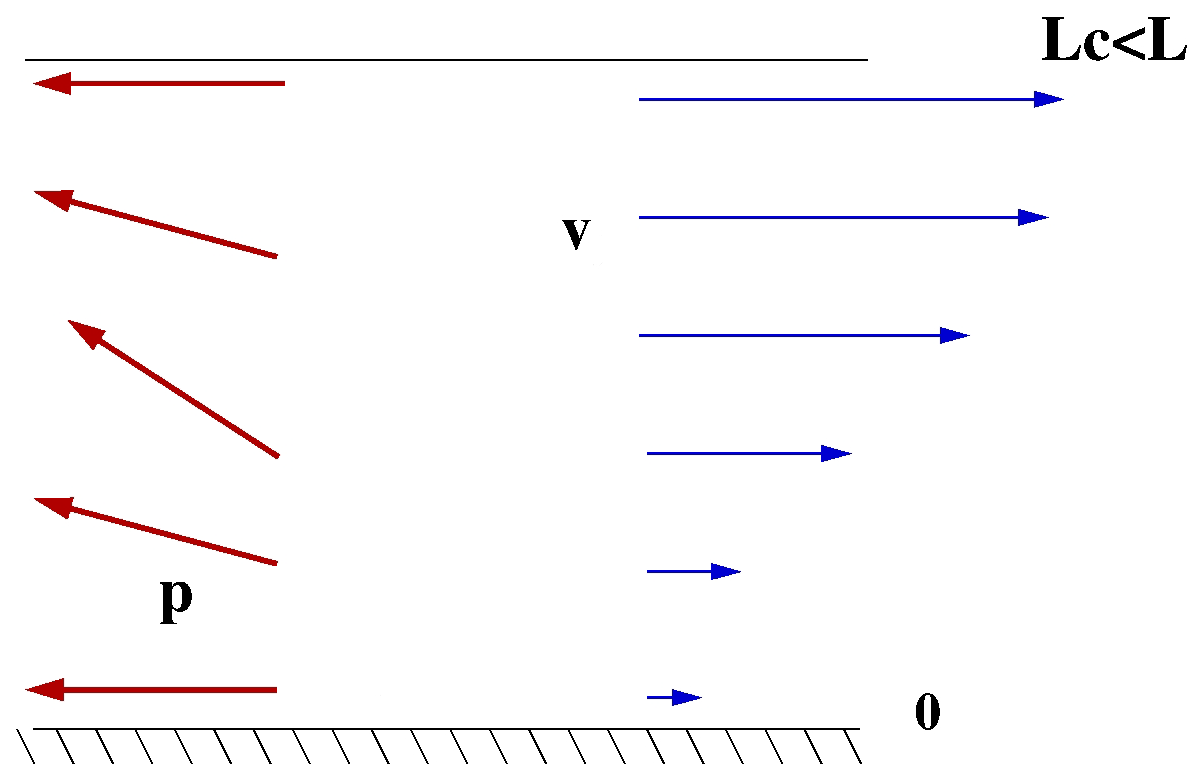}
\caption{Active nematic film of thickness $L > L_c$ on a solid substrate displaying splay distortion of the director with half wavelength equal to the film thickness. Parallel anchoring is imposed at the surfaces. A tilt of $\bf p$ in the center creates an orientation gradient, causing an active stress gradient. This is balanced by a viscous stress gradient, implying a velocity gradient. A typical shear-flow-like velocity profile ensues. }
\label{Frederiks}
\end{figure}

In our setup, we consider a two-dimensional geometry, with transverse ($x$) and longitudinal ($y$) directions, normal and parallel to the film, respectively.  A thin active nematic film of thickness (height) $L$ is placed on a horizontal solid substrate (see Fig.\ref{Frederiks}). The director is anchored in planar alignment on the solid, at $x=0$, as well as at the interface against air, at $x=L$ (parallel anchoring conditions), so that the polarization is strictly along $y$ at $x=0,L$. 

In the homogeneous state the director is uniformly aligned along $y$ and there is no flow. We now examine the stability of this static state. We focus on low Reynolds number systems, with negligible inertia, so that the force balance reduces to Stokes' equation
\begin{equation}
\partial_{\beta} \bar{\sigma}_{\alpha \beta}=0.
\label{Stokes}
\end{equation}
Is it possible to nucleate a flowing state which is translationally invariant along $y$, with ${\bf v} = v_y(x) {\bf e}_y \neq 0$, and with a polarization that displays, e.g., a wave-like distortion from the parallel configuration? We perform a linear stability analysis. The growth rate $\Omega$ of orientational perturbations about the uniform state, with wave vector ${\bf q}$ at an angle $\phi $ away from the director, obeys \cite{Duclos}
\begin{equation}
    \Omega(q)= \frac{\zeta \Delta\mu\,\cos2\phi (1-\nu \cos2\phi)}{2\left(\eta +\gamma\nu^2 \sin^2 2\phi/4\right)} -\frac{\kappa q^2}{\gamma} \frac{\eta + \gamma/4(\nu^2-2\nu \cos2\phi+1)}{\eta +\gamma\nu^2 \sin^2 2\phi/4}.
        \label{instabrate}
\end{equation}
Note that it turns positive for long wavelengths (small $q$) in a manner that depends on $\phi$ \cite{AlertCJ}. For example, for rod-like particles ($\nu < 0$) the quiescent uniform state is unstable to splay distortions ($\phi = \pi/2$) of the director provided the active stresses are contractile ($\zeta <0$). The instability is, in general, caused by the active forces produced by a director perturbation, which drive flow. The active stress is equivalent to a magnetic field in the $x$-direction, in the Fredericks transition analogy. At small $q$ the active stress overcomes the restoring elastic stress of the nematic. The crossover ``critical" length is $L_c$, which for $\phi = \pi/2$ is found, from \eqref{instabrate}, to be
\begin{equation}
    L_c = \left(
-\frac{\pi^2 \kappa (\frac{4\eta}{\gamma}+(\nu+1)^2)}{2{\zeta}\Delta\mu(\nu+1)}\right)^{1/2}.
\label{criticalL}
\end{equation}
Since distortion wavelengths larger than the twice the film thickness $L$ do not satisfy the boundary conditions, the instability can arise only for $L > L_c$.  

\begin{figure}[h!]
\centering
\includegraphics[width=0.4\linewidth]{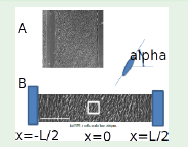}
\caption{Bi-dimensional layer of RPE1 cells confined to a stripe of width $L$. Note that the geometry is rotated by 90$^{\circ}$ relative to that in Fig.\ref{Frederiks}.}
\label{RPE1}
\end{figure}
\begin{figure}[h!]
\centering
\includegraphics[width=0.30\linewidth]{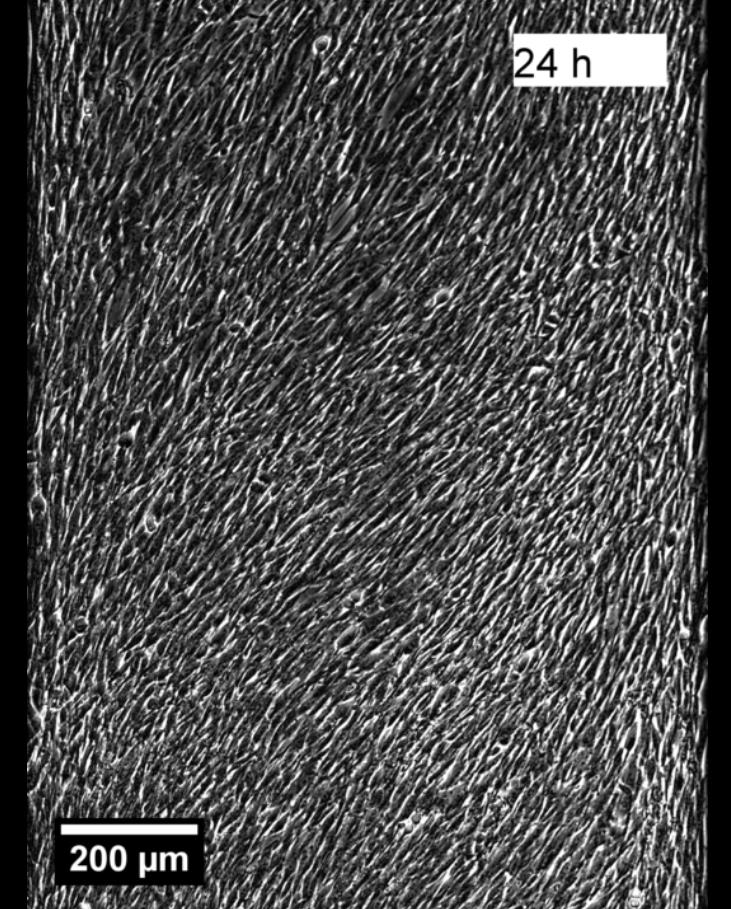}
\caption{Confined RPE1 cells align together with a tilt angle in a sufficiently wide stripe ($L>L_c$). Note that the geometry is rotated by 90$^{\circ}$ relative to that in Fig.\ref{Frederiks}. Reproduced from \cite{Duclos}.}
\label{RPE1tilt}
\end{figure}
\begin{figure}[h!]
\centering
\includegraphics[width=0.6\linewidth]{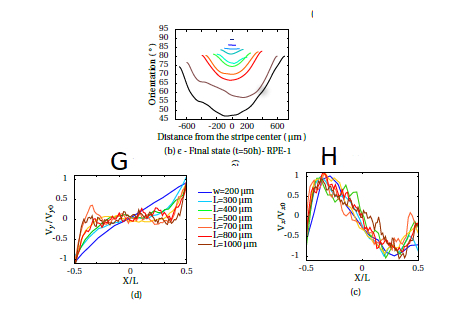}
\caption{Cell orientation and parallel and transverse cell velocity components across the stripe. }
\label{velocities}
\end{figure}

Illustrative experiments investigated the orientation and dynamics of elongated retinal pigment epithelial (RPE1) cells within confining stripes on micropatterned glass substrates (see Fig.\ref{RPE1}). 
The adhesive stripe widths typically vary between $L=10\mu$m and $L=1000\mu$m \cite{Duclos}. For a wide stripe ($L > L_c$) a net tilt angle appears near the center: the cell population organizes in a nematic phase with a director at a finite angle with the main direction of the stripe (Fig.\ref{RPE1tilt} and upper panel of Fig.\ref{velocities}). 

\begin{figure}[h!]
\centering
\includegraphics[width=0.6\linewidth]{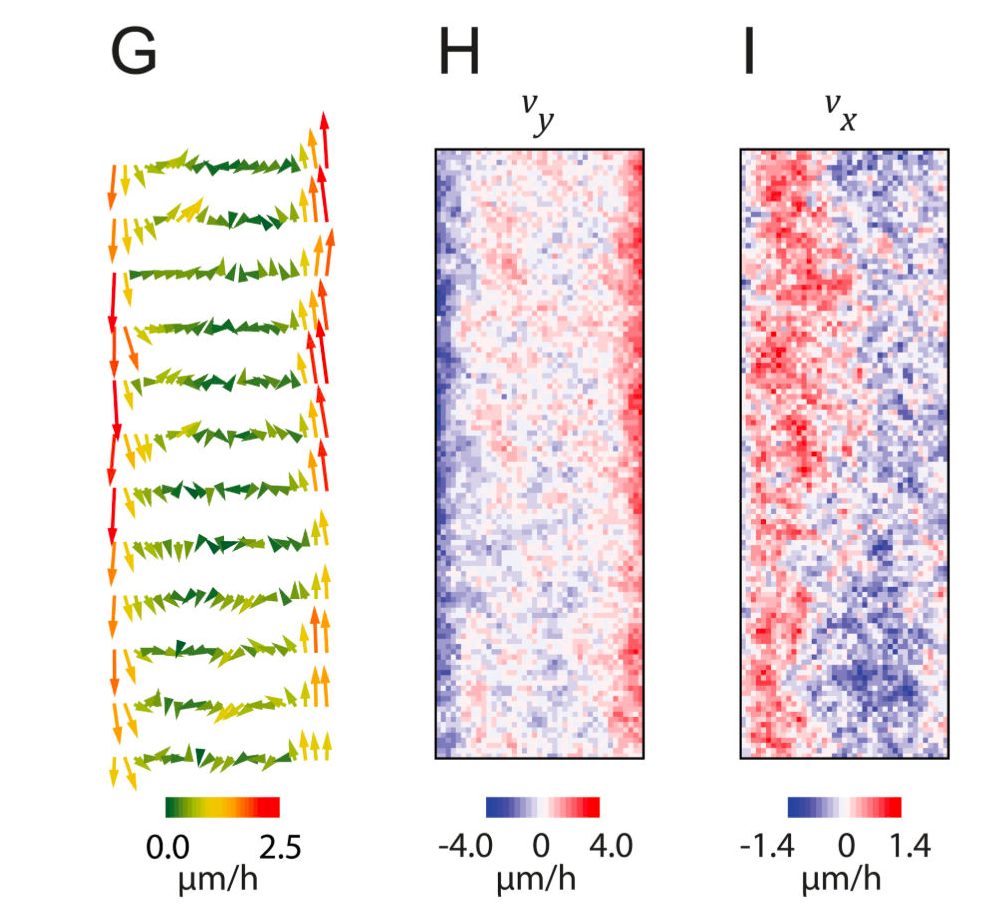}
\caption{Velocity vector and parallel and transverse velocity components. Reproduced from \cite{Duclos}.}
\label{velocityDetail}
\end{figure}

Cells spontaneously develop complex flows with shear and transverse components. The velocity field was mapped by particle image velocimetry (PIV) analysis. The average velocity vector of a small subvolume of particles is shown pictorially in the left panel of Fig.\ref{velocityDetail}. The shear component of the flow (velocity component $v_y$ parallel to the stripe) is represented in the lower left panel of Fig.\ref{velocities} and in the middle panel of Fig.\ref{velocityDetail}, while the transverse component $v_x$ is displayed in the lower right panel of Fig.\ref{velocities} and in  Fig.\ref{velocityDetail} on the right. The existence of a transverse flow convergent towards the center of the stripe is noteworthy.  The cell number is not conserved due to cell extrusion, a process that is stronger near the center than near the boundaries.


\begin{figure}[h!]
\centering
\includegraphics[width=0.7\linewidth]{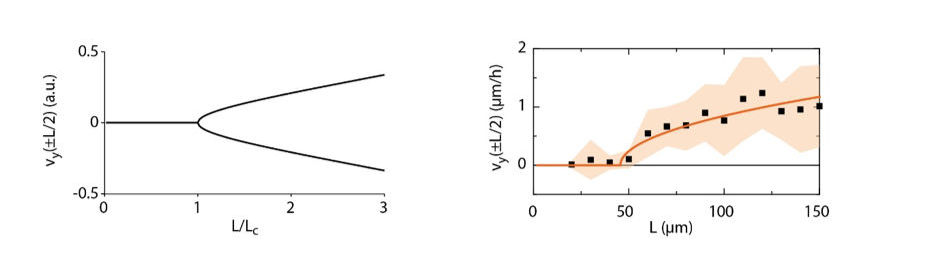}
\caption{Bifurcation of the parallel velocity component at the critical stripe width. Reproduced from \cite{Duclos}.}
\label{velobifur}
\end{figure}

As a function of stripe width the parallel velocity component is observed to undergo a bifurcation (Fig.\ref{velobifur}). For $L < L_c$ cells orient in the direction of the stripe and no net shear flow develops. For $L>L_c$ spontaneous shear flow sets in, in a direction determined by spontaneous symmetry breaking (for non-chiral cells) or by cell chirality. Experiments with several cell types confirm the generic character of this ``active Fredericks transition" scenario. 

The theory of active nematics has been further refined by allowing for friction forces arising from cell-substrate adhesion. Thus, a friction coefficient $\xi$ is introduced in the force balance,
\begin{equation}
 \partial_{\beta} \bar\sigma_{\alpha \beta}=\xi v_{\alpha}.   
\end{equation}
The physical implication is that the active Fredericks transition is postponed to larger stripe widths. This can be quantified by means of a 
length $\lambda$ which characterizes the crossover between dissipation dominated by viscosity (at smaller length scale) and by friction against the boundaries (at larger length scale). This length is found to be
\begin{equation}
    \lambda= \left( 4 \eta + \gamma (\nu + 1)^2/ \xi \right)^{1/2},
\end{equation}
and the increase of the critical width then satisfies
\begin{equation}
   1/L_c^2=1/L_c^2(\xi = 0) - 1/\lambda^2. 
\end{equation}
In our example, data for the experimental velocity profile $v_y (x)$ indicate $\lambda \approx 40\mu$m, which is comparable to $L_c$. Clearly, viscous-like friction forces are important.

\subsection{Active turbulence}
The spontaneous flows in active matter we have encountered so far could be described by a velocity field and associated streamlines characteristic of a steady laminar regime. Can active flows become turbulent and has this been observed? The answer is positive and this may at first glance seem surprising because turbulence in molecular fluids is a high Reynolds number adventure, with inertial forces dominating over viscous ones. Hence ``active turbulence" must be fundamentally different from inertial turbulence. Understanding this difference is the physical challenge to which we now turn.

Spontaneous chaotic flows at low Reynolds number have been observed in various active fluids, at high activity. These include bacterial suspensions, sperm swarms, suspensions of microtubules and molecular motors, monolayers of tissue cells and suspensions of self-propelling particles. Figure 1 in \cite{AlertCJ} illustrates this panoply of examples, providing experimental images, velocity field maps and energy spectra. The latter differ dramatically from the universal Kolmogorov spectrum $E(q) \propto q^{-5/3}$ so well-known for inertial turbulence (in three dimensions). This should not come as a surprise because the energy driving active turbulence is not externally injected at some (macroscopic) scale by stirring, shaking or shearing, but, instead, is produced internally and autonomously in a self-organized manner, by the (microscopic) particles themselves. What happens to concepts like energy cascade, inertial range and self-similarity, when we go from ``inertial" to ``active"?

Our exploration starts with the observation, in simulation and theory, of a universal scaling $E(q) \propto q^{-1}$, for small $q$, in active nematic (but nonpolar) fluids \cite{AlertNP}. Using the concepts discussed hitherto in our lectures, a hydrodynamic theory of a two-dimensional active nematic fluid is developed, neglecting inertial effects and, for simplicity, neglecting flow alignment coupling ($\nu = 0$). So one arrives at equations describing Stokes flow stirred by a spatio-temporal vorticity source, which depends in part on the active driving. An activity number, $A \equiv (L/L_c)^2$, is defined, which compares the system size $L$ to, essentially, the critical length $L_c$ defined in \eqref{criticalL}. One may suspect that $A$ is the active analogue of the Reynolds number. Increasing $A$ (beyond a critical value) offers room to the Fredericks-like instability, hence spontaneous laminar flow, and active turbulence might set in at some (much) larger $A$. 
\begin{figure}[h!]
\centering
\includegraphics[width=0.4\linewidth]{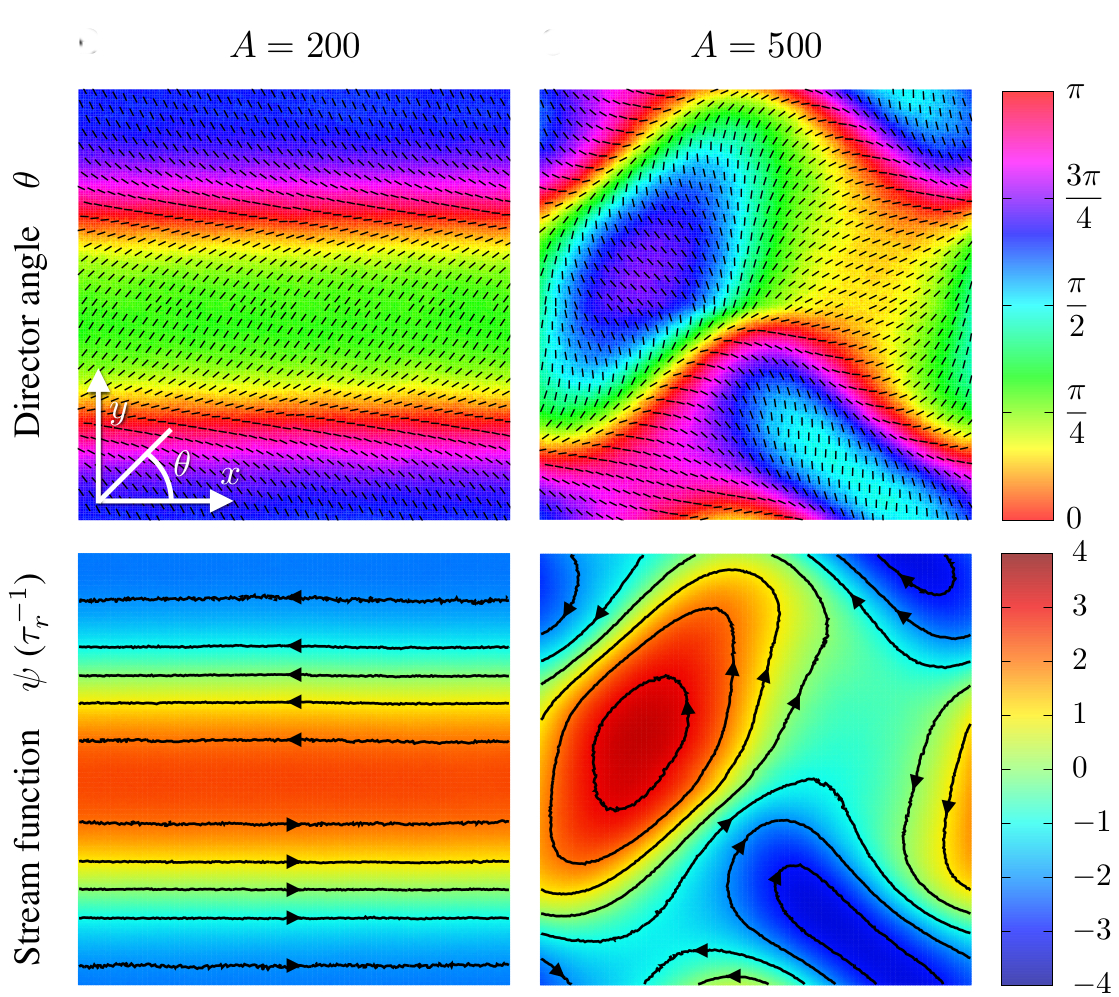}
\caption{Director angles (upper panels) and streamlines (lower panels) in the laminar regime past the spontaneous flow instability. The activity numbers are $A = 200$ (left) and $A=500$ (right). }
\label{statpat}
\end{figure}

\begin{figure}[h!]
\centering
\includegraphics[width=0.4\linewidth]{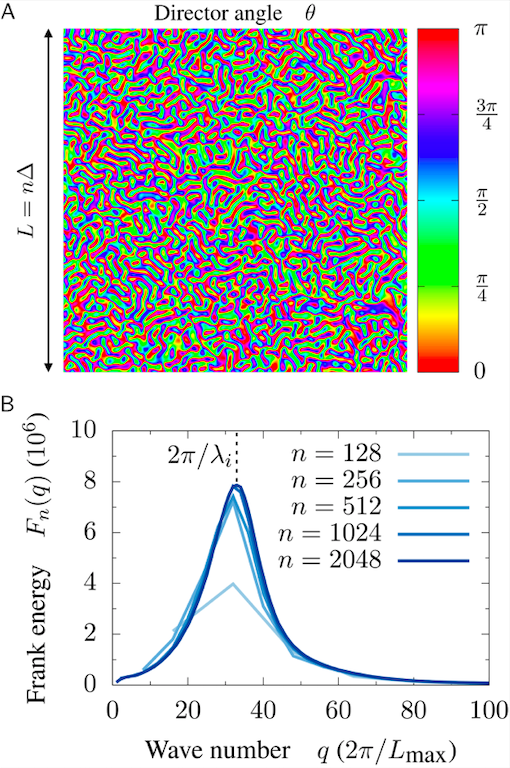}
\caption{Upper panel: Director domains of characteristic linear size of about $L_c$ in the turbulent regime. The activity number is $A \approx 10^5$. Lower panel: Spectrum of the Frank elastic energy, displaying a peak at $q \approx \pi/L_c$, independent of system size $n\propto L$. } \label{turbul}
\end{figure}

The stationary flow patterns in the laminar regime past the instability threshold are illustrated in Fig.\ref{statpat} (left). At higher $A$ the striped pattern develops a zig-zag breaking of translational symmetry, apparent in Fig.\ref{statpat} (right). At still higher $A$ a vortex pattern emerges. Next, for $A \gg 1000$, disorder suggestive of spatio-temporal chaos is observed (see Fig.\ref{turbul}, upper panel). In this regime the Frank elastic energy spectrum features a sharp maximum at a (critical) wavelength independent of system size and selected by the nonlinear dynamics of the director field (Fig.\ref{turbul}, lower panel).

\begin{figure}[h!]
\centering
\includegraphics[width=0.5\linewidth]{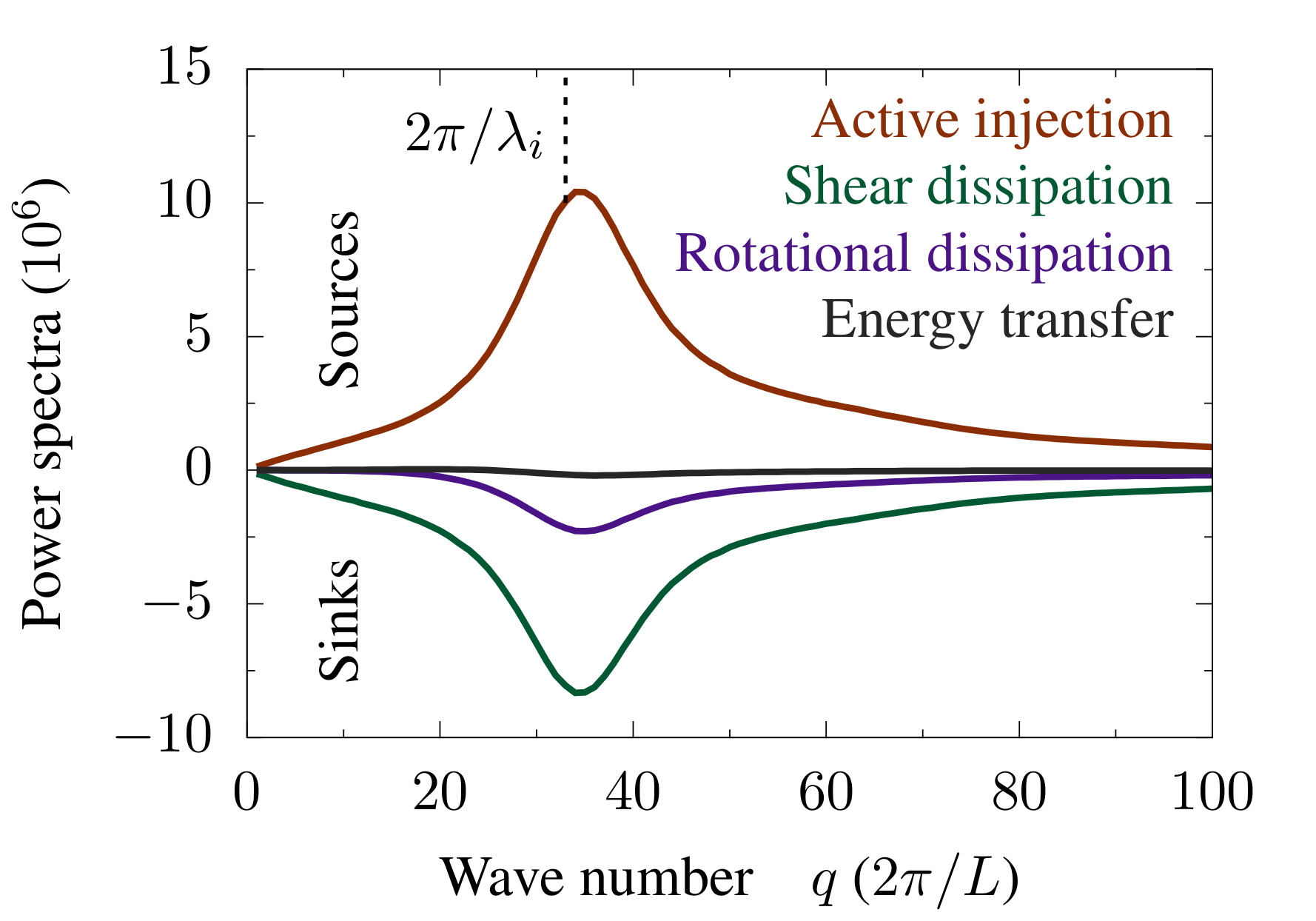}
\caption{Spectra of the four contributions to the energy balance.}  \label{PowerSp}
\end{figure}

Let us examine the energy and power spectra in some more detail. We already mentioned that in linear nonequilibrium thermodynamics the rate of entropy production $\dot S$ can be written in terms of products of fluxes and forces. At constant temperature, $T\dot S = - \dot F$, with $F$ the free energy. For our two-dimensional active nematic fluid, with Frank energy $F_n$ (and negligible kinetic energy at vanishing Reynolds number), one obtains the global energy balance
\begin{equation}
  -\frac{dF_n}{dt} = \int_{\mathcal{A}}
\left[ 2\eta\, v_{\alpha\beta} v_{\alpha\beta} + \frac{1}{\gamma}\, h_\alpha h_\alpha - \zeta\, q_{\alpha\beta} v_{\alpha\beta} \right] d^2\vec{r}  
\label{rateofF}
\end{equation}
Here we identify, in order, the different sources of entropy production, shear viscous dissipation $D_s$, rotational viscous dissipation $D_r$ and active energy injection $I$. Now, since the rate of change of the average energy is zero in a statistically stationary state, the spectra of the quantities in \eqref{rateofF} satisfy
\begin{equation}
    \frac{{dF}_n(q)}{dt} = - D_s(q) - D_r(q) + I(q) + T(q) = 0,
    \label{spectr}
\end{equation}
where $T(q)$, whose integral over reciprocal space vanishes, pertains to the spectrum of the energy transfer between scales. The derivations of \eqref{rateofF} and \eqref{spectr} are given in \cite{AlertNP}. The four spectra, as obtained from simulations (with $\nu =0$) are illustrated in Fig.\ref{PowerSp}. The contributions, in absolute value, all peak at the same self-selected wavelength, which is on the order of the critical length discussed above. Clearly, injected energy by the active stress is dissipated at the same scale length scale, in sharp contrast with inertial turbulence for which energy transfer across length scales is paramount. In these simulations of active turbulence, although there is an instability cascade from laminar to turbulent, there is no energy cascade, and one can prove that $T(q) = 0, \forall q$ (for $\nu =0$).

\begin{figure}[h!]
\centering
\includegraphics[width=0.4\linewidth]{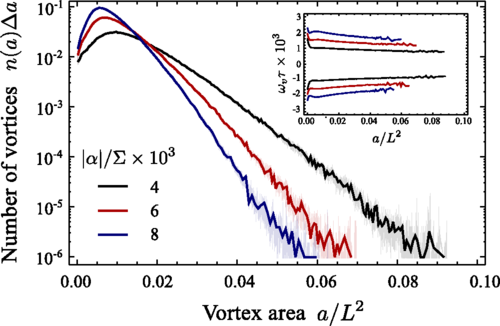}
\caption{Number of vortices $n(a)\Delta a$ (with $\Delta a/L^2= 1.5\times 10^{-5}$) with area between $a$ and $a + \Delta a$, as a function of $a$, in extensile systems. An exponential distribution is apparent in the range $a_{min}<a<L^2$, with $a_{min}$ the area of the smallest active vortex. Inset: The vorticity of a vortex is virtually independent of its area. Reproduced from \cite{Giomi}.}
\label{Gvortex}
\end{figure}

It is interesting to examine the role of vortices in active turbulence in more detail, following the analysis of the mean-field theory and  simulations (in two dimensions) by Giomi \cite{Giomi}. There exists a characteristic area $a^*$ for the vortices, which is featured in the exponentially decaying distribution of their area $a$, $n(a) \propto \exp (- a/a^*)$, where $a^* \approx L_c^2$ (see Fig.\ref{Gvortex}). Not surprisingly, the characteristic linear size of a vortex is on the order of the critical length $L_c$, aka ``active length". 

The vorticity of an individual vortex $i$ is found to be fairly independent of its radius $R_i$ (see Fig.\ref{Gvortex}), so that we can define a vorticity field as follows, 
\begin{equation}
    \omega({\bf r})=\sum_i \omega_0 f(\frac{{\bf r}-{\bf r}_i}{R_i}),
\end{equation}
with $f$ a normalized shape function (e.g., a simple step). The Fourier transform reads
\begin{equation}
    {\tilde \omega}({\bf q})=\sum_i \omega_0 \,R_i^2 \,{\tilde f}(q R_i)\,\exp i{\bf q}\cdot {\bf r}_i 
\end{equation}

 The vorticity is a very useful quantity for determining the statistical properties of the flow. By taking the curl of the force balance \eqref{Stokes} and taking the high-activity limit, we obtain a Poisson equation for $\omega$ with a source term which accounts for the active driving,
 \begin{equation}
     \nabla^2\omega  = \frac{\zeta \Delta \mu}{\eta} \left[\frac{1}{2} \left[\partial_x^2 - \partial_y^2\right] \sin 2\phi - \partial_{xy}^2 \cos 2\phi \right]
     \label{Poisson}
 \end{equation}
 where $\phi$ is the angle between the director and the $x$ direction.
 The prefactor here is a characteristic time scale which is referred to as ``active time", 
 \begin{equation}
  \tau_a \equiv \eta / (|\zeta\Delta \mu|). 
  \label{taua}
 \end{equation}
 Now, on length scales large compared to $L_c$ director correlations (angle correlations) are short ranged. However, the propagator (the Green function) of the Laplacian in \eqref{Poisson} is long ranged ($\propto \log r$ in two dimensions), expressing the long-range character of hydrodynamic interactions. So, short-range orientational correlations can cause long-range correlations in the flow.

\begin{figure}[h!]
\centering
\includegraphics[width=0.7\linewidth]{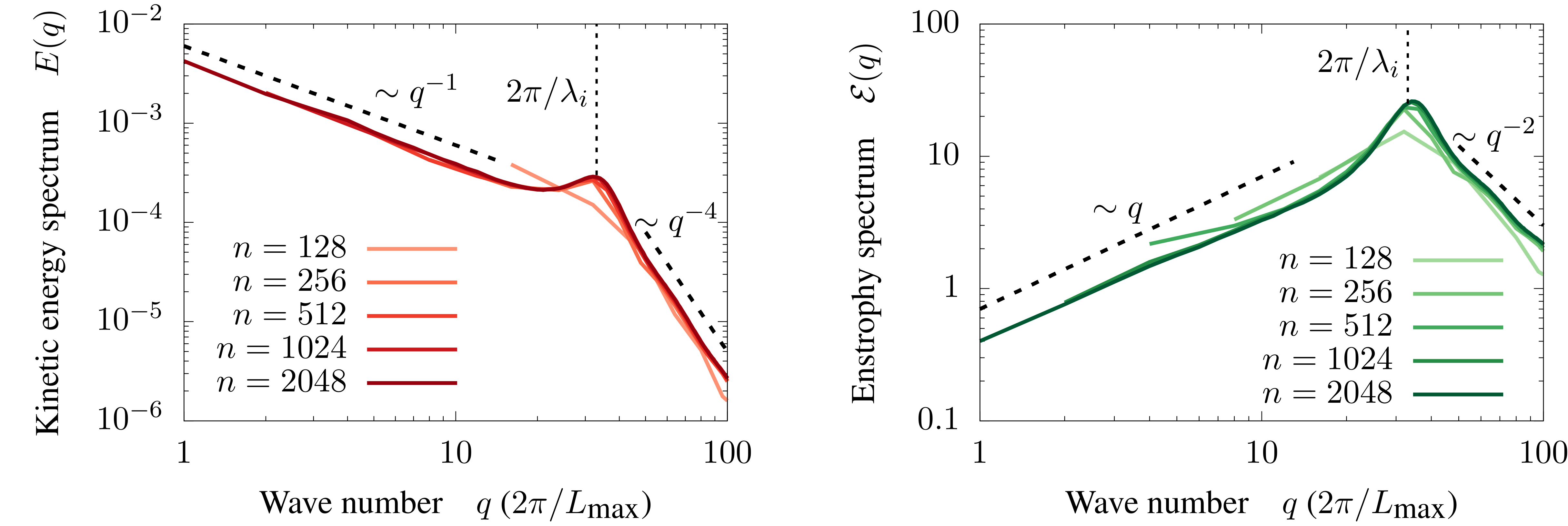}
\caption{Kinetic energy spectrum (left) and enstrophy spectrum (right). }
\label{EEPower}
\end{figure}

Assuming that the absolute value of the vorticity $\omega$ is uniform and equals $\omega_0$, and neglecting vortex-vortex correlations (mean-field approximation for independent vortices), one can calculate \cite{Giomi} the spectrum of the {\it enstrophy} $\Omega$, defined as
\begin{equation}
    \Omega=\frac 1 2 \int d{\bf r} \,\omega ({\bf r})^2, 
\end{equation}
by taking the average of the enstrophy over the vortex area distribution, 
\begin{equation}
    \langle \Omega \rangle \propto \int dq \, {\tilde \Omega(q)} 
\end{equation}
This spectrum, $\tilde \Omega(q)$, given by
\begin{equation}
    {\tilde \Omega}( q) \propto \frac{N \omega_0^2}{q a^*} \int dr \, r^3 \,J_1^2(qr) \exp - (\pi r^2/{a^*}),
\end{equation}
features two scaling regimes. At small wave number, $qL_c \ll 1$, we encounter $\tilde \Omega(q) \propto q$ and at large wave number, $qL_c \gg 1$, we find $\tilde \Omega(q) \propto q^{-2}$ (see Fig.\ref{EEPower}, right panel).

Next, let us examine the kinetic energy power spectrum. The kinetic energy per unit mass density is 
\begin{equation}
  E=\frac{ 1}{ 2} \int d{\bf r} \,{\bf v}({\bf r})^2  
\end{equation}
Note that this $E$ does not correspond to the energy stored in the system, since our system is practically at zero Reynolds number. In fact, upon lowering the Reynolds number the entire kinetic energy
spectrum decreases \cite{MPrat}. Averaged over the vortex area distribution, we obtain the kinetic energy spectrum through
\begin{equation}
    \langle E \rangle = \int dq \, {\tilde E}(q),
\end{equation}  
The $\tilde E(q)$ are related to the velocity Fourier transform and the enstrophy spectrum in the manner,
\begin{equation}
  {\tilde E}(q) = 2\pi q \vert {\tilde {\bf v}}(q) \vert ^2  \propto {\tilde \Omega}( q)/q^2,
\end{equation}
implying the following scaling regimes: at 
    small wave number, $qL_c \ll 1$, ${\tilde E} \propto q^{-1}$, and 
   at large wave number, $qL_c \gg 1$, ${\tilde E} \propto q^{-4}$ (see Fig.\ref{EEPower}, left panel). This is all for two dimensions. In three dimensions the predicted exponents go up by 1, to 0 and $-$3, respectively.

From $\tilde E(q)$ one can obtain \cite{Giomi} the velocity correlation function
\begin{equation}
\langle {\bf v}({\bf 0})\cdot {\bf v}({\bf r}) \rangle \propto -\frac{\zeta\Delta \mu \, \kappa}{\eta^2} \frac{\log r/L}{\log L/L_c} = - \left (\frac{L_c}{\tau_a}\right )^2  \frac{\log r/L}{\log L/L_c}.
\label{correl}
\end{equation}
Expression \eqref{correl}  was derived for $\nu = 0$ and viscosity ratio $\eta/\gamma \approx 1$.

 \begin{figure}[h!]
\centering
\includegraphics[width=0.7\linewidth]{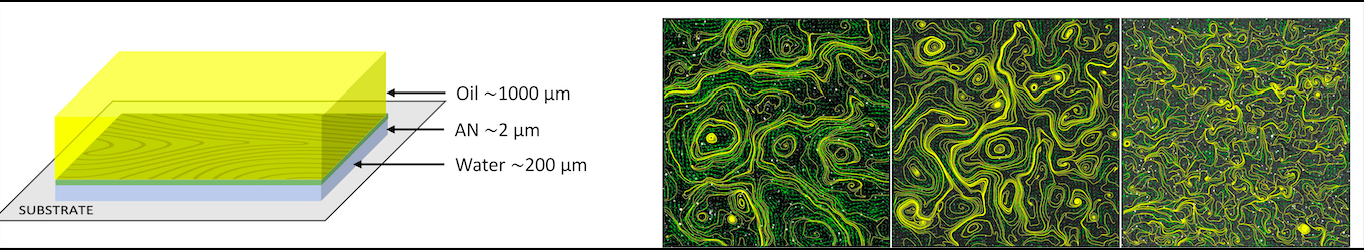}
\caption{Left: Schematic of the experimental system: a quasi-two-dimensional active nematic kinesin-microtubule film, AN, between a water layer and an oil layer. Right: representative microtubule fluorescence micrograph, in order, from left to right, of increasing oil viscosity. }
\label{SubstrFric}
\end{figure}

We have seen that two-dimensional active nematic turbulence is predicted to obey scaling with universal exponents. Experimental verification is complicated by boundary conditions pertaining to the overall three-dimensional nature of the system. One such boundary condition is frictional interaction with substrates, or with external fluid layers. In a recent experiment an active nematic (denoted by AN) was observed between a water (bottom) and oil layer (top), as depicted in Fig.\ref{SubstrFric}, and its kinetic energy spectrum was measured \cite{MPrat}. 
\begin{figure}[h!]
\centering
\includegraphics[width=0.4\linewidth]{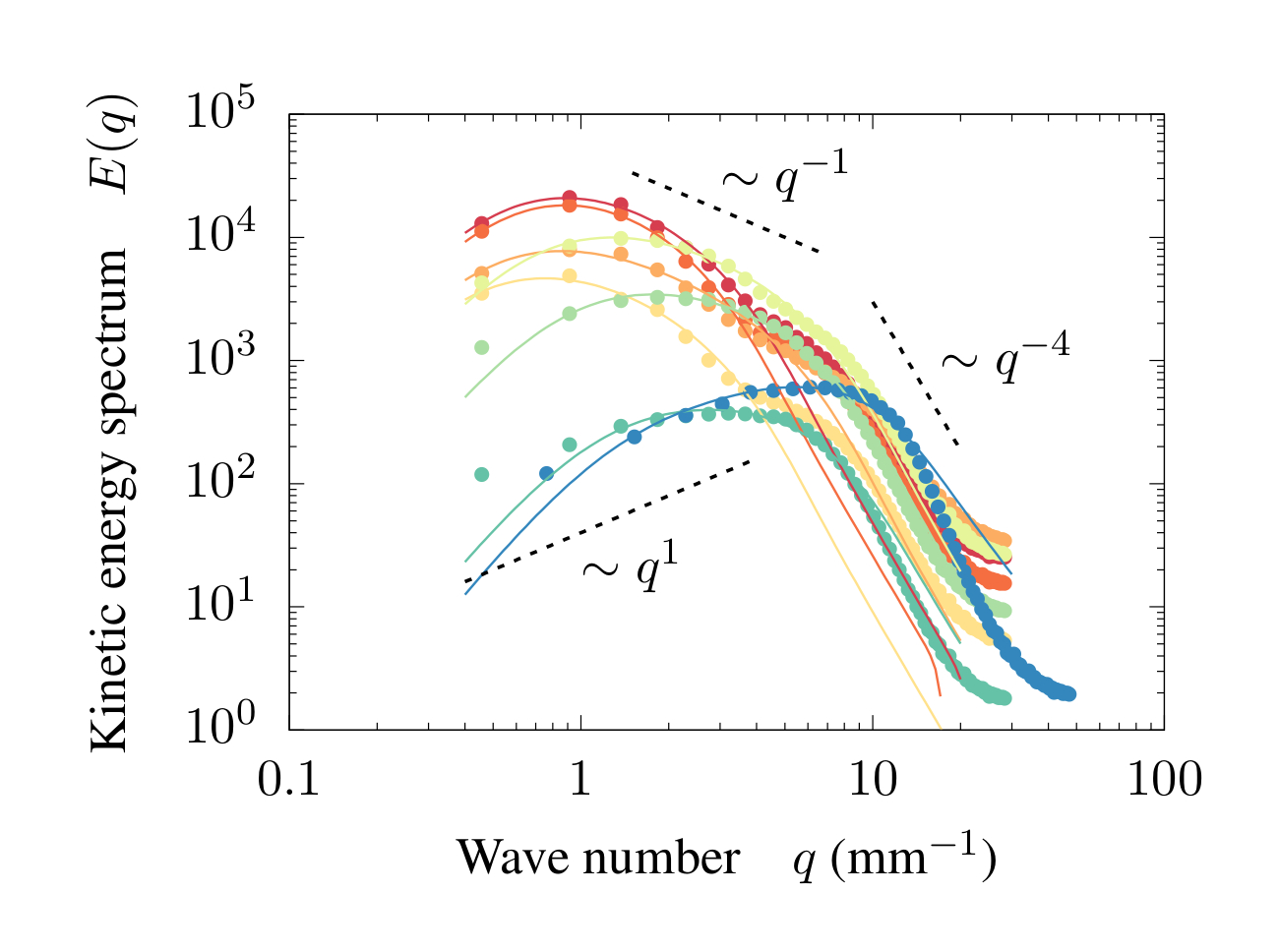}
\caption{Oil viscosity modifies the kinetic energy spectrum of active nematic turbulence. Kinetic energy spectra of turbulent flows
in an active nematic film in contact with an oil layer of variable viscosity. At high viscosity, the new regime $\tilde E(q) \propto q$, for small $q$, is more pronounced (lower curves; blue color). }
\label{EFricPower}
\end{figure}

The data, displayed in Fig.\ref{EFricPower}, confirm the hitherto identified scaling laws for small and large wave number. However, at very small wave number a new scaling regime $\tilde E(q) \propto q$ sets in. This can be understood in terms of a crossover to a regime in which the dissipation is dominated by friction against the oil layer. This crossover can be tuned by varying the oil viscosity experimentally.

The dominant contributions (the effect of the water layer is relatively small) to the kinetic energy spectrum at small wave number are contained in

\begin{equation}
    {\tilde E}(q) \propto \left (\frac{\zeta \Delta \mu}{\eta}\right)^2 \frac{q^3}{\left(q^2 +(\eta_{oil}/\eta ) \tanh qH_{oil} \right)^2}\,,
    \label{crossoverF}
\end{equation}
where $\eta$ is the viscosity of the active nematic, $\eta_{oil}$ that of oil, and $H_{oil}$ the thickness of the oil layer. The crossover, for $q \rightarrow 0$, from a $q^{-1}$ to a $q$-dependence is conspicuous in \eqref{crossoverF}. This prediction of how the spectrum is affected by external dissipation, is verified experimentally (Fig.\ref{EFricPower}).

\section{Cell division and homeostatic pressure}
Biological tissues are an excellent non-equilibrium arena for showcasing active matter physics. The main ideas and tools outlined in our previous Chapter can be applied here. At a microscopic level, in a tissue energy is consumed locally in each cell. Cell division implies growth and programmed cell death (``apoptosis") causes tissue to shrink. Consequently,  a tissue can pull or push on boundaries, quite similarly to what a muscle would do, exerting an extra pressure on them that can be of either sign. The cells can also flow past one another like volume elements in a fluid. At large length scales compared to the cell size and at long times compared to the cell division period, a hydrodynamical description is appropriate, featuring constitutive equations, stress tensor and shear rate tensor. 

\subsection{Homeostatic pressure}

Defining a tissue growth rate $k \equiv k_d - k_a$, with $k_d$ the division rate and $k_a$ the rate of apoptosis, which depend on the biochemical state of the cell and on the mechanical environment, we can write the continuity equation for the cell number density $\rho$,
\begin{equation}
    \frac{\partial \rho}{\partial t}+ \nabla \cdot (\rho {\bf v})=
(k_d(\rho)-k_a(\rho))\rho,
\label{continuity}
\end{equation}
in a mean-field approximation (after averaging over noise in a stochastic process description).

\begin{figure}[h!]
\centering
\includegraphics[width=0.5\linewidth]{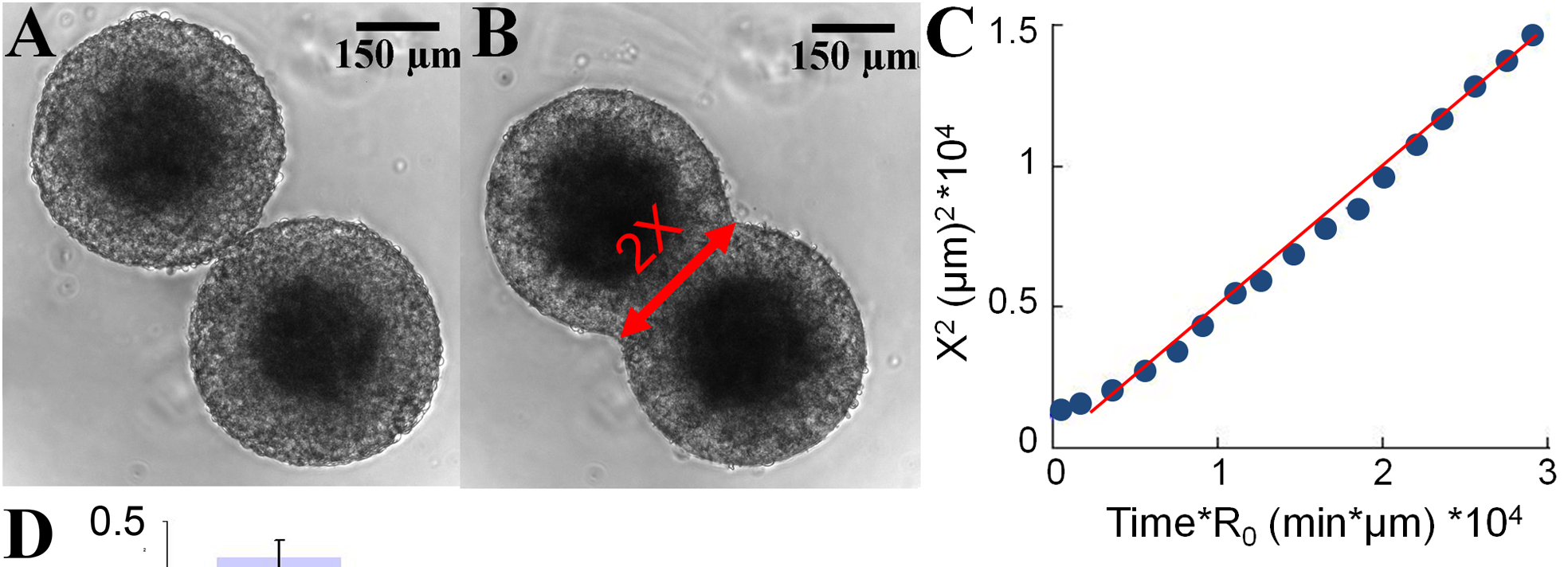}
\caption{Fusion of cells driven by surface tension and resisted by viscosity. (A-B) Images of two fusing embryonic cell aggregates, shown in red is the neck diameter $2X$. (C) Plot of $X^2$ versus $t\times R_0$ (time $\times$ aggregate initial radius). Blue points represent experimental data and the red line is a linear fit the slope of which defines the visco-capillary velocity $v_p \equiv \sigma/\eta$, with $\sigma$ the tissue surface tension and $\eta$ the tissue viscosity. Reproduced from \cite{Mgharbel}. }
\label{2Cells}
\end{figure}

Tissue organization and movement is {\it inter alia} governed by fluid mechanical parameters, such as surface tension and viscosity that are experimentally measurable (see Fig.\ref{2Cells}). If we consider the cell density as a constant, a tissue flows as an incompressible fluid, but in the presence of sources (by division) and sinks (by apoptosis or fusion), so 
\begin{equation}
    \nabla \cdot {\bf v}= k_d - k_a
    \label{incompk}
\end{equation}

\begin{figure}[h!]
\centering
\includegraphics[width=0.5\linewidth]{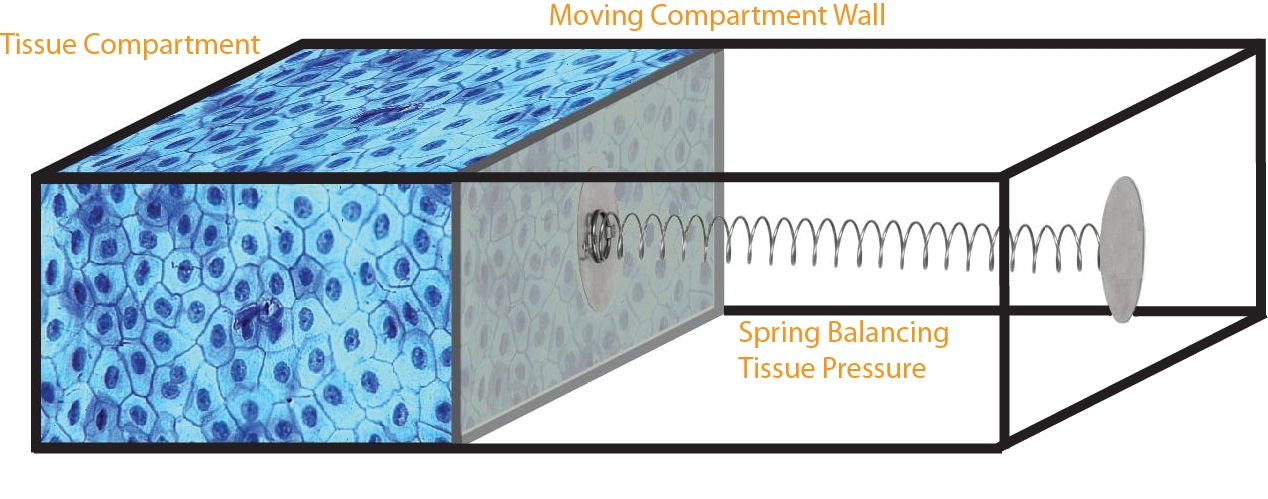}
\caption{Mechanical analogy for defining cell pressure and homeostatic pressure. Reproduced from \cite{Basan}. }
\label{tissuecompart}
\end{figure}

The ``quasi-muscular" force per unit area exerted by tissue on confining walls, over and above the hydrostatic pressure of the extra-cellular or interstitial environment, is the cell pressure $P$ due to cell deformation. The following thought experiment, illustrated in Fig.\ref{tissuecompart}, clarifies its meaning. he tissue is confined by semi-permeable walls, which let molecular liquids and nutrient pass, and exerts a force on the piston, balanced by a spring. When the cell number increases (decreases) one can envisage the extension of the spring to decrease (increase). Now, the rates $k_d$ and $k_a$ depend on the cell pressure $P$ (through their dependence on $\rho$) and a steady state is reached when $P$ equals the {\it homeostatic pressure} $P_h$, for which
\begin{equation}
   k(P_h) = k_d (P_h) -k_a (P_h) =0, 
\end{equation}
and the spring is in mechanical equilibrium. Consequently, the tissue grows for $P > P_h$ and shrinks for $P < P_h$. Like cell number, also $P$ is a stochastic variable which fluctuates in time.

\subsection{Fluidization of tissues by cell division}
Each dividing (or dying) cell exerts forces on its environment, due to its growth (or collapse). Before, during and after the process, the cell is in mechanical equilibrium with its surrounding, so that the total force exerted by the cell, when initially zero, remains zero. The cell exerts thus a force dipole on the surrounding tissue, composed of two opposite forces, ${\bf f}$ and -${\bf f}$, separated by a displacement ${\bf d}$. Mathematically this force dipole $m$ is a symmetric tensor, $m_{\alpha \beta}=f_{\alpha} d_{\beta}$, which can conveniently be separated into an isotropic part and a traceless remainder. Note that the tensor $m$ must be symmetric because there can exist no torques; this means that the forces and the vector between the points where the forces are applied, are aligned. For example, when the direction of cell division is along $x$,
    \begin{equation}
   m_{\alpha \beta}=
  \begin{pmatrix}
  m & 0 & 0\\
  0 & 0 & 0\\
   0 & 0 & 0\\
  \end{pmatrix}= \frac m 3 \delta_{\alpha \beta}+ 
  \begin{pmatrix}
  2m /3& 0 & 0\\
  0 & -m/3 & 0\\
   0 & 0 & -m/3
  \end{pmatrix}\equiv \frac m 3 \delta_{\alpha \beta}+\tilde m\,q_{\alpha \beta},
\end{equation}
using the traceless local orientation tensor $q$ defined in \eqref{nematic tensor}. Note: in this example $\tilde m=m$. 

The total local stress in the tissue, which is featured in the force balance equation, has an elastic component and an internal one due to division and apoptosis. The internal component is proportional to the force dipole density. Both components are also conveniently separated into an isotropic and traceless part,
\begin{equation}
    \sigma_{\alpha \beta}= \sigma \delta_{\alpha \beta} +{\tilde \sigma}_{\alpha \beta}, \;\mbox{with} \;{\tilde \sigma}_{\alpha \alpha}=0
\end{equation}
We first discuss the traceless components. For the time derivative of the traceless part of the internal stress, ${\tilde \sigma}^i_{\alpha \beta} $, we suppose that it has two contributions (from cell division and apoptosis) and that each is proportional to the traceless part of the force dipole density and the rate of the process. This leads to the following dynamical equation,
\begin{equation}
    \frac{d{\tilde \sigma}^i_{\alpha \beta} }{dt}= -\rho \,q_{\alpha \beta}(k_d {\tilde m}_d+k_a {\tilde m}_a)
    \label{internevol}
\end{equation}
Note that during division the cell pushes on its environment (positive force dipole), so that we must have $m_d = \tilde m_d >0$, which is consistent with the sign in \eqref{internevol}. Also note that cell death typically is isotropic so that $\tilde m_a = 0$. 

Interestingly, when an external stress is applied to the tissue, the orientation of cell division is mostly along the principal axis of the (local) stress field \cite{Fink}. It is therefore reasonable to assume the proportionality $q_{\alpha \beta} = {\tilde \sigma}_{\alpha \beta}/\sigma_0$. In the presence of spontaneous orientational order, i.e., $q^0_{\alpha \beta} \neq 0$, one may generalize this coupling between stress and cell division to the linear relationship 
\begin{equation}
    q_{\alpha \beta}=q^0_{\alpha \beta} + \frac{\tilde \sigma_{\alpha \beta}}{\sigma_0}
    \label{stressdivision}
\end{equation}

Let us now proceed and compose the constitutive equation for an isotropic tissue, $q^0_{\alpha \beta} = 0$. For the traceless stress, we already have identified the internal contribution, so let us now add the elastic part. This is given by 
\begin{equation}
    \tilde\sigma_{\alpha\beta}^{el}=2G \tilde u_{\alpha\beta},
\end{equation}
with $G$ the shear modulus and $\tilde u_{\alpha\beta}$ the traceless part of the strain tensor. Together with Eqs. \ref{stressdivision} and \ref{internevol} we can then write down the constitutive equation for the traceless part of the total stress $ {\tilde \sigma}_{\alpha \beta}=\tilde\sigma_{\alpha\beta}^{el}+\tilde\sigma_{\alpha\beta}^i$,
\begin{equation}
    \frac{d\tilde\sigma_{\alpha\beta}}{dt}+\frac{\tilde\sigma_{\alpha\beta}}{\tau_a} =2 G \tilde v_{\alpha\beta}, 
    \label{constitless}
\end{equation}
where $v_{\alpha\beta}$ is the strain-rate tensor of \eqref{vgrad}.
A tissue hence behaves as a Maxwell viscoelastic fluid with relaxation time $\tau_a=\sigma_0/( \rho (k_d {\tilde m}_d+k_a {\tilde m}_a)) \propto 1/k_d$, typically 1 day long, and shear viscosity $\eta\propto G k_d^{-1}$.

Next, we complete our discussion by deriving the constitutive equation for the isotropic stress component. Let us suppose that the tissue has an equation of state of the form $\sigma = \sigma (\rho)$ and compression modulus $K \equiv - \rho \,d\sigma/d\rho$. The total time derivative of the isotropic stress is then
\begin{equation}
    \frac{d \sigma}{dt} = \frac{d \sigma}{d \rho} \frac{d \rho}{dt}= -\frac{K}{\rho} \frac{d \rho}{dt}
\end{equation}
The total time derivative of the cell density is found from the continuity equation \eqref{continuity} and the cell conservation equation then reads
\begin{equation}
    \frac{d \rho}{dt} = \rho (k_d  - k_a ) - \rho v_{\gamma \gamma}
    \label{cellcont}
\end{equation}
For pressures near the homeostatic value, the difference $k_d-k_a$ is small and in linear approximation becomes, given that $\sigma = -P$,
\begin{equation}
    k_d-k_a = (\sigma + P_h)/\bar \eta,
\end{equation}
in which $\bar \eta$ has dimension of viscosity. The constitutive equation describing the pressure relaxation can now be written as
\begin{equation}
    \tau \frac{d {\sigma}}{dt} + {\sigma + P_h} = \bar \eta \, v_{\gamma \gamma},
\end{equation}
with pressure relaxation time $\tau = \bar \eta / K$. In the limit $K \rightarrow \infty$ of an incompressible tissue this reduces to 
\begin{equation}
  \bar \eta \,v_{\gamma \gamma} =    P_h - P.
\end{equation}

When the tissue is composed of anisotropic, e.g., elongated cells, the physics of active stress in fluids with nematic or polar order comes into play. For polar order, cells have a spontaneous ``motility", their velocity vector being proportional to the polarization. Suppose we have a nonzero nematic order parameter $q^0_{\alpha \beta}$ in the absence of external mechanical stress, so that Eq.~\eqref{stressdivision} applies. The constitutive equation \eqref{constitless} is now modified due to the presence of an active stress term coupling to the intrinsic nematic order,
\begin{equation}
    (1 + \tau_{\rm a}(d/dt)){\tilde \sigma_{\alpha\beta}}=2\eta \,{\tilde v_{\alpha\beta}} - \zeta \Delta\mu \,q_{\alpha\beta}^0, 
\end{equation}
where the stress $\sigma_0$ in Eq.~\eqref{stressdivision} is now represented by the active stress $\zeta \Delta\mu = \tau_a  \rho \,(k_d {\tilde m}_d+k_a {\tilde m_a})$. 

Is the active stress in a cellular monolayer contractile or extensile? Single isolated cells generate contractile force dipoles. The net forces pulling on adhesion sites on the substrate constitute a pair of approximately equal and opposite forces acting inwards along the cellular long axis. However, in a monolayer cells can be contractile, or extensile, according to the direction of pushing or pulling forces exerted by surrounding cells or against the substrate. For example, a monolayer of fibroblasts is a contractile system whereas epithelial or neural progenitor monolayers are extensile \cite{Yeomans1}.

\begin{figure}[h!]
\centering
\includegraphics[width=0.7\linewidth]{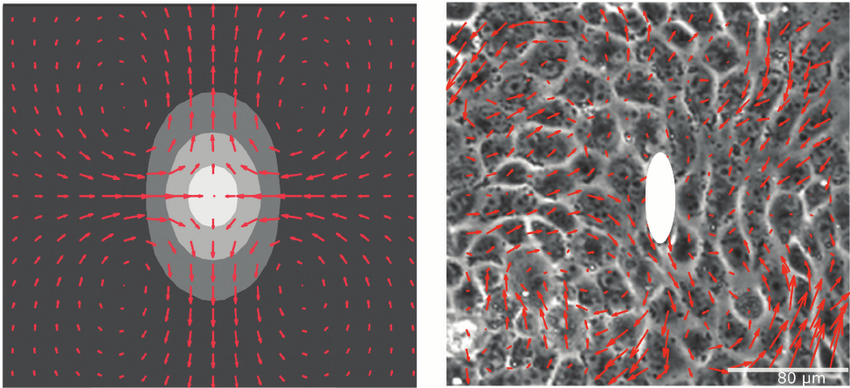}
\caption{Simulated dipole-like flow field due to a cell division event (left) and experimental measurements of the flow field around a dividing cell (right). Red arrows denote velocity fields.  The dividing cell in the experiment is marked by the white ellipse. Reproduced from \cite{Yeomans2}.}
\label{contrac}
\end{figure}

Cell division, which is a source of energy injection and stress generation, modeled as a force dipole, can induce activity, even when intrinsic activity is absent ($\zeta = 0$). It generates extensile active stress that drives dynamical patterns in tissues, and contributes to coordinated motion, e.g., collective migration of cells, as was demonstrated recently \cite{Yeomans2}. Cell division leads to a local dipole-like flow field (see Fig.\ref{contrac}).

Numerical simulations of tissues using Dissipative Particle Dynamics, including cell division and apoptosis, have been performed \cite{Ranft,Basan2}, illustrating the hydrodynamic theory outlined so far. In these simulations intra- and intercell interactions, repulsive at short and attractive at larger distances, are featured, as well as dissipation due to friction, and noise. In this stochastic process, the cell division rate depends on local pressure and density and the death rate is taken constant. In the simulations the shear viscosity is measured of a tissue grown between two walls until it reaches the homeostatic state. The tissue is sheared, as usual, by moving the top wall relative to the bottom wall, while keeping their distance fixed. The shear rate is determined from the measured velocity profile and the stress exerted on the walls is measured.

\begin{figure}[h!]
\centering
\includegraphics[width=0.4\linewidth]{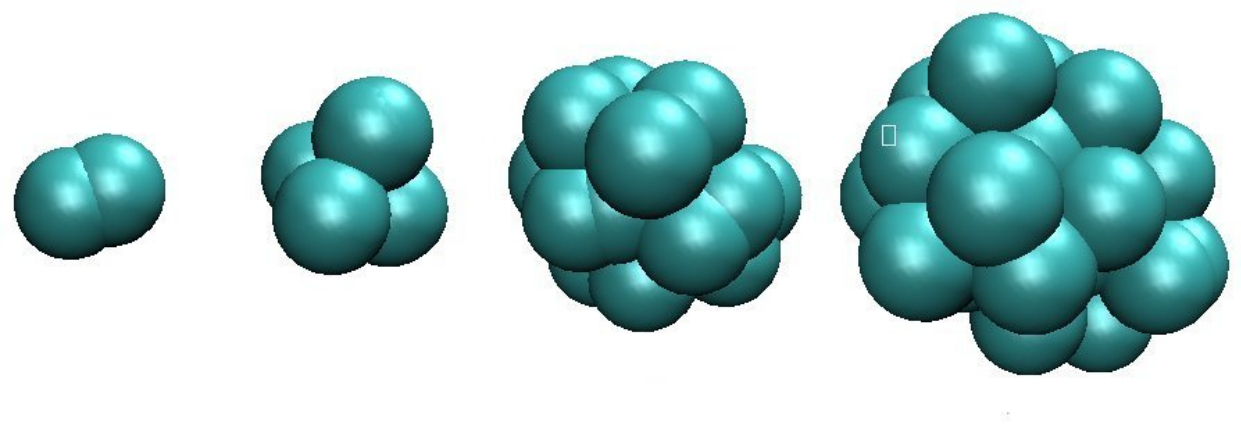}
\caption{Numerical simulation of tissue growth. Shown are the first cell division and the early stages of tissue growth. Each cell is represented by two spheres. Reproduced from \cite{Ranft}.}
\label{celldiv}
\end{figure}
\begin{figure}[h!]
\centering
\includegraphics[width=0.5\linewidth]{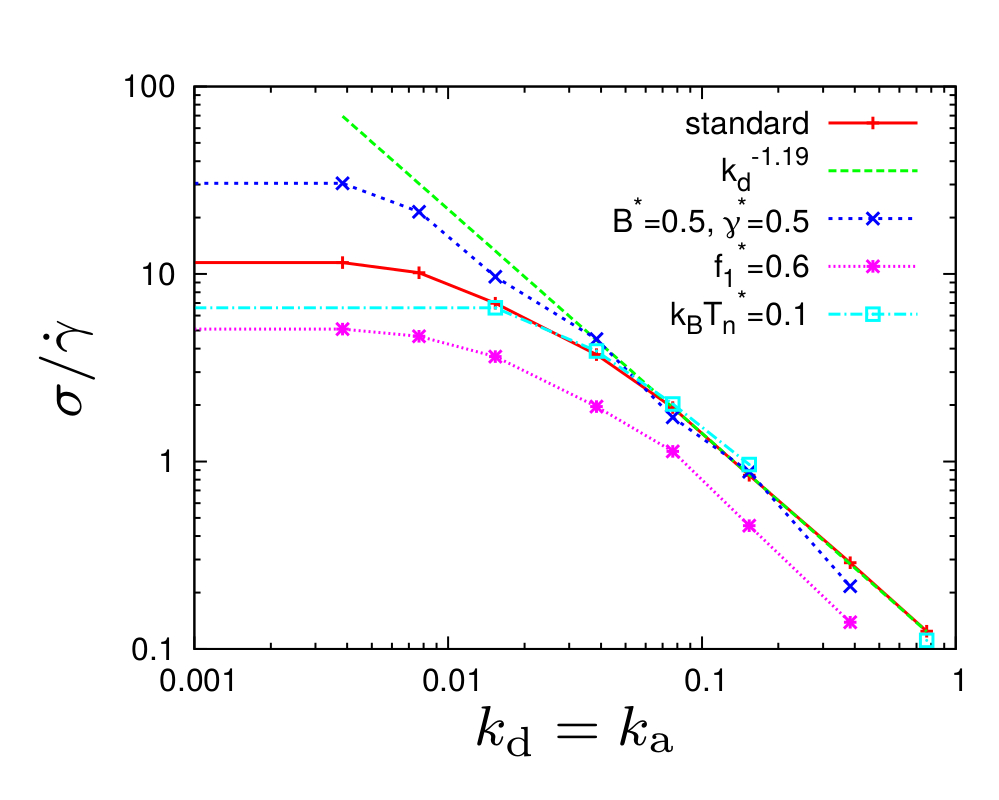}
\caption{Double logarithmic plots of the effective viscosity, i.e., the ratio of the shear stress $\sigma$ and the velocity gradient $\dot\gamma$ versus division rate in simulations of shear flow between plates, for different tissue parameters in the homeostatic state. Reproduced from \cite{Basan2}. }
\label{sigma(k)}
\end{figure}

In Fig.\ref{sigma(k)} the ratio of (a pertinent component of) the shear stress tensor, in our notation $\tilde \sigma_{\alpha\beta}$, and (a pertinent component of) the strain rate or velocity gradient tensor, in our notation $\tilde v_{\alpha\beta}$ is plotted versus division rate $k_d = k_a$. The effective viscosity of the tissue is then defined as $\eta = \tilde \sigma / \tilde v$ in compact notation. For large $k_d$ one observes approximately the theoretically predicted $\eta \propto 1/k_d$ behavior for a Maxwell visco-elastic fluid, so the shear viscosity is inversely proportional to the cell turnover rate. Cell turnover naturally
fluidifies tissues on long timescales, while tissues are elastic (like a solid) on short time scales.

\begin{figure}[h!]
\centering
\includegraphics[width=0.6\linewidth]{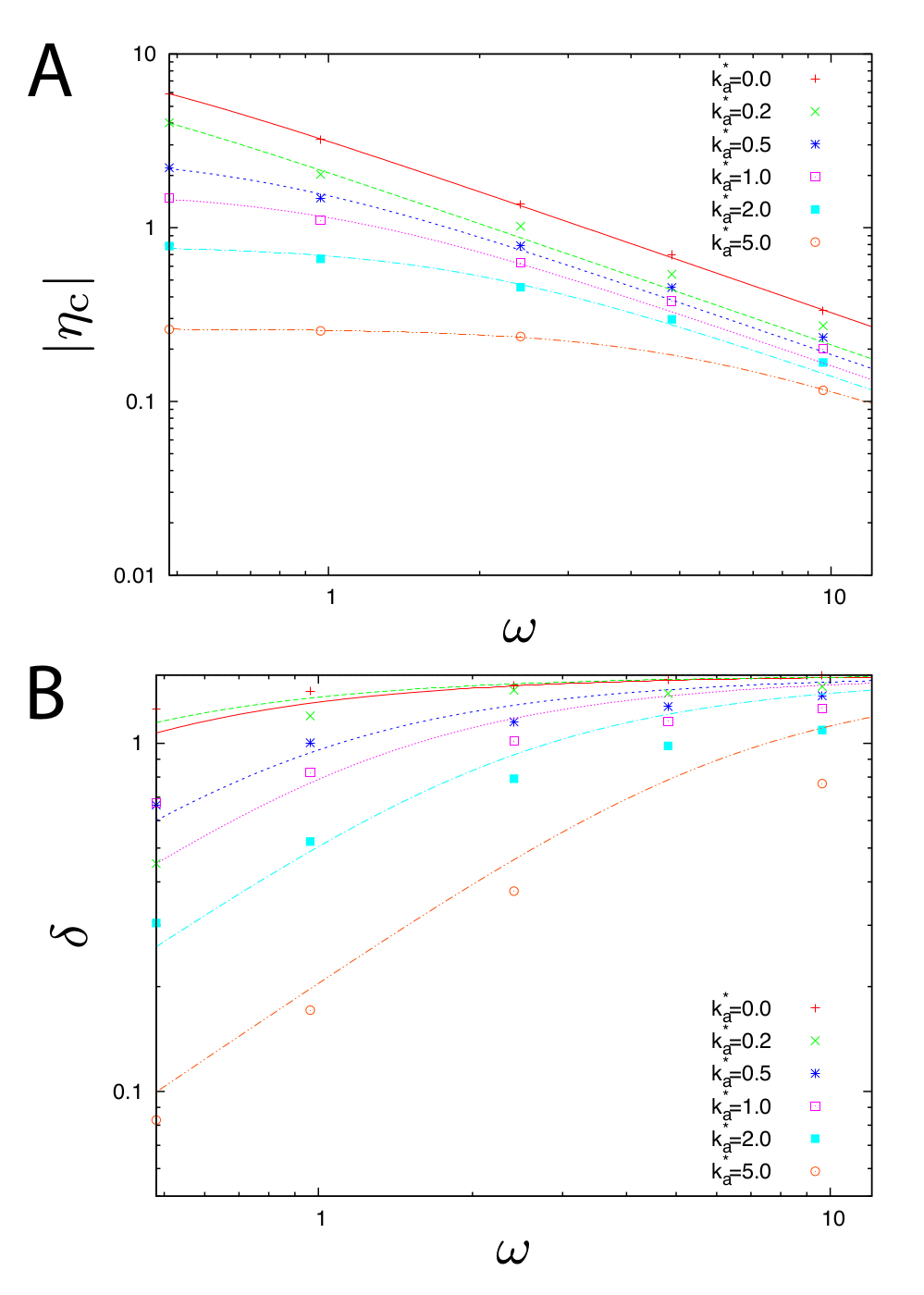}
\caption{Modulus $|\eta_c|$ (panel A) and phase $\delta$ (panel B) of the complex viscosity measured in the oscillatory shear simulation as a function of frequency $\omega$ for various apoptosis rates $k^*_a = k^*_d $. Larger viscosity signifies more solid-like behavior. Phase $\delta = \pi$/2 characterizes a tissue in the elastic limit, and $\delta = 0$ one in the fluid limit. The scale for $\delta$ (in radians) is logarithmic and the value $\delta = \pi$/2 is indicated by a (red) tickmark on the ordinate axis. We again see fluidization with increasing cell turnover rate. Lines are fits to the Maxwell model. Reproduced from \cite{Basan2}.}
\label{eta(omega)}
\end{figure}

To elucidate visco-elastic behavior at different
timescales, oscillatory shear simulations are carried out. An oscillating force is applied and a shear profile for the velocity arises at driving frequency $\omega$. The  complex viscosity $\eta_c$ is measured. 
Its modulus and phase are plotted as a function of
shear frequency $\omega$ and for different apoptosis rates $k_a$ (see Fig.\ref{eta(omega)}). The viscosity (modulus) decreases with increasing shear frequency and the phase $\delta$ increases towards $\pi/2$. In a Stokes fluid the velocity follows the force
(i.e., $\delta = 0$), whereas in an elastic material the displacement
follows the force (i.e., $\delta = \pi/2$). Thus, increasing
the frequency (shortening the time scale) the tissue behaves more and more elastically.

\begin{figure}[h!]
\centering
\includegraphics[width=0.4\linewidth]{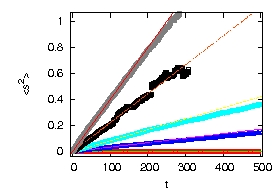}
\caption{Mean squared displacement versus time with typical random walk signature: $\langle s^2 \rangle \propto t$. 
}
\label{msqdis}
\end{figure}

\begin{figure}[h!]
\centering
\includegraphics[width=0.4\linewidth]{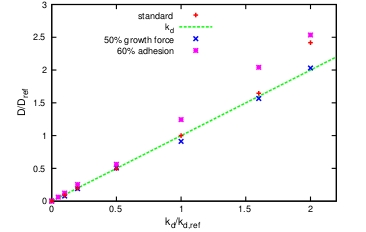}
\caption{Diffusion coefficient of cells $D$ in a tissue simulation. $D$ is essentially linear in $k_d$ in the homeostatic state ($k_d=k_a$). 
Reproduced from \cite{Ranft}.}
\label{diffu}
\end{figure}

Next, we discuss the diffusion coefficient $D$ of cells inside an aggregate. In experiments it has been observed that cells perform a random walk.  In simulations $D$ is found by measuring the mean squared displacement versus time, tracking  individual cells over time until they die and averaging the result on several cells (see Fig.\ref{msqdis}). Importantly, the cell turnover rate (division and apoptosis) affects $D$ a lot. Fig.\ref{diffu} shows that $D$, in the homeostatic state, is proportional to the cell turnover rate. Further, this behavior is not much modified by thermal noise. One may conclude that the diffusion of cells is a random drift due to cell division and apoptosis. 

Let us now draw some attention to the role of fluctuations in the number of cells (in the homeostatic state). Cell division and apoptosis are stochastic processes. What is the effect of associated noise on the mechanical properties of a tissue? 
The cell number balance equation, \eqref{cellcont}, is endowed with division and apoptosis noise in the manner, 
\begin{equation}
    \frac{d\rho}{dt}= \rho (k_d - k_a) - \rho v_{\gamma\gamma} + \xi({\bf r},t),
    \label{cellcontnoise}
\end{equation}
with $\xi$ an additive random noise with zero average and delta-correlated as follows,
\begin{equation}
    \langle \xi({\bf r}, t)\xi({\bf r'}, t') \rangle=(k_a+k_d) \rho \delta (t-t')\delta ({\bf r}-{\bf r'}).
    \label{noisecorr}
\end{equation}
In addition, the constitutive equation \eqref{constitless} is also endowed with a white noise term describing random cell deformations at some effective temperature. Taking together the two noise sources, the diffusion constant can be calculated from the velocity correlation function \cite{Ranft}. The result is that in the homeostatic state, the diffusion constant $D$ is proportional to $k_d=k_a$  for small $k_d$. 

We have so far understood that self-propulsion, on the one hand, and cell division and apoptosis on the other, are capable of fluidizing a confluent cell assembly, which otherwise is arrested in a glassy state.
Our final task in this section is to look in more detail into the nonlinear rheology of tissues, as it emerges in simulations (in two dimensions) of a particle-based model, which incorporates activity in the form of cell division and apoptosis, and in mean-field theory \cite{Barrat}. The mechanical response of cell aggregates under deformation exhibits elastic behavior, elasto-plastic crossover and viscous flow, depending on the applied forces and on the observation time scales. 

\begin{figure}[h!]
\centering
\includegraphics[width=0.4\linewidth]{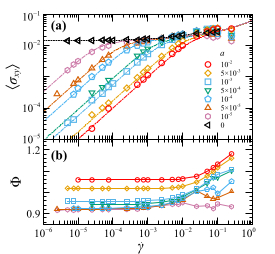}
\caption{Simulation and mean-field theory for passive and active systems.
(a) Steady-state average shear stress $\langle \sigma_{xy}\rangle$ versus the applied external shear rate $\dot\gamma$ for different apoptosis rates $a$. Symbols indicate microscopic simulations results. The dashed line is a Herschel-Bulkley fit for the passive system ($a=0$). The dash-dotted lines are the mean-field model fits. (b) Shear-rate dependence of the corresponding cell packing fraction for the same values of the apoptosis rates. Reproduced from \cite{Barrat}.}
\label{svsdg}
\end{figure}

\begin{figure}[h!]
\centering
\includegraphics[width=0.4\linewidth]{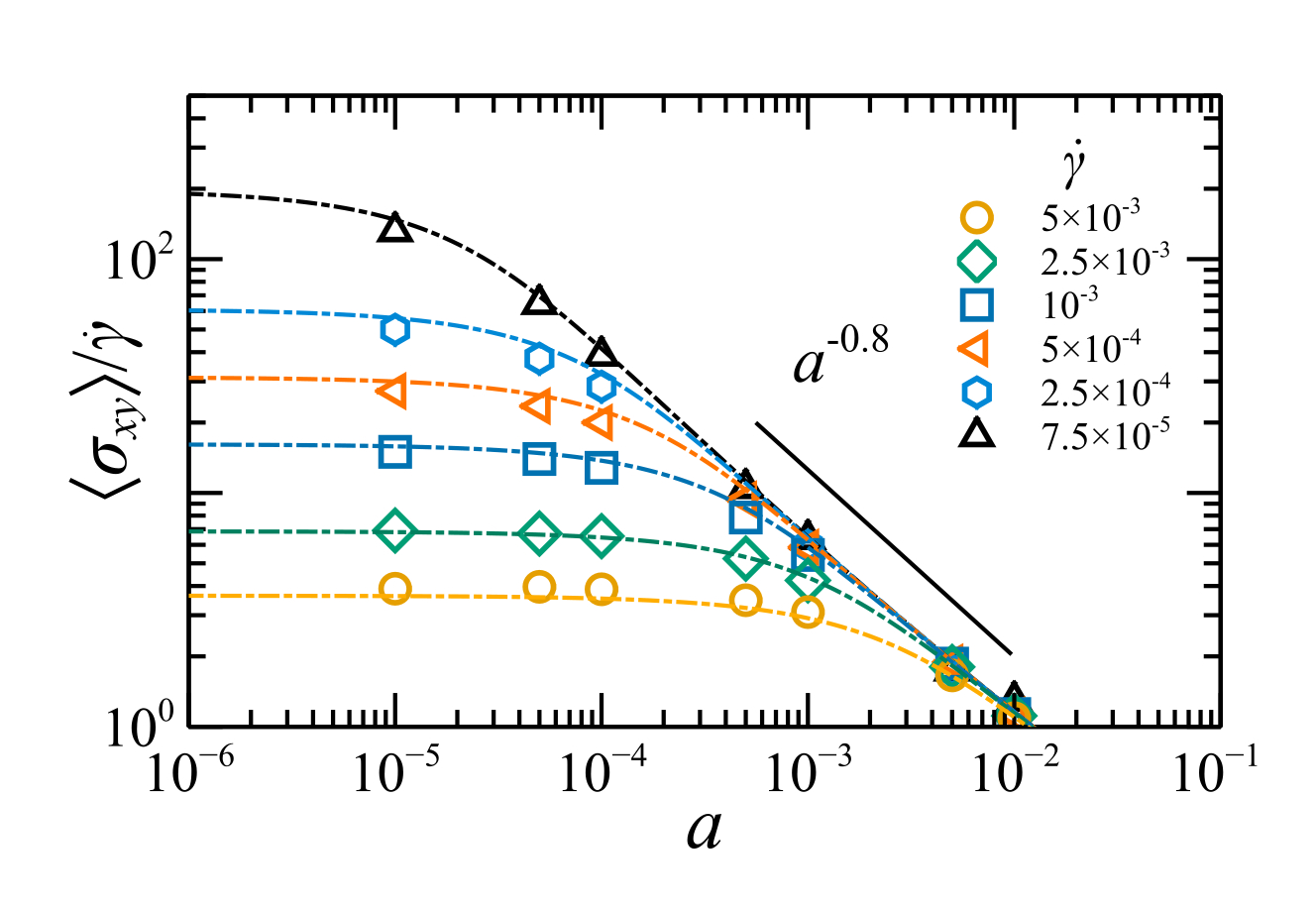}
\caption{Viscosity $\langle \sigma_{xy}\rangle/ \dot\gamma$ as a function of the apoptosis
rate $a$ for different small values of the shear rate (symbols for
the simulation data and dash-dotted lines for the mean-field
theory). Reproduced from \cite{Barrat}. }
\label{vvsa}
\end{figure}

The activity is measured by the apoptosis rate, now denoted by $a$, which, in homeostatic equilibrium, is balanced on average by the configuration dependent local division rate. In the absence of activity ($a = 0$) the system exhibits a nonlinear rheology as observed in foams, with a non-zero yield stress $\sigma_Y$ (plastic behavior). The average shear stress, here denoted by $\langle \sigma_{xy} \rangle $ is then well fitted by the Herschel-Bulkley model, with free parameters $\sigma_Y$, $k_{HB}$ and $n$,
\begin{equation}
   \langle \sigma_{xy}\rangle =\sigma_Y + k_{HB} \;\dot\gamma^n,
\end{equation}
where $\dot \gamma$ here denotes the shear rate. The exponent $n$ is approximately $0.5$. In contrast, a finite activity, $a > 0$, already
prevents the system from having a finite yield stress and a 
linear behavior is encountered at low shear rates, with a viscosity
that decreases when $a$ increases. A remarkable feature is the crossover, at a shear rate denoted by $\dot \gamma ^*$ and controlled by the activity $a$, from activity-driven
fluidization (liquid-like) to a plastic regime with yield stress.  This diversity of behavior is illustrated in Fig.\ref{svsdg} for passive and active systems. The observation that there is no glassy tissue as soon as there is cell division and apoptosis is also well illustrated in Fig.\ref{vvsa}, which shows viscosity versus cell turnover activity. Note that the proportionality $\eta \propto 1/a$ is consistent with the dependence found in Fig.\ref{sigma(k)}, and again demonstrates the concept of fluidization through activity. The figure also shows the variation of the cell density with the shear rate and the cell apoptosis rate in the homeostatic state. At small shear rate, the cell density is independent of shear rate and shows an increase  with the cell apoptosis rate. At larger shear rate, the cell density increases with shear rate. 

\section{Active behavior of tissues}
\subsection{Topological defects in nematic tissues}
In this part of the lectures we deal with tissues composed of elongated cells that organize into nematic domains. We think in particular of spindle-shaped cells that occur in smooth muscles and in some cancer tissues. The main feature about these active cellular nematics is that their ordering displays intrinsic topological defects, i.e., singularities in the orientation field. Those defects are very useful for us because they may be utilised to probe the physical properties \cite{Silber}.

We have studied how activity in active non-cellular nematics can lead to complex chaotic flows. Now, with the cellular system, we ask, experimentally and theoretically, in which regime of activity they operate and what is the role and importance of cell-substrate friction. The non-cellular active systems discussed previously
are in a regime with large activity and viscous
damping dominates cell–substrate friction. In contrast, cell activity is damped a lot by friction against the substrate, and 
the interaction between defects is governed by the 
elastic nematic energy much like in an equilibrium liquid crystal. These cell-based systems thus are more stable than other active nematics, which may be necessary for their biological function.

\begin{figure}[h!]
\centering
\includegraphics[width=0.5\linewidth]{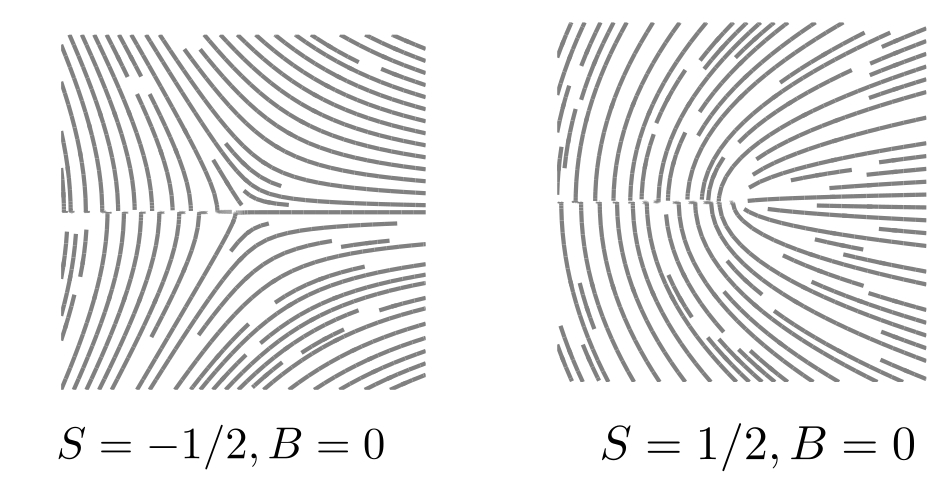}
\caption{Topological defects. (Left) Disclination with topological charge $-1/2$, and (Right) Disclination with topological charge $+1/2$. }
\label{cartoon}
\end{figure}

In the general theory of defects, a topological defect can be characterized by its topological charge.  The topological charge depends on how much the director field
winds around the defect core in one loop. In two dimensions, if we call
$\phi$ the local polar angle of the director, then $\hat {\bf n} = (\cos\phi, \sin\phi)$ and the topological charge $s$ is defined by 
\begin{equation}
    \oint d\phi = 2\pi \;s,
    \label{cont}
\end{equation}
where the integral is along an arbitrary contour encircling the defect core in the $(r,\theta)$-plane. Since, after making a loop around the defect, the director should be in the same direction, the topological charge is an integer or a half-integer number. In passive liquid crystals disclination defects with a charge $s=\pm 1/2$ have the lowest energy and are by far the most commonly observed. We will see below that it is also the case in active tissues and in the following we only study these defects. A cartoon of the director pattern in these defects is given in Fig.\ref{cartoon}.

The nonequilibrium aspect to be appreciated here is that cell activity entails the defects to self-propel, come together and pairwise
annihilate until a dynamically arrested glassy state results, as the cell density increases due to cell proliferation. This frozen configuration consists of perfectly ordered large domains separated by the remaining, trapped, topological defects. Typically these are disclinations, with topological charge $+1/2$ or $-1/2$. In the final steady state the remaining defects are almost $1000\mu$m apart. The following figures \ref{topo} and \ref{spedo} illustrate such domains and defects. 

\begin{figure}[h!]
\centering
\includegraphics[width=0.5\linewidth]{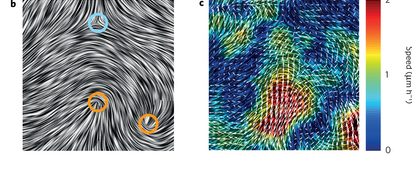}
\caption{Orientation field in a monolayer of proliferating
spindle-shaped NIH 3T3 mouse embryo fibroblasts. (b) Line integral convolution representation showing the orientation field lines. Defects are marked with coloured circles ($-1/2$ = cyan, $+1/2$ = orange). (c) Complex velocity flows develop in the monolayer near the +1/2 defects. Reproduced from \cite{Silber}. }
\label{topo}
\end{figure}

\begin{figure}[h!]
\centering
\includegraphics[width=0.4\linewidth]{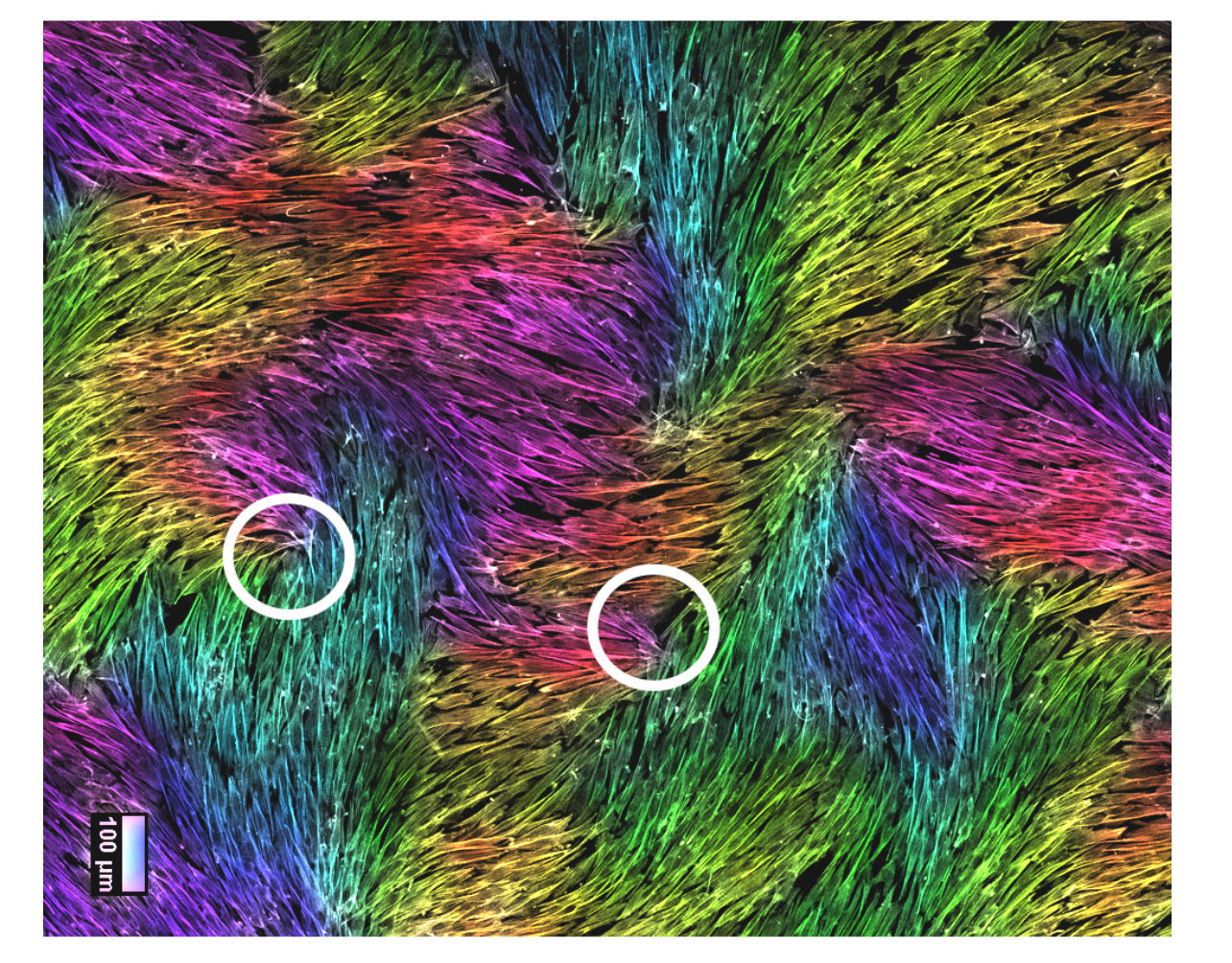}
\caption{C2C12 murine myoblasts cells at confluence self-organize in well-aligned domains between which topological defects position themselves. Reproduced from the Supplementary Material of \cite{Sarkar}. }
\label{spedo}
\end{figure}

\begin{figure}[h!]
\centering
\begin{subfigure}{0.99\textwidth}
\includegraphics[width=0.6\linewidth]{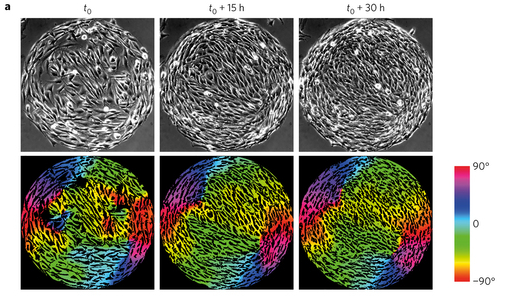}
\end{subfigure}
\begin{subfigure}{0.99\textwidth}
\includegraphics[width=0.7\linewidth]{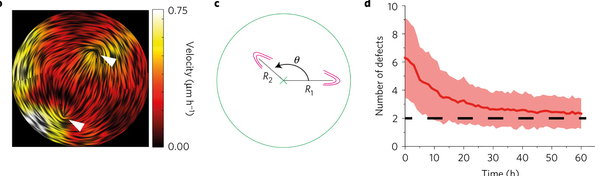}
\end{subfigure}
\caption{Cell tissue confined in a disk. (a) With time, the number of defects decreases and the defects reach a stable position. (b) Identification of the two $+1/2$ defects in the line-integral-convolution representation (white triangles). (c) Sketch of the defect positions and convention. (d) At long times, there remain only two defects in the disk. (Coloured area is the standard deviation, number of disks = 150.) All panels: NIH 3T3 cells, disk radius $R_0 = 350 \mu$m. Reproduced from \cite{Silber}.}
\label{dish}
\end{figure}

It has been observed that $-1/2$ defects do not move significantly, consistently with symmetry arguments. In contrast, the more ``pointed" $+1/2$ defects are mobile and exhibit directed motion with their ``comet tail" forward. This animated behavior is the signature of an active contractile system. (For an extensile system the motion would be ``comet head" forward.) 

For a better control of the defect localization and the associated cellular collective motion, the cells were confined to micropatterned
adhesive disks of various radii. Results for that configuration are shown in Fig.\ref{dish}. The observed defect relaxation evokes the picture outlined above. Several randomly distributed defects occur initially in each domain when the cells touch to form a continuum assembly. Cells near a boundary or a domain wall align tangentially. As the density increases further, the defects annihilate pairwise until two $+1/2$ defects remain.  These position themselves preferentially along a disk diameter, as
expected on the basis of symmetry. 

To understand this behaviour, we minimized the Frank–Oseen
free energy, which, in the one-elastic constant approximation, is given by \eqref{Frank2D}. Circular boundary conditions and a perfect alignment with the boundary are imposed. Activity is ignored, in a first step. Two defects are placed in a passive nematic disk, at polar positions $(r,0)$ and $(r, \theta)$ (see Fig.\ref{dish}c). The positions of the two defects then shift as a result of a balance
between the alignment of the cells at the boundary and the repulsion
between these two defects of identical charge $+1/2$. 

\begin{figure}[h!]
\centering
\includegraphics[width=0.35\linewidth]{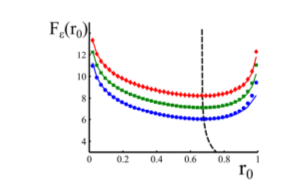}
\caption{Energy of the nematic drop as a function of the defect position,
for three different values of the length cut-off $\epsilon$. The dashed line goes through the minima found upon varying $\epsilon$. }
\label{FO}
\end{figure}

The system
attains minimal energy for $\theta = \pi$ and $r_0  \equiv r/R_0 \approx 0.67$ (see Fig.\ref{FO}). In the calculation a length cut-off $\epsilon$ is introduced so that a small area around a defect is excluded from the integration domain (to avoid a divergence). The energy, of course, depends strongly on this cut-off, but, remarkably, the optimal reduced radial distance $r_0$ does not.
Importantly, this most likely position of the defects is predicted to
be independent of both the activity and the Frank elastic constant,
explaining why the same position was experimentally observed 
for different cell types.

In a second step, an explicit test of the robustness to activity is the following strategy. Since cells are active particles, the defect positions are distributed
about the most probable ones. The widths
of these distributions can be parametrised by an ``effective temperature" $T$, which is defined here to have the same units (of energy) as $\kappa$. Now, fitting the model to experimental data one obtains 
small ratios $ T/ \kappa $, typically $ 0.1 < T/\kappa < 0.2$, independently of the size of the disk.

In sum, we conclude that the activity of these cellular nematics, although responsible for the self-induced motion of the $+1/2$ defects, is negligible because it is damped by cell–substrate friction.

\begin{figure}[h!]
\centering
\includegraphics[width=0.6\linewidth]{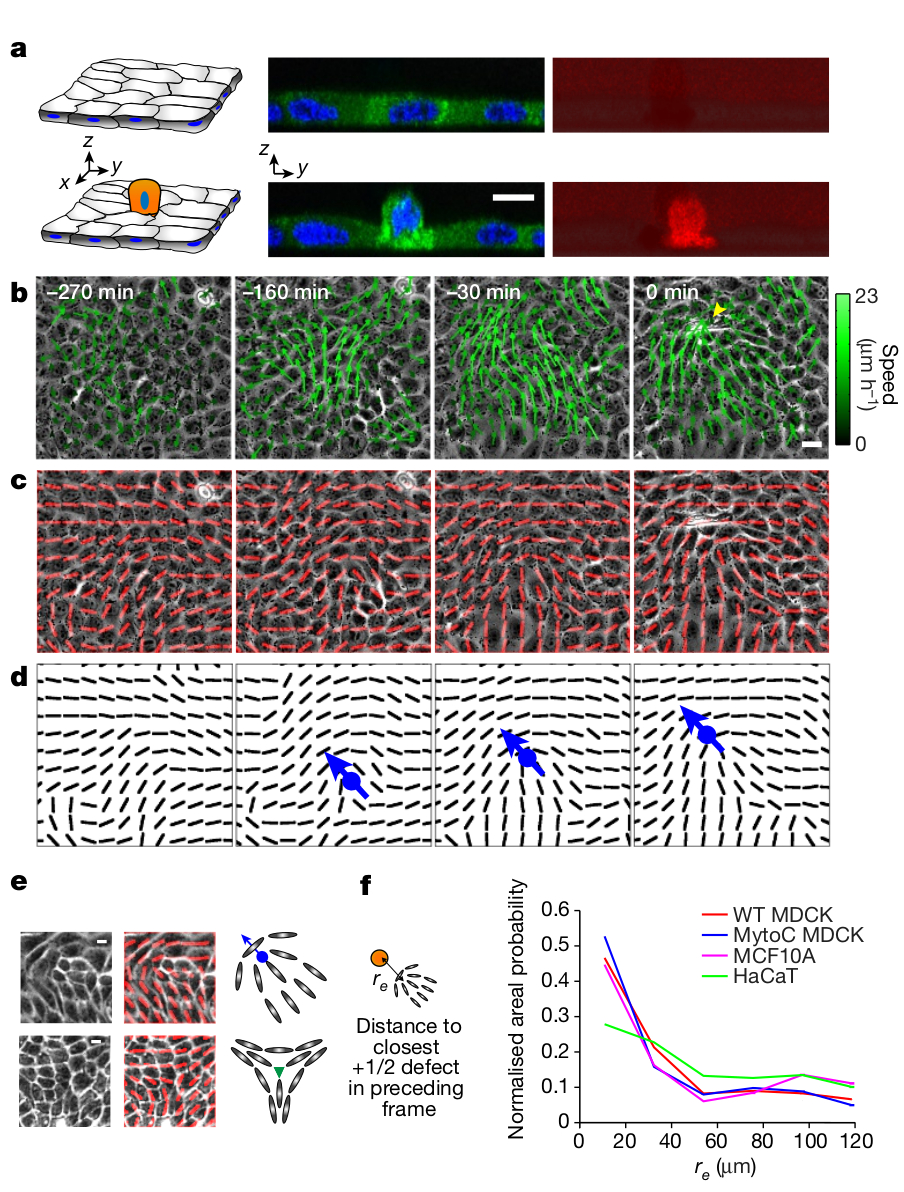}
\caption{Extrusion correlates with defects with a $+1/2$ topological
charge in the epithelia. (a) Left: confluent monolayer (top) and extruding
cell (bottom). Middle: confocal image of confluent MDCK monolayer
(top) and extruding cell (bottom). Right: activation of caspase-3 signal (red): top, monolayer; bottom, extruding cell. (b) Phase-contrast images showing monolayer dynamics before extrusion (yellow arrowhead),
overlaid with velocity field vectors. 
(c), (d) Average local orientation of cells. The group
of cells moving towards the extrusion forms a comet-like configuration
(e) Experimental (left) and schematic (right) images of $+1/2$ defect (top,
comet configuration) and $-1/2$ defect (bottom, triangle configuration).
(f) Correlation between
$+1/2$ defects and extrusions. Scale bars, 10 $\mu$m. Reproduced from \cite{Saw}. }
\label{topex}
\end{figure}

Next we comment on the role that defects in nematic orientational order play in the cell {\it extrusion} process in, e.g., epithelial tissue. Extrusion consists of seamlessly removing unwanted or dying cells while maintaining the integrity of the epithelial barrier. It has been shown that apoptotic cell extrusion is provoked by singularities in cell
alignments precisely in the form of the comet-shaped topological defects that carry charge $+1/2$. Fig.\ref{topex} illustrates this. The epithelium monolayer of cells behaves as an extensile active
nematic, since the flow along the elongated
axes of the cells moves towards the head region of the ``comet", whereas, as we have encountered above, contractile flow is directed towards the tail. Also, unlike in the relaxation process in our previous system, the epithelium is highly active, and defects incessantly and spontaneously form by active stresses in the monolayer. High compressive isotropic stresses occur near the $+1/2$ defects. These induce cell death signals, which in turn trigger extrusion.

\begin{figure}[h!]
\centering
\includegraphics[width=0.6\linewidth]{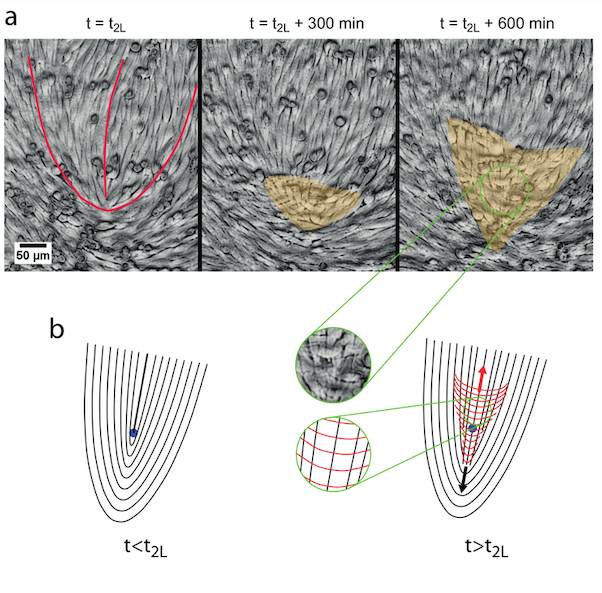}
\caption{Bilayering in C2C12 myoblasts mediated by $+1/2$ defects. (A) Snapshots illustrating the progression of the bilayering (orange region). 
(B) Sketch of the bilayering process. Red lines: orientation field lines of layer 2; Black: those of layer 1. The blue dot shows the initial position of the core. Reproduced from \cite{Sarkar}.}
\label{comet}
\end{figure}

Another interesting role played by topological defects is to provoke {\it multilayering} of cell tissue. For example, in hydrostatic skeletons (in invertebrates, e.g., jellyfish), in which two stacked perpendicular layers of muscle cells first assemble and
then organize with crisscross orientation \cite{Sarkar}. In the experiment initially cells organize in a monolayer, behaving as a contractile active liquid crystal, in which most $+ 1/2$ defects self-propel but, importantly, a fraction of these comet-shaped defects remains stationary. 
At the cores of these defects, perpendicular bilayering occurs by collective migration. The monolayer splits into two stacked layers that 
migrate actively in antiparallel directions, one on top of the other, reminiscent of plate tectonics in geology, but here on a micrometer scale.
Because the cells keep their orientations, top and bottom
cells end up oriented perpendicularly. Fig.\ref{comet} illustrates this process. 

We proceed our discussion with a closer analysis of the motion of  defects.  In free monolayers
of nematic tissue, they are the only observed defects. For a defect at the origin, the director orientation that minimizes the energy \eqref{Frank2D} and satisfies \eqref{cont} is the linear function
\begin{equation}
    \phi = s \,(\theta +\theta_0),
    \label{phitheta}
\end{equation}
where $\theta_0$ us the angle of orientation of the disclination (called $\psi$ in Fig.1a in \cite{Vromans}). The corresponding orientation tensor \eqref{nematic tensor} (for saturated nematic order $S =1$ and $\theta_0 =0$) then reads, e.g., for $s=1/2$,
\begin{equation}
    q=\frac 1 2 \begin{pmatrix}
 \cos\theta & \sin\theta  \\
 \sin\theta & -\cos\theta
      \end{pmatrix} 
      \label{qmatrix}
\end{equation}
The associated defect energy is, 
\begin{equation}
    F_d =\int d{\bf x} \,\frac{\kappa}{2} (\nabla \phi)^2 = \pi s^2 \kappa \log(L/r_c) + {\rm const.},
\end{equation}
where the logarithm comes from integrating the square of the angular component of the gradient over the plane, from the core radius $r_c$ out to the system size $L$. The cutoff $r_c$ represents the tiny distance across which the nematic order $S$ rises from zero (at $r=0$) to unity (at $r=r_c$).  

In a next step, let us consider the active force around a defect. The active force per unit area ${\bf f}$ is the gradient of the active stress \eqref{actstress}, so that
\begin{equation}
    f^a_{\alpha} =- \partial_{\beta}(\zeta \Delta\mu \,q_{\alpha \beta}),
\end{equation}
with components, using \eqref{qmatrix},
\begin{equation}
    f^a_x=-\frac{1}{2}\zeta\Delta\mu \,\left( -\sin\theta \frac{\partial \theta}{\partial x} +\cos\theta \frac{\partial \theta}{\partial y} \right) =-\zeta\Delta\mu \,\frac{{\bf e}_{\theta}\cdot \nabla \theta}{2 }, 
 \end{equation}
and
\begin{equation}
   f^a_y = -\zeta\Delta\mu \,\frac{{\bf e}_r\cdot \nabla \theta}{2 } = 0
\end{equation}
Notice that for a comet-shaped defect oriented along $x$ (recall $\theta_0 = 0$ in \eqref{phitheta}) the force is also along $x$. Furthermore, for contractile stress ($\zeta <0$) the force causes the defect to move in the direction from head to tail, at constant limiting velocity due to viscous dissipation.

\begin{figure}[h!]
\centering
\begin{subfigure}{0.49\textwidth}
\includegraphics[width=0.8\linewidth]{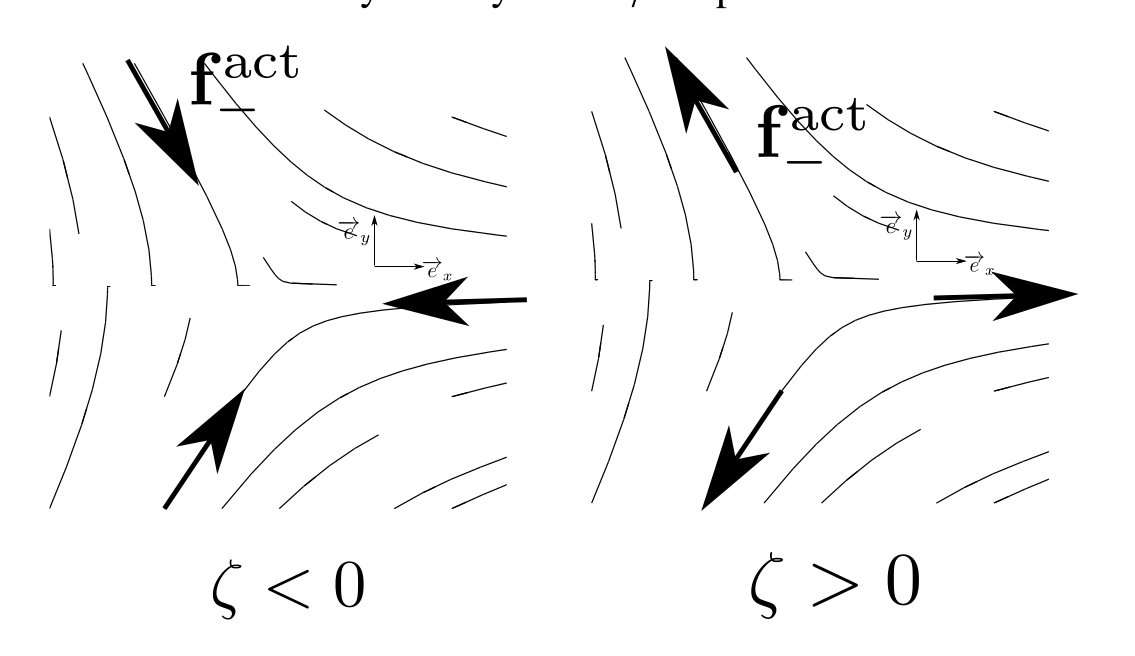}
\end{subfigure}
\begin{subfigure}{0.49\textwidth}
\includegraphics[width=0.8\linewidth]{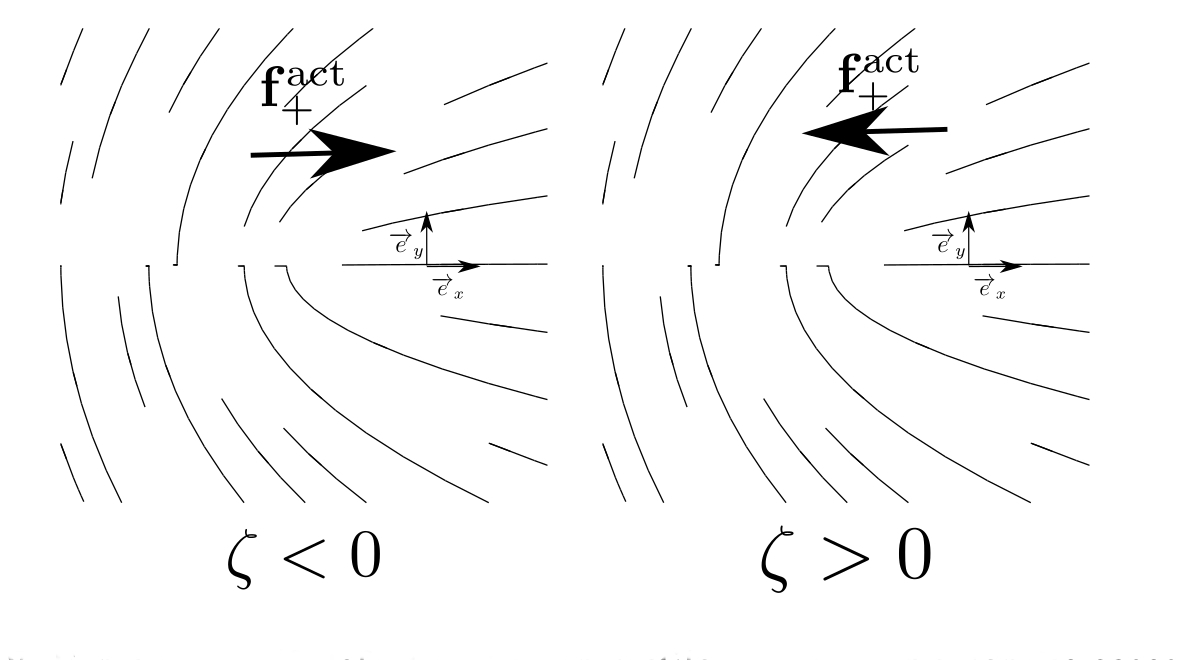}
\end{subfigure}
\caption{Passive orientation of the director field (solid lines)
and resulting active force density (arrows) in contractile ($\zeta < 0$) and extensile cases ($\zeta >0$), for $s=-1/2$ (left) and $s=+1/2$ (right) topological
defects. 
Inspired by \cite{Brezin}.}
\label{triacom}
\end{figure}

What can we say quantitatively about the flow and stress patterns around $s=\pm 1/2$ topological defects in unbounded, active nematics in $d=2$?  Let us recall a couple of facts. We know that the active stress, proportional to $q$, displays strong gradients, and therefore strong active forces act, around defects. These forces induce cell flow. Defects in general are associated with vortices of the flow field. For $s=-1/2$ defects, which are threefold symmetric, there is no preferred direction of motion and these defects are effectively passive and just diffuse. In contrast, $s = +1/2$ defects have polar symmetry and move actively in the direction determined by the sign of $s$ (see Fig.\ref{triacom}).

\begin{figure}[h!]
\centering
\begin{subfigure}{0.49\textwidth}
\includegraphics[width=0.8\linewidth]{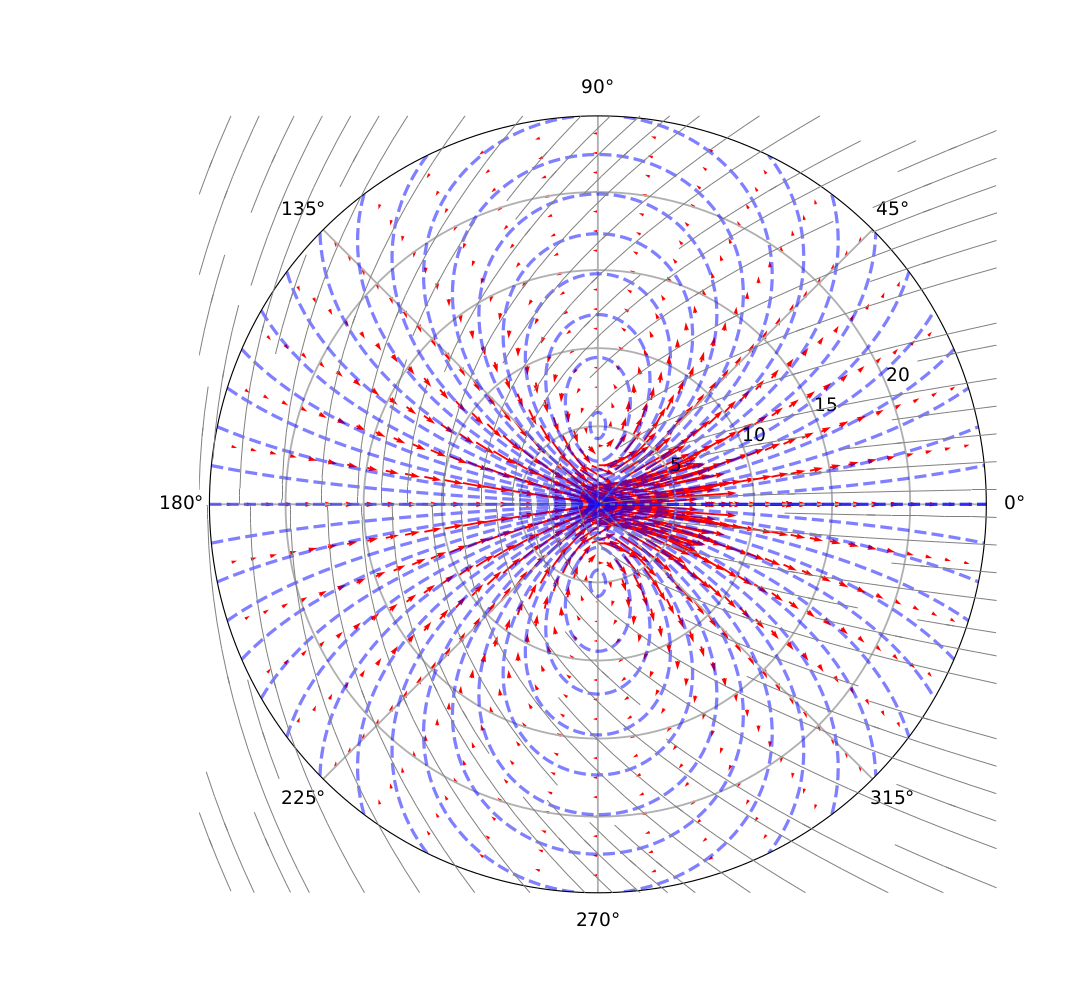}
\end{subfigure}
\begin{subfigure}{0.49\textwidth}
\includegraphics[width=0.8\linewidth]{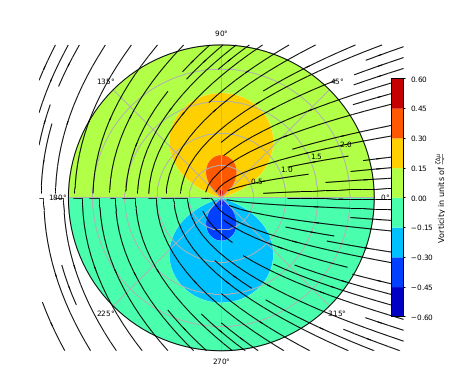}
\end{subfigure}
\caption{ (Left) Flow structure
around an active $+1/2$
topological
defect. The director field
orientation is represented by faint black solid lines (in the background). Blue dashed
lines are the streamlines of the velocity
field. Red, arrows, dashed lines are the velocity vectors. (Right) Colormap of the dimensionless vorticity of
the flow around a $+1/2$ defect. The director field
orientation is represented by thin black lines.  Inspired by \cite{Brezin}.}
\label{cometflo}
\end{figure}

The orientation field \eqref{orienfield} is parallel to ${\bf p}$
in a non-flowing steady state. In a state with a moving $s=+1/2$ defect core it is parallel to the comet axis, so $h_{\perp} = 0$. In the limit of zero rotational viscosity $\gamma$, also $h_{\parallel}$ must vanish \cite{Brezin}. Force balance then entails the limiting self-advection velocity ${\bf v}_0$ of the defect (relative to the substrate),
\begin{equation}
    {\bf v}_0 =- \frac{\pi \zeta \Delta \mu}{8(\xi\eta)^{1/2}}{\bf e}_x,
    \label{selfadvec}
\end{equation}
where $\xi$ is the coefficient of friction (with dimension of viscosity per unit area) against the substrate. The ratio $L_s \equiv \sqrt{\eta/\xi}$ is the hydrodynamic screening which quantifies the relative importance of viscous dissipation and frictional damping. Further, we recognize here the time scale \eqref{taua}. The velocity field ${\bf v}$ (and its vorticity) is illustrated in Fig.\ref{cometflo}, left panel (right panel).

Some active $s=+1/2$ defects, exceptionally, do not move, as for example those that promote multilayering discussed above \cite{Sarkar}. One can compute the force necessary to balance the active force and to stall (or pin) an active topological defect of this kind. A calculation \cite{Brezin} provides the stalling force (or pinning force) ${\bf f}_s = - f_s {\bf e}_x$ on a defect with core radius $r_c$, with
\begin{equation}
    f_s=\left (\frac{\eta}{\xi}\right )^{1/2}\frac{\pi^2 \zeta\Delta\mu}{\log \{ (\frac{\eta}{\xi})^{1/2}/r_c \}}, \;\; \mbox{for} \;\; r_c \ll L_s.
\end{equation}

\begin{figure}[h!]
\centering
\includegraphics[width=0.4\linewidth]{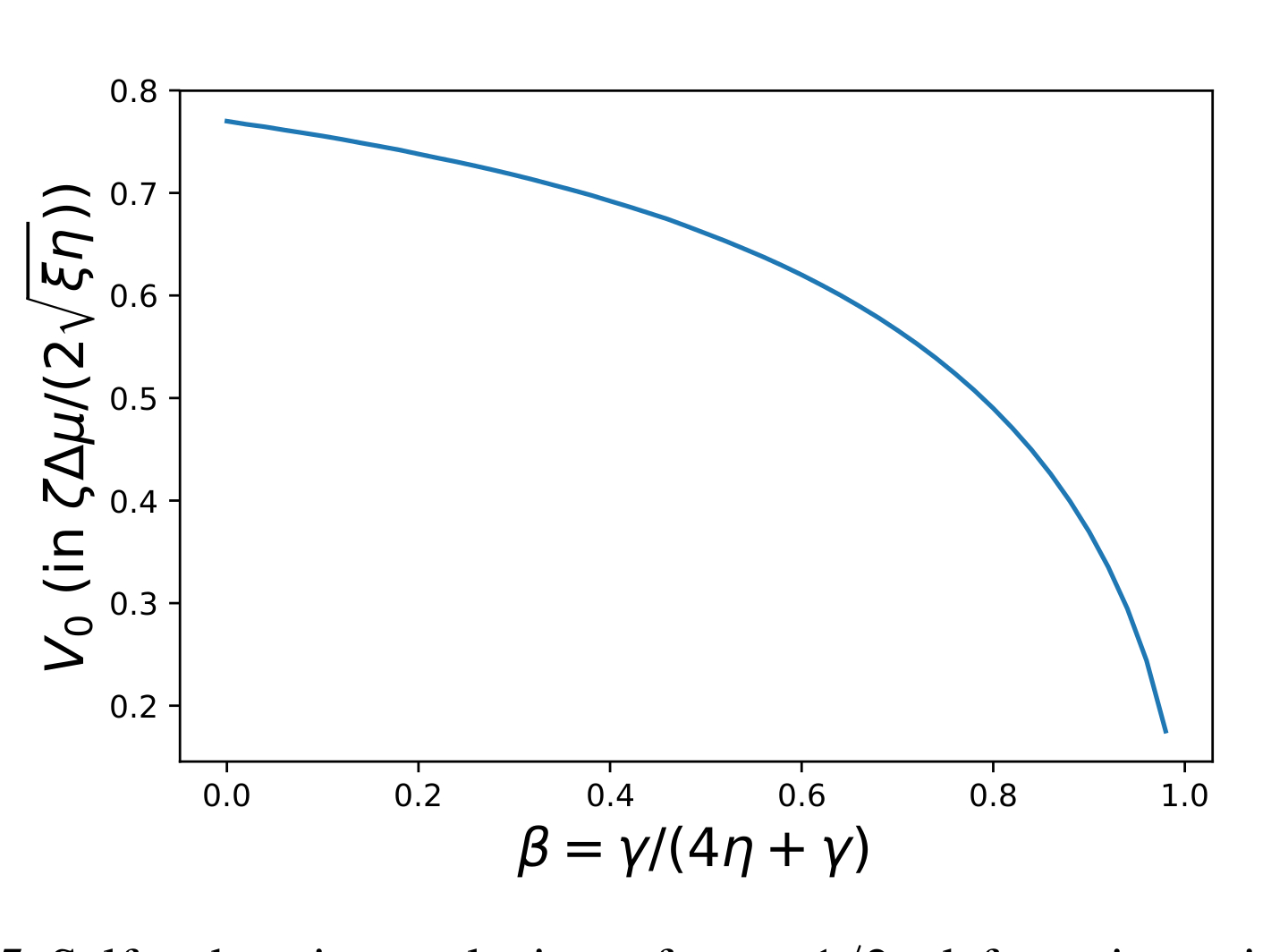}
\caption{Decrease of the self-advection velocity of a $+1/2$ defect with increasing rescaled rotational viscosity $\beta$. Inspired by \cite{Brezin}.}
\label{velovis}
\end{figure}

What is the effect of (small) rotational dissipation? Up to now, the velocity field was computed with fixed director orientation \eqref{phitheta}, corresponding to the director field of a passive
defect at equilibrium. Can we stick to this aproximation, in a first step, but allow for a non-zero rotational viscosity $\gamma$, which couples the
velocity field to the orientation gradients of the nematic
director? We consider the case of vanishing
flow-alignment parameter, $\nu =0$. The general idea now is to perform a perturbation expansion in the activity to linear order. A first observation is that the velocity field up to first order in the activity $\zeta$ can be computed while still keeping the passive defect orientation \eqref{phitheta}. The first-order orientational field takes a particularly intuitive form,
\begin{equation}
    {\bf h} = \gamma\, [ ( {\bf v}\cdot \nabla) {\bf p} + {\bf \omega}\cdot {\bf p}],
    \label{orifield}
\end{equation}
where $\omega_{\alpha\beta}$ is the vorticity tensor. Now the non vanishing perpendicular field $h_{\perp} \neq 0$ contributes to the force balance. The self-advection velocity gets refined to the following expression
\begin{equation}
    {\bf v}_0 =- \frac{ \zeta \Delta \mu}{2(\xi\eta)^{1/2}} f(\beta) {\bf e}_x,
\end{equation}
with $\beta \equiv \gamma/(4\eta +\gamma)$. The effect of rotational viscosity on the velocity is shown in Fig.\ref{velovis}. The pinning force $f_s$ is also affected and the logarithmic dependence on the core radius $r_c$ is replaced by a power law with an exponent that depends on $\beta$. Still, one recovers $f_s \rightarrow 0$ for $r_c \rightarrow 0$ (vanishing core size).

In a second step we take into account the feedback of the flow on the orientation of the director, to first order in activity. This approach entails a correction to \eqref{phitheta}, which in our setting reads
\begin{equation}
    \phi = \frac{\theta}{2} + \delta \phi,
    \label{phithetacor}
\end{equation}
The result of the calculation, to first order in $\gamma/\eta$, is
the following distortion of the defect orientation field by activity, 
\begin{equation}
    \delta \phi =\frac{\gamma \sin\theta}{4\kappa} r \,v_{0,x} \log (r/r_0)
    \label{deltaphi},
\end{equation}
where $r_0$ is an arbitrary length. The radial domain of validity of \eqref{deltaphi} is discussed in \cite{Brezin}.

\begin{figure}[h!]
\centering
\includegraphics[width=0.4\linewidth]{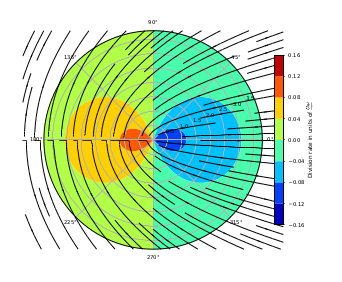}
\caption{Colormap of the (net) cell division rate, in the cellular flow field around a $+1/2$ defect, for non-conserved cell number. In the background: black solid lines showing the director field. Inspired by \cite{Brezin}.}
\label{divr}
\end{figure}

We now examine the role of non-conserved
cell number (owing to cell division, apoptosis, extrusion, ...) 
on the active flow created by a 
defect. We introduce an effective
cell-proliferation or turnover rate $k$, which accounts for these
three processes. With incompressible cells, the continuity
equation then reads:
\begin{equation}
    \nabla \cdot {\bf v} = k_d-k_a = (P - P_k)/\bar \eta \equiv k,
\end{equation}
which signifies a pressure-dependent cell turnover rate. In here, $\bar \eta$ is an effective viscosity. The result for the self-advection velocity of the
defect reads,
\begin{equation}
    {\bf v}_0=-\frac{\pi\zeta \Delta \mu}{8(\xi\eta)^{1/2}}\left(1+\left (\frac{\eta}{\eta+\bar\eta}\right)^{1/2}\right){\bf e}_x,
\end{equation}
from which, by comparison with \eqref{selfadvec}, we learn that cell turnover stimulates defect motion. We recover the incompressible limit (cell number conservation) for $\bar \eta = \infty$, implying $k=0$. We can interpret this as follows. An incompressible fluid with a nonzero divergence of the velocity field, is  peculiar to living systems. Division and death act, respectively,
as sources and sinks. In a contractile system the active forces 
create a low-pressure environment at the head of the
defect, which promotes cell division, and the converse takes place in the tail: extrusion is assisted in the high-pressure zone. Fig.\ref{divr} displays a colormap of the cell turnover rate, which essentially is the same as the divergence of the velocity field. 

\begin{figure}[h!]
\centering
\includegraphics[width=0.35\linewidth]{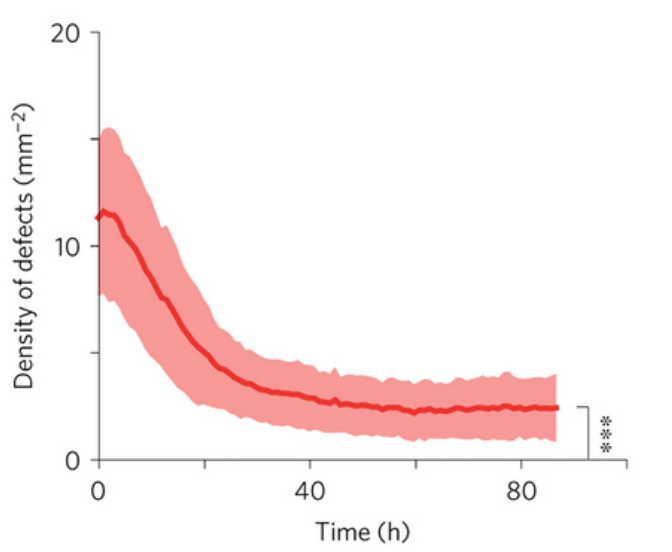}
\caption{Time evolution of the number of topological defects in a monolayer of spindle-shaped NIH 3T3 mouse embryo fibroblasts.  A finite density is reached at long times. The small positive fluctuations are measurement artefacts. No defects are created in the experiments. Reproduced from \cite{Silber}.}
\label{Nvst}
\end{figure}

Let us close this subchapter by drawing attention to experimentally measured velocity profiles of $s=+1/2$ defects and showing some remarkable defects textures. We expect to see pronounced and unambiguous velocity profiles for the motile fraction of $s=+1/2$ defects, on their way towards annihilation with their $s=-1/2$ counterparts. Recall that motile defects cannot provide nucleation sites for multilayering. Indeed, ``microtectonics" is reserved for the non-motile fraction, which diffuses rather passively and does not engage in annihilation processes.
Fig.\ref{Nvst}, which is similar to Fig.\ref{dish}d, shows an observed decay of the number of $s=+1/2$ defects in a cellular monolayer.

\begin{figure}[h!]
\centering
\includegraphics[width=0.7\linewidth]{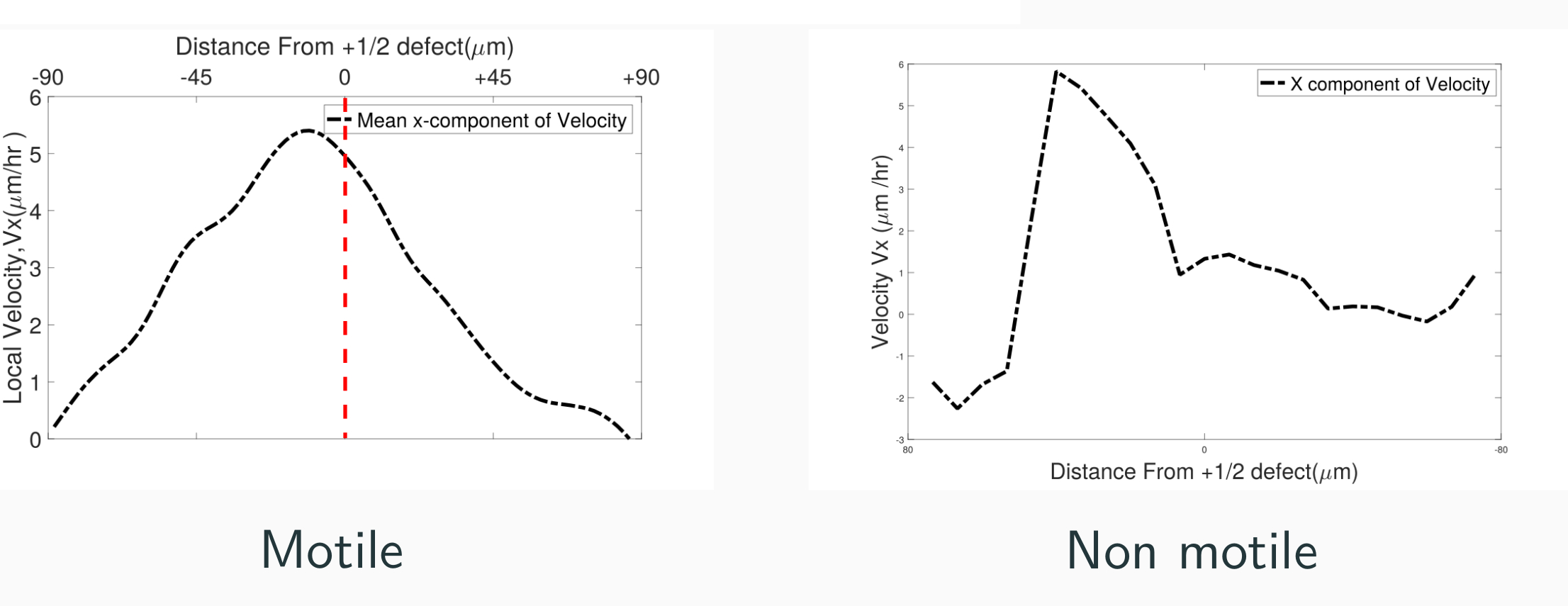}
\caption{Defect core velocity profiles for $s=+1/2$ defects, measured along the head-to-tail axis. (Left) Active motion. (Right) Diffusive quasi-passive motion with no net velocity. Reproduced from \cite{Duclos}.}
\label{vpro}
\end{figure}

The difference in motion between motile ad non-motile $s=+1/2$ defects is conspicuous in the measured velocity profiles along the head-tail axis \cite{Duclos}. The motile defects display a clear net velocity, peaked in magnitude about their core, whereas the non-motile defects have no preferred direction. This is shown in Fig.\ref{vpro}. 

\begin{figure}[h!]
\centering
\includegraphics[width=0.6\linewidth]{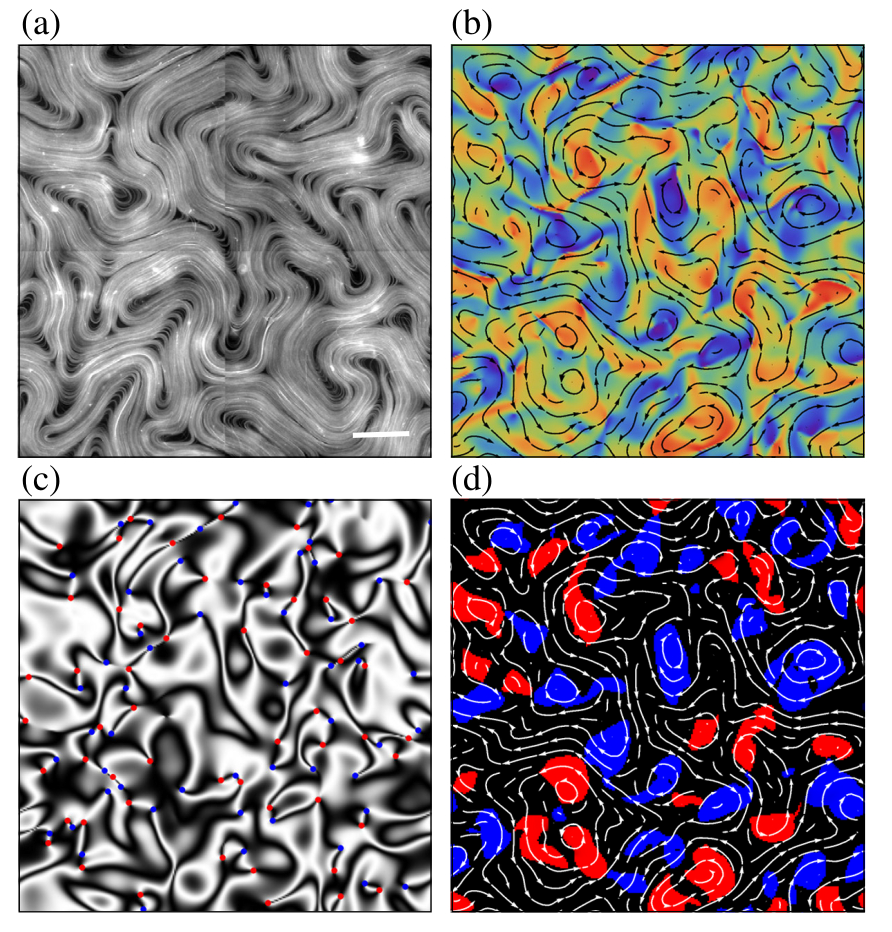}
\caption{Active nematics and defects textures. (a) An active nematic suspension, effectively in $d=2$, of
microtubule bundles and kinesin at the water-oil interface. White scale bar: 100 $\mu$m. (b)–(d) Numerical simulations of an extensile active
nematic. (b) Flow velocity
(black streamlines) and vorticity (background color). (c) Schlieren
texture constructed from the director field ${\bf n}$. The red and blue
dots mark, respectively, $s=+1/2$ and $s=-1/2$ disclinations.
(d) Clockwise rotating (blue) and counterclockwise rotating
(red) vortices, detected by measuring the Okubo-Weiss field.
Reproduced from \cite{Giomi}.}
\label{texture1}
\end{figure}

\begin{figure}[h!]
\centering
\includegraphics[width=0.6\linewidth]{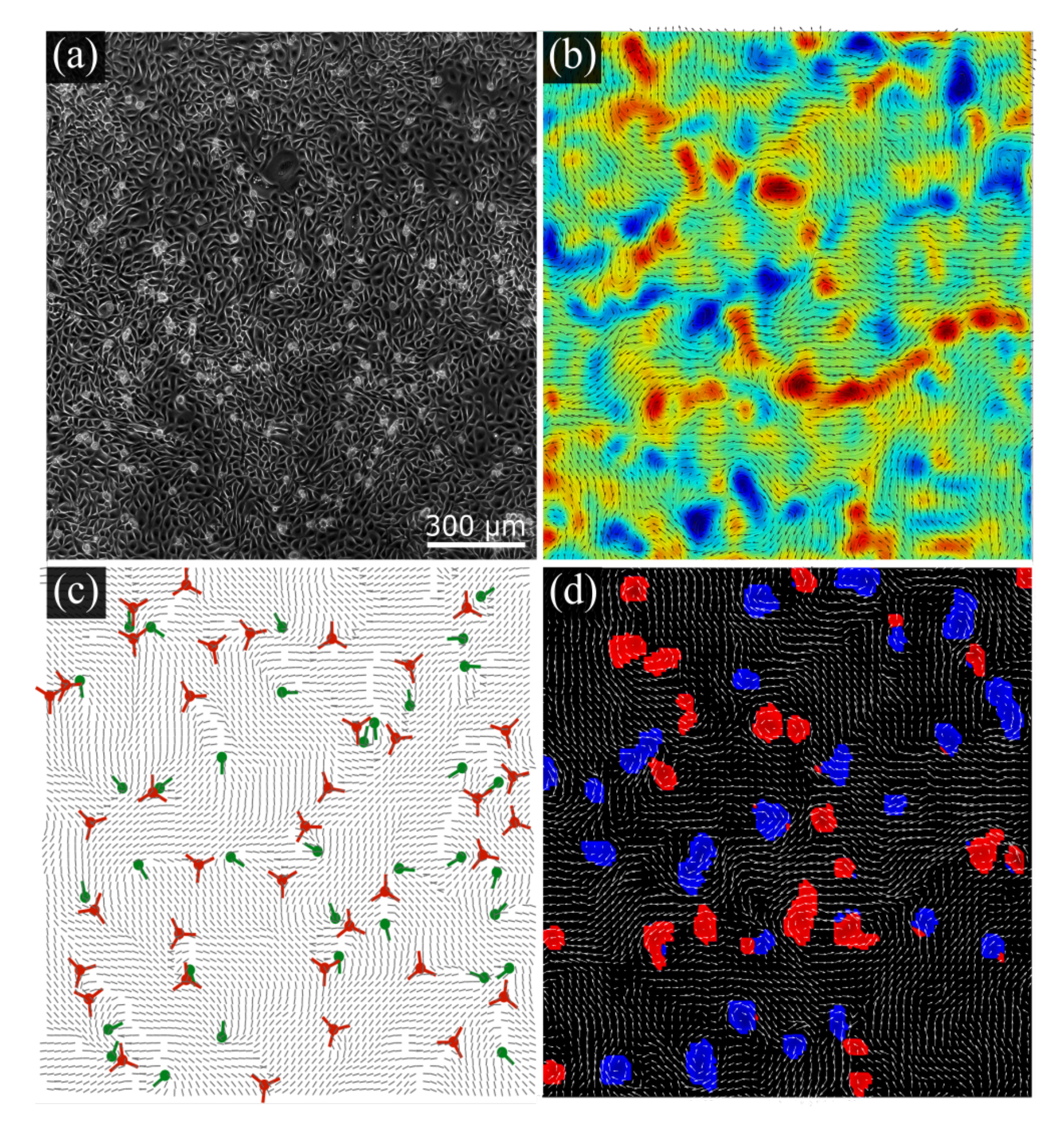}
\caption{Nematodynamics of human bronchial epithelial cells. (a) Phase-contrast image. (b) Normalized vorticity map with the Particle Image Velocimetry (PIV) flow field
showing collective behaviour. (c) Cell orientation map with
green (red) dots marking the position of $s=+1/2$ ($s= - 1/2$) defects.
(d) Map of the Okubo-Weiss field with blue (red) domains indicating
clock-wise (counter clock-wise) rotating vortices. Scale bar is 300 $\mu$m.
Reproduced from \cite{Blanch}.}
\label{texture2}
\end{figure}

With Fig.\ref{texture1} and Fig.\ref{texture2} at hand, which illustrate experimental and simulational active nematics flow fields and defects textures, we formulate some qualitative conclusions. Cells
tend to self-organize into transient mesoscopic domains
with a well-defined nematic order, separated by pairs of
$\pm 1/2$ defects. In active nematics, which as we have seen can exhibit turbulence at low Reynolds number, there is a deep connection between the topological structure of the orientationally ordered constituents and the flow dynamics \cite{Giomi}. It appears that active turbulence is mediated by
unbound pairs of topological defects. The total number of topological defects is proportional to the number of vortices. The nematic director pattern near a $\pm 1/2$ disclination determines the local
vortex structure. The interplay between the nematic defects
and activity gives rise to the spontaneous formation of
transient swirls. Topological defects serve as a
template for the turbulent flow, which in turn advects the defects themselves, leading to chaotic mixing. A typical order of magnitude of the velocity of a motile defect is 1$\mu$m/min
 \cite{Blanch}.

\subsection{Multicellular spheroids}
In this part of the lectures, we are concerned with three-dimensional active matter and discuss {\it mechanical} factors that can influence the time evolution of tumors in biological tissue. Loosely speaking, the research we expose here favors the idea that direct mechanical effects can have strong implications in cancer proliferation. 

Tumors mostly grow subject to limitations of space and compete for space with the surrounding tissue. To grow against the surrounding tissue, cells have to exert mechanical stress on the neighboring cells. This can lead to a mechanical ``crosstalk", which we analyze. The ``model tumors" for this purpose, grown experimentally in a liquid medium (DMEM), are aggregates named multicellular spheroids (MCS). When a controlled mechanical stress is applied to MCS, a biomechanical sequence unfolds which is studied over long time scales (10 days and more) \cite{Delarue,MontelNJP}. 

Before describing this response we emphasize that there are essentially two ways of communicating between a tumor and its micro-environment. One is biochemically and the other is mechanically. We focus here on the latter. Competing for space results in a two-way mechanical interaction between the tumor and the stroma (i.e., the supportive tissue). The growing neoplastic (abnormal) tissue compresses
the stroma, and, an active stroma, e.g., containing contractile
myofibroblasts, can inflict mechanical stress on the tumor. MCS are ideal to study this because they do not crosstalk biochemically
with their surrounding, but only mechanically.

\begin{figure}[h!]
\centering
\includegraphics[width=0.5\linewidth]{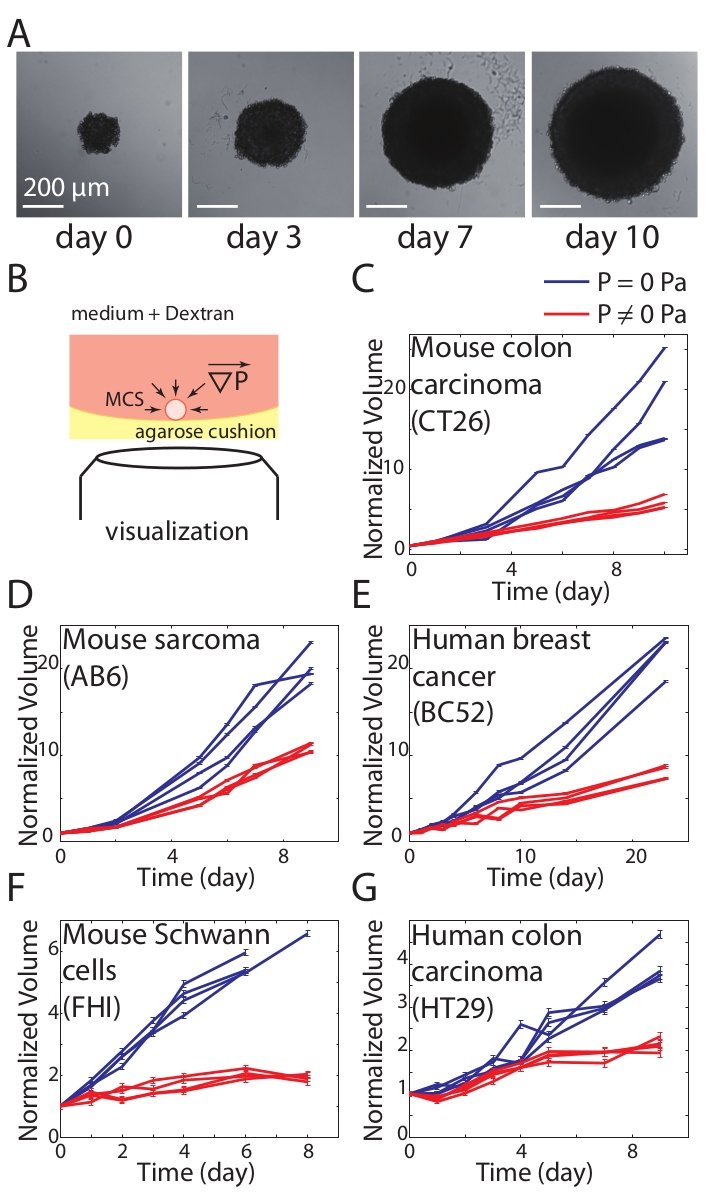}
\caption{Multicellular spheroids and compressive mechanical stress. (A) Growth of a CT26 MCS from day 0 to day 10. (B) Dextran is added to the culture medium, causing 
moderate osmotic stress solely on the outermost layer of cells, which is mechanically transmitted to inner
cells. (C–G) MCS volume normalized to
the initial volume, versus time, with and without applied stress: (C–F) 0 Pa (blue) and 5 kPa (red); (G) 0 Pa (blue)
and 10 kPa (red). 
Reproduced from \cite{Delarue}.}
\label{groexp}
\end{figure}

In early studies it was observed that a compressive stress applied to MCS grown from the mouse colon carcinoma cell line CT26 drastically
and reversibly reduces their growth rate \cite{Montel,MontelNJP}, mainly through decreasing the cell division rate in the center
of the MCS. Fig.\ref{groexp} illustrates a more recent setup in which a controlled stress is applied, and 5 different cell lines are used. 
In these experiments a biocompatible polymer, Dextran, is added to the culture medium to exert a mechanical stress. The
biopolymer does not penetrate single cells and does not
affect single-cell growth, but is in direct contact with the MCS. Indirect experiments, in which a dialysis membrane (or bag) impermeable to Dextran is interposed between the Dextran and the MCS, with the Dextran osmotically pushing on the bag and the bag pushing on the MCS, have the same effect on the growth \cite{Montel}. The applied compressive stress is in the range from 5 to 10 kPa only, which is comparable to the MCS Young modulus (which can be measured with a parallel-plate compression test). For larger stress, the effect saturates. Notwithstanding quantitative differences between different cell lines used, all spheroids are found to react to applied compressive stress by
drastically reducing their growth rate. 

\begin{figure}[h!]
\centering
\includegraphics[width=0.35\linewidth]{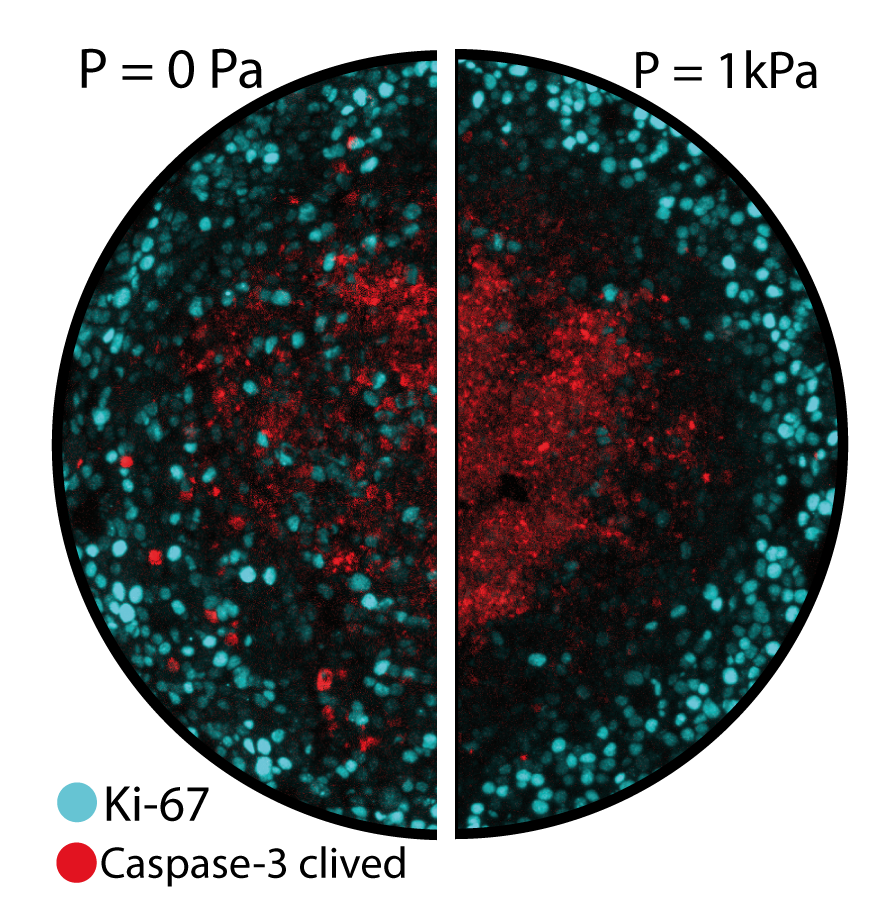}
\caption{Effect of stress on the (radial) distribution of
proliferation and apoptosis. Cryosections and immunofluorescence
of the spheroids are used to label the cell divisions. 
Using a classical immunostaining protocol,
fluorescent labeling is performed, of dividing cells (in cyan), with an anti-Ki-67 antibody, and of apoptotic cells (in red), with an anticleaved Caspase-3 antibody.
The MCS radius is $R \approx 400\mu$m. (Left) Half section
of an MCS grown in a normal medium for 4 days. (Right) Half
section grown under a stress of 1 kPa for 4 days.
Reproduced from \cite{Montel}.}
\label{casp}
\end{figure}

Inside the MCS one can probe the (radial) spatial dependence of cell division and apoptosis using cryosections and
immunofluorescence, and compare the cases without and with applied stress. Typical results are shown in Fig.\ref{casp}. Naively, under stress, one would rather expect cells to get crushed and die at the surface of the MCS, where the osmotic stress impacts directly. This, however, is opposite to what is observed. Cellular response to stress is small at the periphery and large at the center. Division of cells in the deep interior is arrested and apoptosis gets concentrated there.

In order to better understand this stress dependence of
cell division numerical simulations were performed. The agreement with experiment is good. A steady state is observed that depends on the applied stress. Under increasing stress, the division rate depends strongly on the distance to the surface of the MCS. The decrease is strongest
in the core. A two-rate model is proposed to explain this. The core of the spheroid is mostly undergoing apoptosis, whereas its periphery is proliferating (``surface growth") because cell division is facilitated by available space for deformation inside a, presumably, relatively soft stroma. The net growth (death) rate scales with the area (volume) of the MCS. The net bulk growth rate $k=k_d-k_a$, as defined in \eqref{continuity}, is stress dependent.  Within a radial distance $\lambda$, on the order of 10 to 100$\mu$m, from the surface, the net growth rate is supposedly augmented to $k_d-k_a+\delta k_s$, where the surface growth rate increment $\delta k_s > 0 $ also depends on stress, but weakly. Combining surface and volume effects, the rate of volume increase is found to obey, for MCS with large radius $R \gg \lambda$, 
\begin{equation}
    \partial_t V=(k_d -k_a)V+ 4\pi{(\frac{3}{4\pi})}^{2/3}\delta k_s\lambda V^{2/3} 
\end{equation}

\begin{figure}[h!]
\centering
\includegraphics[width=0.55\linewidth]{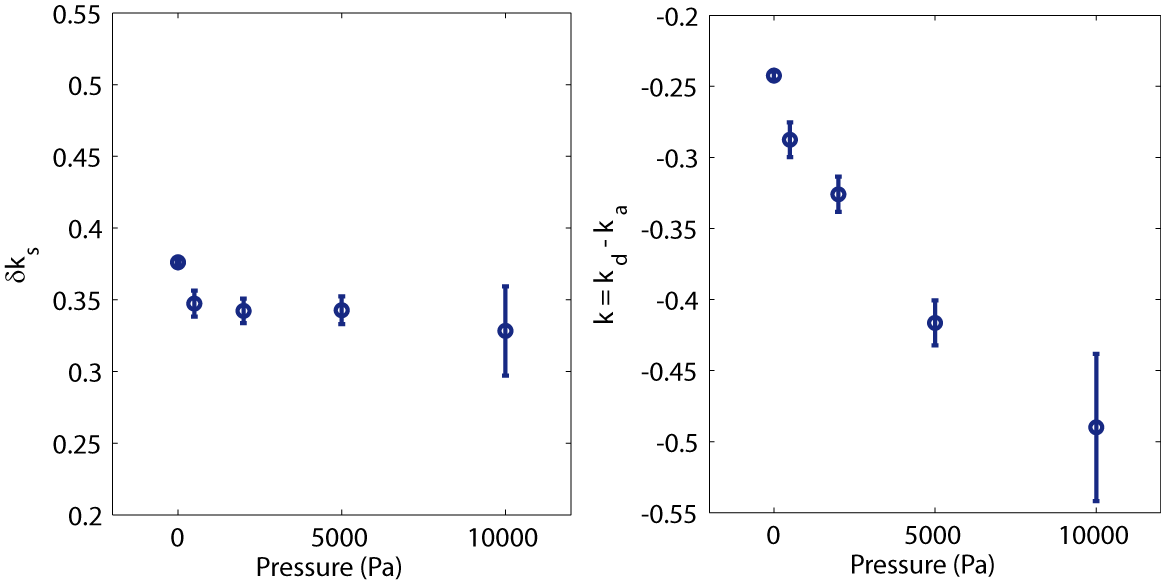}
\caption{Two-rate model simulation results for stress-dependent growth rates. (Left) Surface growth rate increment $\delta k_s$ versus stress.
(Right) Bulk net growth rate $k$ versus stress. 
Reproduced from \cite{Montel}.}
\label{predep}
\end{figure}
The stress dependence of $k$ and $\delta k_s$ in the simulations is displayed in Fig.\ref{predep}. We add that, although $k_d-k_a <0$, the surface proliferation dominates in the range of $V$ studied and gets balanced by the volume contribution when the stable steady state volume is reached, $\partial_t V =0$. Also, for small $V$ ($R < \lambda$) the net growth rate is positive. 

Besides mechanical effects, another contribution to an increased division rate at the surface of the MCS is nutrient depletion in the interior, since nutrients are consumed within a penetration depth from the surface, and this might be stress dependent. This has been tested experimentally by measuring the penetration of a growth factor and determining that it is not significantly changed by stress. This supports the main finding that a direct mechanical effect, being the strong decrease of $k$ with increasing stress, is the likely cause of MCS proliferation reduction in the presence of stress.

In the previous chapters we have paid much attention to (spontaneous) flow of active matter. Therefore, we should have expected now to face, and are well prepared to tackle, the problem of cell migration and cell flow in (model) tumors. Again, this type of response can be triggered by biochemical and/or mechanical signals, and we will deal with the latter. Since division (vs. apoptosis) acts as a source (vs. sink) of the velocity field, it is no surprise that the growth rate dichotomy between surface and core can drive cellular flow. We now ask what is the effect of
mechanical stress on this flow. 

Recall that the net growth rate of an MCS is
drastically reduced by external stress, an effect which saturates at high stress ($> 5$kPa). Neither cell death nor cell
density, but especially cell division is strongly impacted by stress and notably much reduced in the core. Can a hydrodynamic model in the manner developed earlier for active particles, and taking into account this gradient of cell proliferation, describe the cellular flow?

\begin{figure}[h!]
\centering
\includegraphics[width=0.8\linewidth]{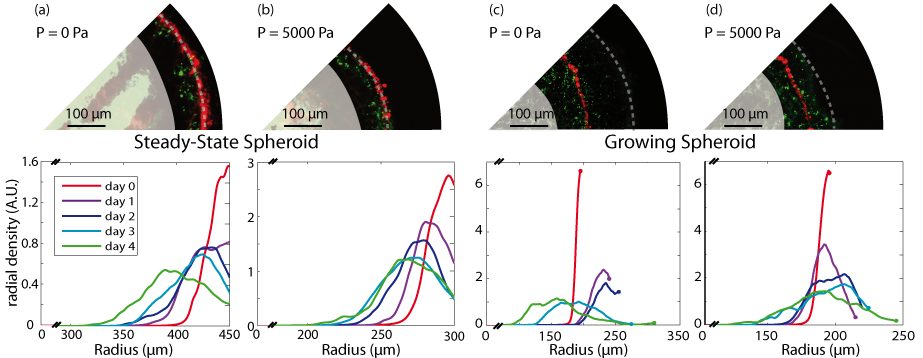}
\caption{(a)–(d), (top): Cryosections of MCS right after staining (red) and four days later (green). The
dashed line marks the radius at day 4. (bottom): Time evolution of the
distribution of fluorescently labeled MCS: red, day 0; purple, day 1; dark blue, day 2; light blue, day
3; green, day 4. (a) and (b): MCS at steady state, without applied stress and radius 450 $\mu$m (a) or under
5 kPa and radius 300 $\mu$m (b). (c) and (d): Growing MCS, with initial radius 200 $\mu$m, without (c) or with 5kPa
mechanical stress (d). The dots indicate the border (surface) of the MCS. 
Reproduced from \cite{DelaruePRL}.}
\label{fluoin}
\end{figure}

In order to measure the cellular flow 
peripheral cells were labeled using nanobeads containing fluorescent dyes. The nanoparticle distribution was monitored in time with and without applied mechanical stress. To distinguish the effect of spheroid growth from 
hydrodynamic cellular flow, two different experiments were conducted. In the first, spheroids 
reached the steady-state radius following growth with or without
stress. In the second, spheroids in their initial
growing phase were used. Fig.\ref{fluoin} shows spheroid sections 
directly after staining, or 4 days later.
In all cases the distribution of the stained cells broadens in time and the maximum moves towards the center. 

\begin{figure}[h!]
\centering
\includegraphics[width=0.6\linewidth]{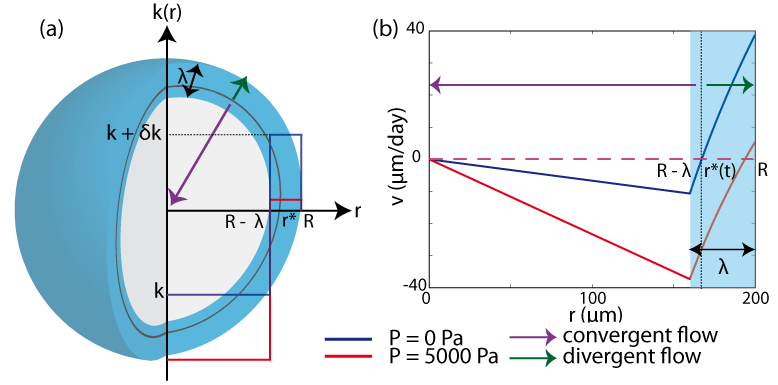}
\caption{Transport model results for growing MCS. (a) Net growth rate $k(r)$ as a step function of (radial) position $r$ inside the spheroid. Blue step, zero applied stress; red step,
5 kPa applied stress. Note the surface layer of thickness $\lambda$ with augmented growth rate. (b) Calculated velocity field. Blue, zero applied stress; dashed-red,
5 kPa applied stress; dashed-purple arrow, convergent flow; green arrow,
divergent flow.
Reproduced from \cite{DelaruePRL}. }
\label{kofr}
\end{figure}

In steady-state MCS, there is a radial convergent flow. Without applied stress, the speed is about 25 $\mu$m per day, compatible with convective motion induced by
cell division. With applied stress (ca. 5 kPa) the flow is reduced by about a factor 2. For growing spheroids, the labeled cell 
distribution moves from the (outward moving) periphery towards the
center. Without applied stress, the maximum
first follows the outward motion (divergent) but later moves inward (convergent). With applied mechanical stress the cellular flow is
reduced and the initial divergent flow is practically absent.
A highly simplified transport model, only taking
into account {\it convection} and using the continuity equation for incompressible flow with sources and sinks, \eqref{incompk}, can capture these observations quantitatively, which
validates the fact that both diffusion and chemotaxis are
negligible. The two-rate model for MCS growth (Fig.\ref{kofr}a) and the hydrodynamic cellular convection model provide characteristic velocity fields and convergent (and divergent) flows illustrated in Fig.\ref{kofr}b.

How does a growing MCS respond in the short term, i.e., on time scales shorter than the cell division time, to a sudden application of compressive stress (a pressure jump) \cite{Delarue,DelaruePhD,DelarueIF}? What is the decrease in cell volume, and is this decrease homogeneous throughout the spheroid? These are the questions to which we turn. Experiments show that on time scales of minutes, a compressive stress causes a reduction of the
MCS volume, largely due to a very important decrease of the cell volume in
the core. In contrast, the cell density does not change
appreciably close to the surface of the spheroid. This radial density profile implies a high pressure in the
core, even higher than the increased pressure at the surface. 

\begin{figure}[h!]
\centering
\includegraphics[width=0.7\linewidth]{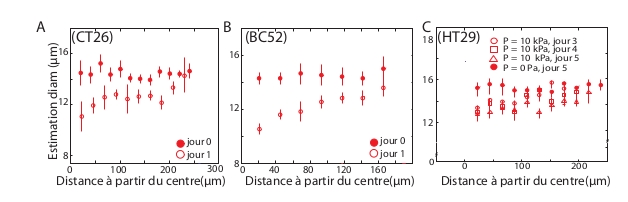}
\caption{Volume reduction in response to a sudden compressive stress. Reduction of the cellular diameter $d(r)$ of MCS from CT26 (A),
BC52 (B) under 5 kPa applied stress, and from HT29 under 10 kPa (C), as a function of the distance $r$ from the MCS center. Reproduced from \cite{DelaruePhD}. }
\label{phd1}
\end{figure}

Fig.\ref{phd1} illustrates this effect for MCS grown from three different cell lines. In the analysis of these experiments, which makes use of the position autocorrelation function of cell nuclei, the local cell diameter $d(r)$ is determined by measuring the distance between nuclei
of adjacent cells as a function of their distance $r$
from the center of the MCS.  Without stress $d$
is approximately constant. Under a compression (of 5 to 10 kPa) an inhomogeneous reduction of $d(r)$ is observed, typically up to 20$\%$ in the center of the MCS, concomittant with the shrinking of the whole MCS. 

\begin{figure}[h!]
\centering
\includegraphics[width=0.8\linewidth]{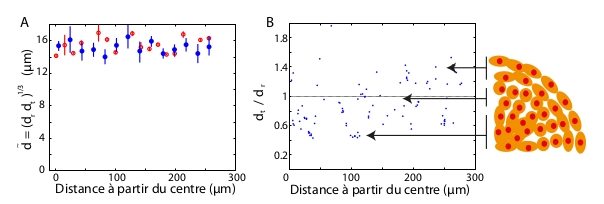}
\caption{Cell-to-cell distance anisotropy in an MCS without applied stress. Radial $d_r$ and tangential nearest-neighbour cell-to-cell distance $d_t$ are measured versus distance $r$ from the MCS center. The ``isotropic mean" diameter $\tilde d = (d_rd_t^2)^{1/3}$ is constant (A).
The ratio $d_t/d_r (r)$ reveals the anisotropy and its radial inhomogeneity (B). A cartoon is added: cells closer to the center are elongated radially while those closer to the surface tend to flatten tangentially to the circumference of the section. 
Reproduced from \cite{DelaruePhD}. }
\label{phd2}
\end{figure}

Not suprisingly, active stresses are important to
account for the observed density (and pressure) profiles. However, there is also a remarkable {\it anisotropy} at play here, which is instrumental to a proper interpretation and understanding of the observations \cite{DelaruePhD,DelarueIF}. The cell shape turns out to be ``polarized" (in the nematic sense). Cells are elongated radially, except in a thin shell close to the spheroid surface (where a tendency towards tangential ordering exists). This effect is already present without applied stress to the MCS, as the data in Fig.\ref{phd2} demonstrate.

\begin{figure}[h!]
\centering
\includegraphics[width=0.4\linewidth]{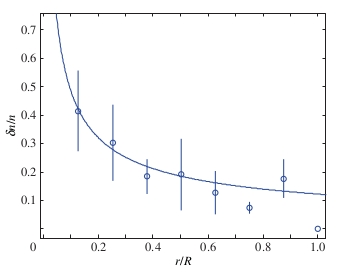}
\caption{Cell compression
profile. Relative cell density increase of HT29 cells 
5 min after an external pressure jump, as a function of radial distance $r$
normalized by the spheroid radius $R$ (experimental data: circles; theoretical power-law fit: solid line). 
Reproduced from \cite{DelarueIF}. }
\label{denvsr}
\end{figure}

Based on active gel hydrodynamics with active stress depending on pressure, featuring polarized cells, and allowing for the dependence of cell division and apoptosis
rates on the local stress, a theoretical treatment is worked out \cite{DelarueIF}. The calculation for this system predicts, for a fixed moment in time after the pressure jump, a power-law decay of the local density with radial distance \cite{DelarueIF}, displayed in Fig.\ref{denvsr},
\begin{equation}
    \frac{\delta n}{n}=\frac{\Delta P (3+\beta_e)}{~3K} \left( \frac{r}{R_0}\right )^{\beta_e}.
\end{equation}
The decay exponent takes the value $\beta_e \approx -0.6$, for the data corresponding to the response of the MCS 5 min after the shock. Here, $\delta n/n = (n(r)-n_0(r))/n_0(r)$ is the relative cell number density increase. At time $t = 0$ the density $n_0 (r)$ is practically constant from $r=0$ up till the instantaneous MCS radius $R_0$. $\Delta P$ is the pressure jump at $t=0$, and $K$ is the bulk compressibility modulus. From this and similar fits to the data $K$ has been estimated to be on the order of $10^4$ Pa.

Next we discuss co-cultures of cancer cells and macrophages and the modeling of multi-cellular aggregates \cite{Benaroch,ABAJ}. Macrophages are a type of white blood cells that do not divide and normally play a favorable innate role in the immune system by killing intruders and cleaning up cellular debris through phagocytosis, and assisting other immune system components. However, there can also be problems related to macrophages because macrophages can enhance cell division and therefore enhance tumor growth.

Noteworthy in the contect of our lectures is that macrophages are plastic cells that can be polarized by their environment. Their polarization state ranges from a proinflammatory,
antitumor phenotype to an anti-inflammatory, protumor phenotype. That is, depending on
their polarization, macrophages 
promote either antitumor or protumor responses. In the TME, macrophages are usually polarized toward
an anti-inflammatory phenotype, favoring tumor growth and invasive
behavior. 

Clearly, interactions between cancer cells and macrophages are important in tumor development. Therefore, recent biophysics research has focused on 
the mixing of different cell types into multi-cellular aggregates \cite{ABAJ}, which are often considered as experimental realizations of solid tumors. 
From a biological point of view, the study of these aggregates {\it in vitro} (cells in culture) and in theory and simulations may help to understand the interactions between different cell types in a growing tumor {\it in vivo}. 

Aggregates may comprise a significant number of cells (around $10^4$) and at this scale statistical physics, soft matter physics, elasticity and transport theory are all relevant. Of crucial importance in a theory is to allow for coupling between the cellular and nutrient densities inside the aggregate. A hydrodynamic description of a single-cell aggregate typically includes only cell division, cell death, and diffusion
and consumption of nutrients. Access to nutrients is necessary for
cell division (even though cancer cells are notorious for their fasting skills) and this is of course much easier at the surface than in the core.  
A continuum mixture model, which leads to the same results as hydrodynamic theories constructed from symmetry arguments, features dynamical equations for the composition and flow fields, obtained by extremization of the dissipation in the system and valid for dissipative systems in which inertia is negligible (low Reynolds number). With two or more components, the model is extended in the spirit of the Cahn–Hilliard theory (first developed in metallurgical context) for the phase separation of fluids, which has proven useful for describing phase transitions in equilibrium systems. Typically one works with three densities, two pertaining to the mixture and one for the outer medium, which can be a passive spectator like a rigid wall or an extracellular matrix, or active like the stroma around tumors {\it in vivo} or an inter-cellular fluid.

The free energy density (for three interacting components) is an extension of the Flory–Huggins free energy used in polymer physics and derived from macroscopic Langevin dynamics and from Fokker–Planck equations. The dynamical equations for the cell concentrations are obtained by inserting the constitutive equation for the flux into the conservation equation, which includes cell proliferation. The cell mixture is unstable when the determinant of the diffusion
matrix becomes negative. This instability is called spinodal decomposition and signals a macroscopic dynamic phase
separation between the cells and the inert component, or between the different cell types.   

\begin{figure}[h!]
\centering
\includegraphics[width=0.5\linewidth]{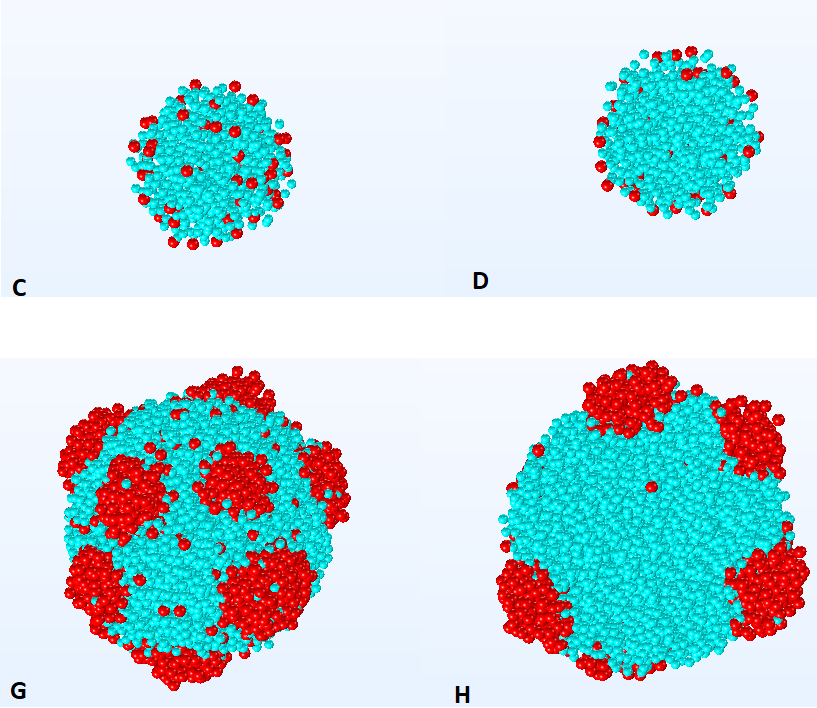}
\caption{Growth of a multi-cellular aggregate containing two cell types T1 and T2. At $t = 0$ one starts with the steady state solution for an aggregate containing
1668 T1 cells of radius 10 $\mu$m and 50 T2 cells of radius 8 $\mu$m. T1 cells divide asymmetrically into T1 and T2 cells, whereas T2 cells do not proliferate. 
At time $t = 0$, panel C: Surface of the aggregate. T1 cells are colored in cyan and T2 cells in red. In panel D: Section of the aggregate.
Below, at $t = 10$ days, panel G: Surface of the aggregate. In panel H: Section of the aggregate. Reproduced from \cite{ABAJ}. }
\label{heteros}
\end{figure}

A great question in mathematical oncology is how microscopic
features at the cell level cause morphological changes at the macroscopic level. This can be studied numerically using an aggregate composed of two cell types (and inter-cellular fluid). The structure of the aggregate depends on the parameters of the Flory interaction potential. For therapeutic purposes, aggressive cells or cells prone to metastasis must be
accessible to the anti-tumor drug. It is thus vital to know their locations inside a tumor. The model employed is inspired by the behavior of cancer stem cells, which either divide symmetrically between two identical daughter cells or asymmetrically into a stem cell and a differentiated cell. The latter may or may not divide. Or, one takes an aggregate formed by cancer cells (which proliferate) and cells of the immune system such as macrophages (which don't). In Fig.\ref{heteros} the growth
of an aggregate containing two cell types is illustrated, T1 (in cyan) and T2 (in red). A complex structure with (partial) phase separation develops over time. Qualitatively, the equilibrium configuration of fully segregated phases can be predicted by comparing the different interfacial tensions as one does in the theory of wetting phenomena. However, the composition of the phases and the dynamics of the phase separation can only be obtained from the full non-equilibrium description. 

Is combining physical descriptions, as we do, a sufficient strategy for capturing the self-assembly of tissue comprehensively? Compared to other self-assembling processes in passive soft matter, the collective behavior of living objects is richer. It cannot be predicted from the 
interactions between assembling constituents alone. The character and the biological function of cells can change upon
assembly and the collective behavior of cells in a tissue can be qualitatively different from that of individual (or a small number of) cells. 

\subsection{Cyst growth}
Human induced pluripotent stem cells (HiPSCs) are a promising cellular source in cell therapies or regenerative medicine. They are used in large numbers to create differentiated cells. However, providing sufficients amounts of high quality stem cells {\it in vitro} (through culture) is difficult. Typically $10^8$ to $10^{10}$ cells are needed and care must be taken to preserve their stemness character (which perhaps might get lost, e.g., under large mechanical stress). 

HiPSCs self-organize into cysts, which are spherical closed epithelia (like the layers of cells that line hollow organs and glands). We discuss how one can monitor their morphology, organization and growth, and sketch
how a morpho-elastic model can quantitatively explain the growth dynamics. In fact, insight in the morphological dynamics,
which is mainly governed by mechanical elasticity and anisotropic proliferation, may help provide new strategies to
obtain sufficient quantities of medical-grade HiPSCs. 

The standard method for cell expansion relies on tissue culture in a flat dish ($d=2$), but better would be to use aggregates ($d=3$) and grow
spheroids that reach
a (sufficient) maximal size of about 300 $\mu$m, corresponding to the limit above. However, in this geometry oxygen and nutrient deprivation below some penetration depth from the surface leads to the
formation of necrotic cores (i.e., cell cemeteries). But do these cells spontaneously form spheroids? No, in a solvent pluripotent stem cells tend to self-assemble into a closed membrane-like monolayer akin to a cyst (thin-walled hollow organ or cavity in animals or plants). This is efficient because inside as well as outside surfaces maintain access to nutrients. 

Subsequently the cyst evolves towards a pseudo-stratified epithelium closed around a lumen (the cavity of a tubular organ or part). At some point, when their outer surface hits the (stiffer) confining wall (of a capsule), it follows this wall and ``confluent" growth starts, during which important internal mechanical stresses develop, resulting in multilayer buildup in the interior until the lumen is filled up and an effective spheroid morphology is reached. There is a danger here that the high stresses may alter the stemness of the cells by inducing unintended mutations or differentiation, so it pays to be able to exploit the non-confluent growth phase and to stop before confluent growth begins. 

\begin{figure}[h!]
\centering
\includegraphics[width=0.7\linewidth]{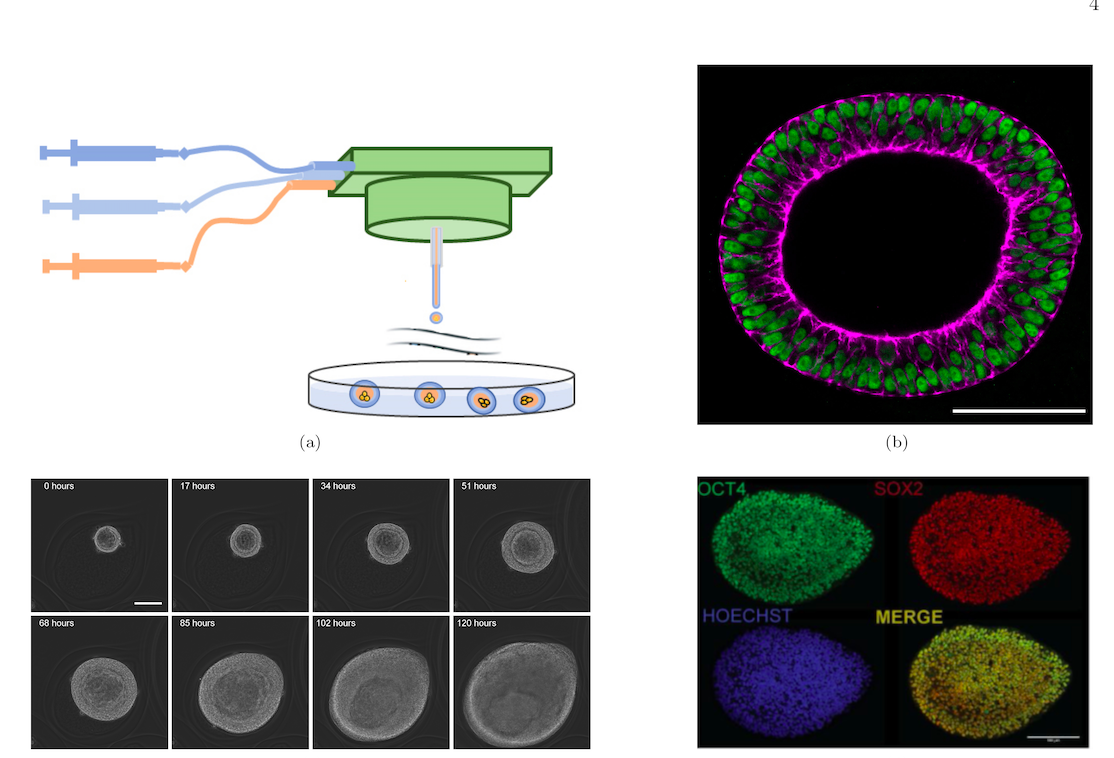}
\caption{Growth, morphological and stemness characterization of encapsulated HiPSC cysts. (a): Drawing of the microfluidic encapsulation platform.  The composite droplets formed upon Rayleigh instability fall into a calcium bath and form the hollow capsules
entrapping the cells. (b): Confocal image of the equatorial plane of a cyst stained for nucleus (DAPI in green) and actin (phalloidin in magenta). Scale bar is
100 $\mu$m. (c): Snapshots of HiPSC cysts growing inside an alginate capsule. Time interval between images is 17 h and scale bar is equal to 100 $\mu$m. (d): Confocal
images of encapsulated HiPSC colonies in the late stages (10 days after encapsulation) stained for two standard stemness markers (OCT4 and SOX2) and nucleus
(HOECHST). Reproduced from \cite{Nassoy}. }
\label{cys}
\end{figure}

\begin{figure}[h!]
\centering
\includegraphics[width=0.8\linewidth]{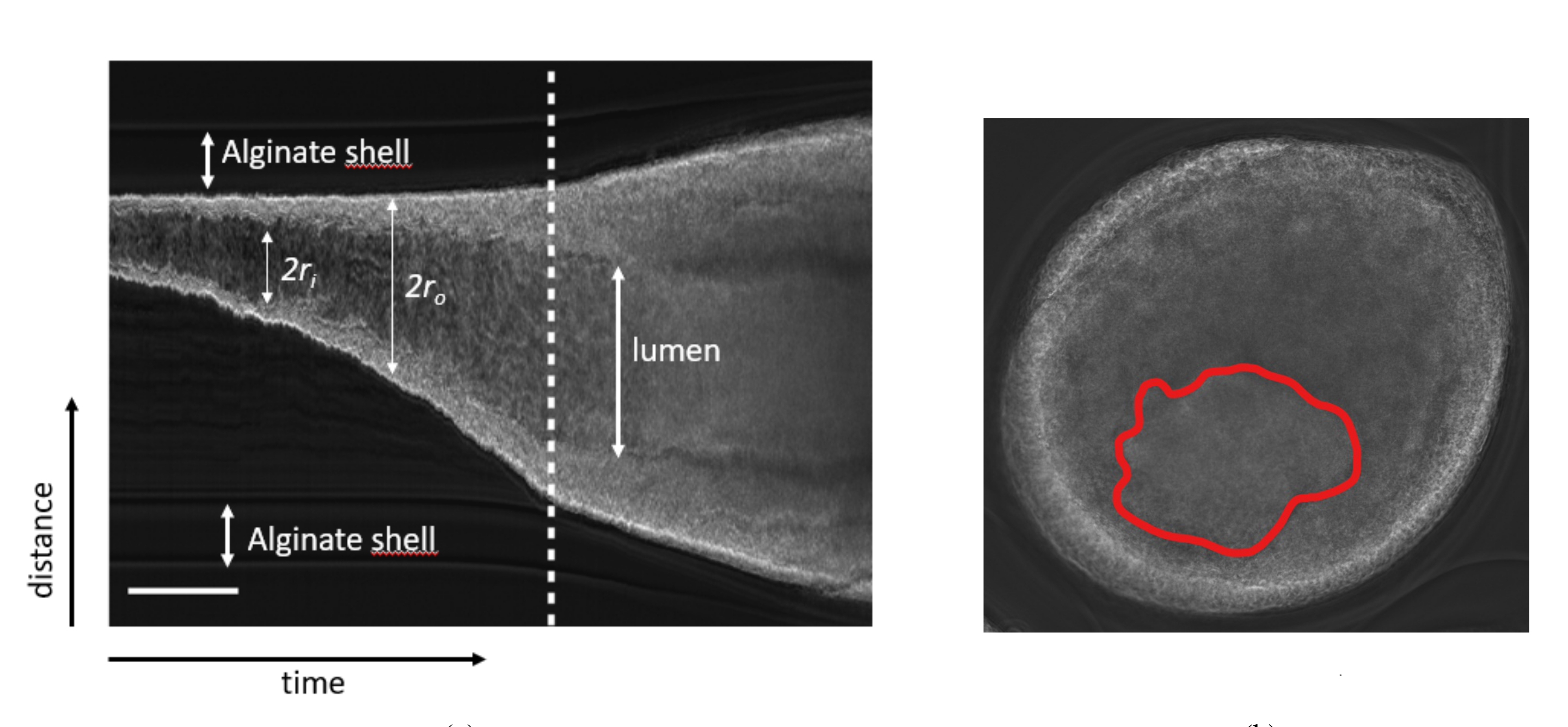}
\caption{Morphological evolution of encapsulated growing cysts. (a): Kymograph distance versus time through an encapsulated HiPSC cyst, monitoring inner and outer radius growth pre and post impact on capsule. The dotted line represents the time at which confluence is reached. The scale bar corresponds to 1 day for the abscissa axis and to 100 $\mu$m for
the ordinate axis. (b): Image and articulation of a Biot instability at the interface between the cells and the lumen. Reproduced from \cite{Nassoy}. }
\label{lum}
\end{figure}

In Figs.\ref{cys} and \ref{lum}, using phase contrast imaging, different growth stages are displayed. HiPSCs, encapsulated in a hollow alginate sphere, self-organize as a cyst around a lumen. As long as the cyst is smaller than the capsule (pre-confluence phase), it grows freely and spherically. Upon contact with the capsule (7 days later),
at confluence, it hugs the near-spherical shape of the capsule. It gets compressed by the elastic capsule, its growth gets slowed down and it grows inwards until, eventually, a lumen-free
aggregate is formed.

A theoretical morpho-elastic description of the cyst growth inside an alginate capsule is proposed, in which the growth (cell proliferation) is coupled to the mechanical stresses (elasticity) in the cyst and in the cyst-capsule system. The anisotropic growth dynamics is represented by a symmetric second-order tensor. Its eigenvalues $\gamma_i$ represent the deformation gradient in direction $i$ (one radial and two equivalent angular directions). We distinguish between an expansive growth deformation gradient $\gamma$ and a volume-preserving elastic deformation gradient $\alpha$, the total geometric deformation gradient being their product. That is to say, the change in length of a line element $dL_i$, due to the accumulated deformation during a time $t$, is $\delta l_i = \alpha_i\gamma_i dL_i$. For example, in the radial direction, $\alpha_r \gamma_r = dr/dR$. Incompressibility constrains the elastic components in the manner $\alpha_r \alpha_{\theta}^2 =1$. Let $\alpha \equiv \alpha_{\theta}$. Volume thus grows by a factor $\gamma_r \gamma_{\theta}^2$. The anisotropic growth rate $k = k_r + 2 k_{\theta} $ at time $t$ is then characterized by the components 
\begin{equation}
    k_i(t)  = \partial_t \log \gamma_i(t)
\end{equation}
Note that for constant growth rates the deformation gradients increase exponentially in time.

Anisotropic growth generates elastic strains and stresses. Indeed, when the growth is slow enough
that elasticity dominates over proliferation, biological tissues deform and buckle consistently with the minimization of elastic energy, respecting the boundary conditions. These are derived allowing for i) the permeability to water of the alginate and cyst membranes implying equal fluid pressure throughout the entire system, and ii) zero surface tension in the spontaneously self-assembling cyst membrane in the non-confluent stage. 

A Neo-Hookean model for incompressible material is assumed, for both cyst and alginate capsule, the latter having the highest Young modulus. For simplicity, in the cyst, it is supposed that the growth rates are linear in the radial Cauchy stress, and, of course, the more compressive the radial stress, the smaller the growth rates. Growth is generally inhibited in compression and favored in tension. A calculation, assuming spherical geometry, of the elastic energy density $W_{\rm NH}$ in this model then provides
\begin{equation}
    W_{\rm NH}=\frac{E}{6} (\alpha^{-4}+2 \alpha^2-3), 
\end{equation}
where $E$ is the Young modulus of the medium concerned (cyst, but post confluence also capsule), and $\alpha = (1/\gamma_{\theta})r/R$, with $r$ the radius at time $t$ of a scout sphere (co-growing inside the medium) whose initial radius was $R$.

Another consequence of the increase of stresses due to growth is the Biot buckling instability occurring at the free surface of the
aggregate, in this case the inner surface of the cyst, in a late stage after confluence. This is shown and made conspicuous in Fig.\ref{lum}b.

\begin{figure}[ht!]
\centering
\begin{subfigure}{0.49\textwidth}
\includegraphics[width=0.85\linewidth]{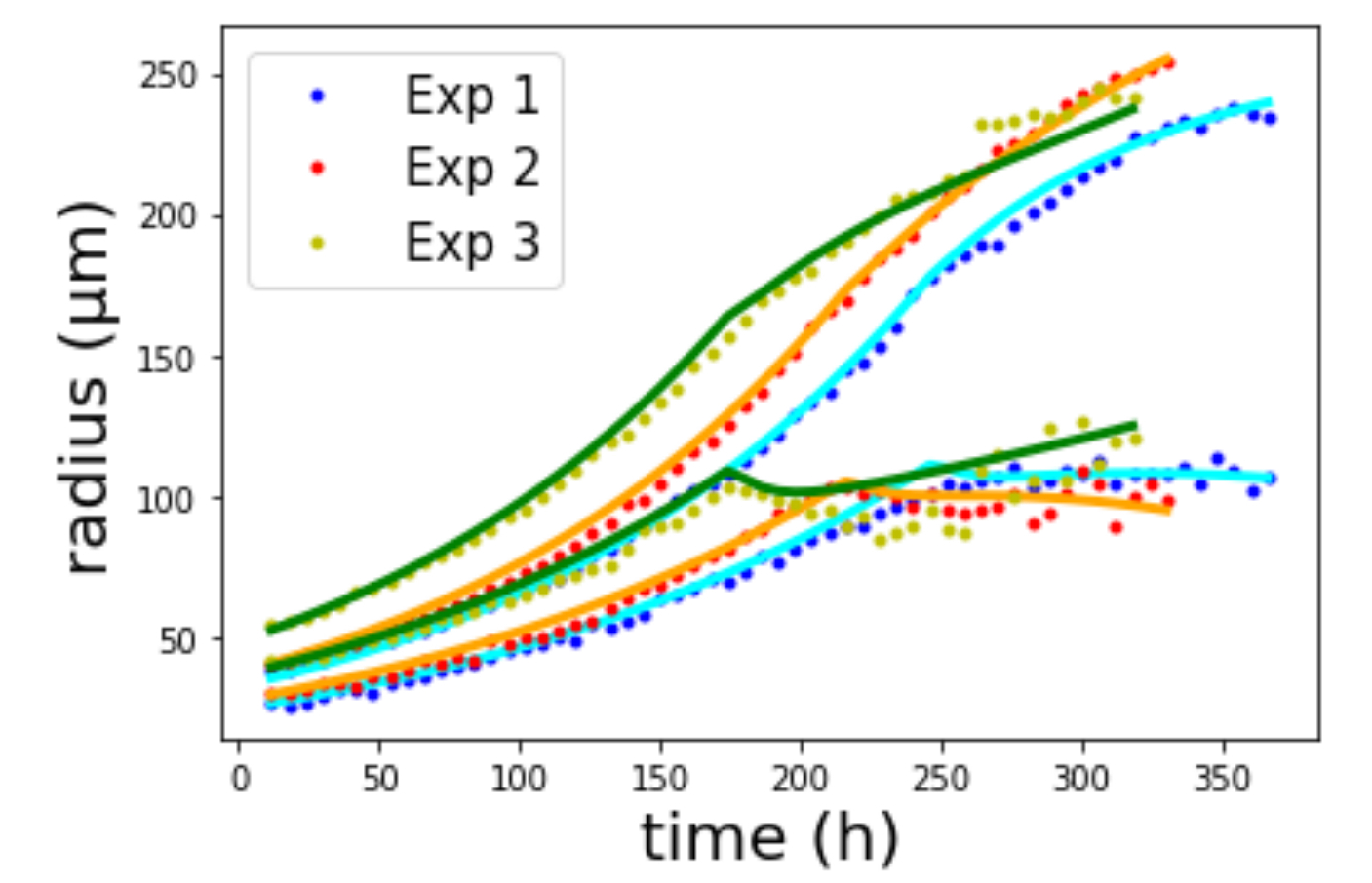}
\end{subfigure}
\begin{subfigure}{0.49\textwidth}
\includegraphics[width=0.85\linewidth]{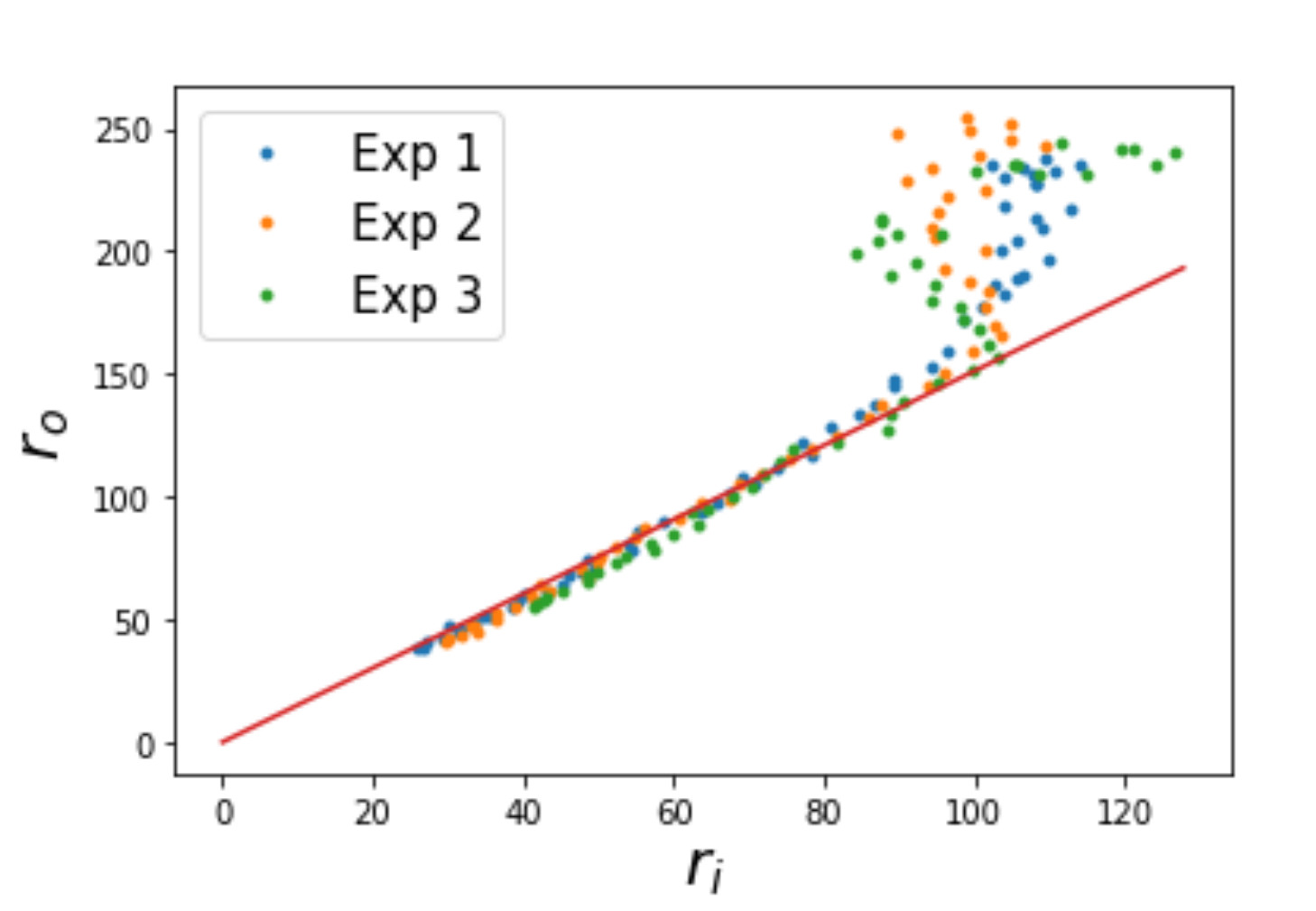}
\end{subfigure}
\caption{Comparison of the theoretical results with three experimental data sets. (Dots and dotted lines: experiment, continuous lines: model). (Left): Inner and outer radii from experiment and theory. Note that confluence does not occur simultaneously for the three different experiments.
(Right): Outer radius versus inner radius. Before confluence, the relation between the outer and inner radii is almost linear due to the weak coupling between elasticity and growth: $r_i/R_i \approx r_o/R_o$. Reproduced from \cite{Nassoy}. }
\label{cysRt}
\end{figure}

The growth anisotropy is apparent from the difference between the values obtained for $k_r$ and $k_{\theta}$. This difference is tiny before confluence, a time interval during which the growth is nearly stress-free and isotropic. The volume growth is illustrated in Fig.\ref{cysRt}, left panel, which shows the inner and outer cyst radii versus time. The relation between the cyst inner and outer radii, $r_i$ and $r_o$, is practically linear in the pre-confluent stage (see Fig.\ref{cysRt}, right panel).

In sum, the cyst versus spheroid configuration provides equal opportunity  access to nutrients for each cell. Homogeneous growth and low
elastic stresses result and persist throughout the pre-confluent stage. Unlike in volumetric growth cells are not subject to mechanical constraints and preserve their stemness. Stresses increase slowly during cyst evolution due to anisotropic growth effects. Post-confluent stages of growth entail capsule inflicted stresses that may ignite uncontrolled differentiation or mutations that will jeopardize the quality of the stem cells for medical applications. 

\subsection{Villi and crypts in the intestine}
In this final part of the lectures we choose a topic that permits us to make a synthesis of various concepts and findings explored hitherto. 
Once again, we scrutinize the two-way coupling between cell proliferation and mechanical stress. On this occasion we deal with a monolayer of epithelial cells found in the small intestine and in the colon (of our body). The interplay between cell division and elasticity of the stroma on which the monolayer resides can have fascinating morphological consequences and can be studied with the force balance considerations and visco-elastic theory at hand. 

\begin{figure}[h!]
\centering
\includegraphics[width=0.5\linewidth]{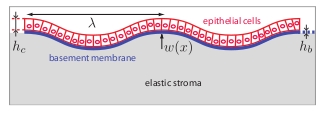}
\caption{Schematic picture of a buckled epithelial monolayer on
top of a basement membrane and a soft stroma. Typical buckling wavelength: 100 $\mu$m. Reproduced from \cite{Hannezo}. }
\label{bucmo}
\end{figure}

When cell division dominates over cell death, the monolayer is out of homeostatic equilibrium and attempts to increase its area. This corresponds to a negative surface tension and provokes a buckling instability at a finite wavelength. This results in the 
formation of periodic arrays of villi (finger-like protrusions) and crypts (deep pits).  Villi are important as they facilitate the
exchange of nutrients. To describe and explain this buckling phenomenon a simple three-layer model is adopted, as sketched in Fig.\ref{bucmo}. The inner part is a monolayer of epithelial cells of thickness $h_c \approx 10 \mu$m. Under it is a ten times thinner basement membrane. The surrounding outer layer is a (thick) soft stroma. The bending modulus $K$ (bending energy) of the monolayer is much larger than that of the membrane. The elasticity of the monolayer manifests itself when its curvature is perturbed, not when it is pushed or pulled in plane, since cell division or apoptosis neutralize lateral extension or compression forces (over long time scales). However, there is a (negative) surface tension, even in homeostatic equilibrium (steady state in which cell division balances cell death).

In homeostatic equilibrium the (negative) surface tension $-\gamma_0 <0$ is related to the homeostatic pressure $P_h$ through $\gamma_0 = P_h h_c >0$. The energy cost of 
a periodic undulation of the monolayer then has a positive bending energy contribution and a positive contribution proportional to the Young modulus $E_s$ of the surrounding material, but a negative contribution due to negative surface tension. For $\gamma$ exceeding a critical value 
\begin{equation}
    \gamma_c = (3 K E_s^2)^{\frac{1}{3}}, 
\end{equation}
which typically does happen in the biological systems concerned, the energy cost for a nonzero wavenumber undulation becomes negative,
yielding a characteristic buckling wavelength 
\begin{equation}
    \lambda_c = 
2\pi \left(\frac{K}{E_s}\right)^{\frac{1}{3}},
\end{equation}
which is on the order of 100 $\mu$m. This value, governed by elastic properties, is quite robust to variations of biological parameters, as is verified by observations in diverse samples {\it in vivo}. 

\begin{figure}[h!]
\centering
\includegraphics[width=0.5\linewidth]{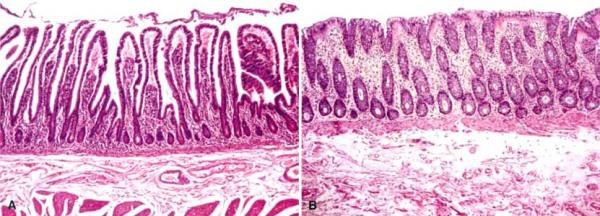}
\caption{Intestine and colon. Reproduced from \cite{Hannezo}. }
\label{intcol}
\end{figure}

\begin{figure}[h!]
\centering
\begin{subfigure}{0.49\textwidth}
\includegraphics[width=0.85\linewidth]{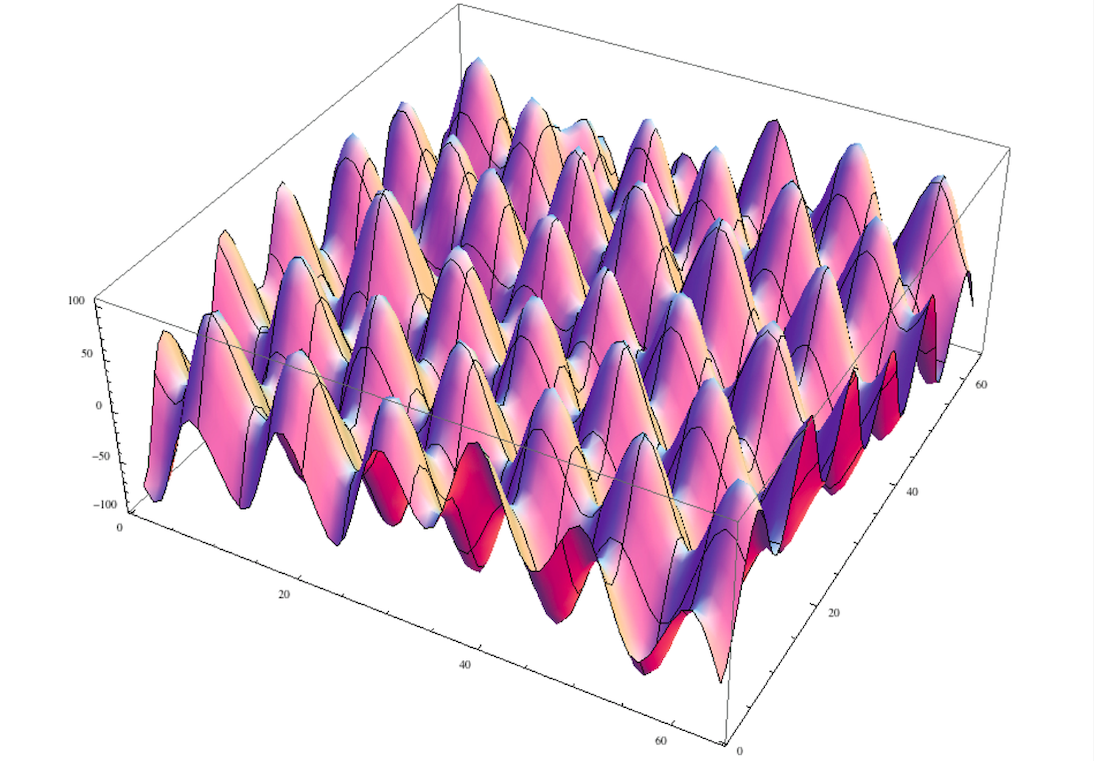}
\end{subfigure}
\begin{subfigure}{0.49\textwidth}
\includegraphics[width=0.85\linewidth]{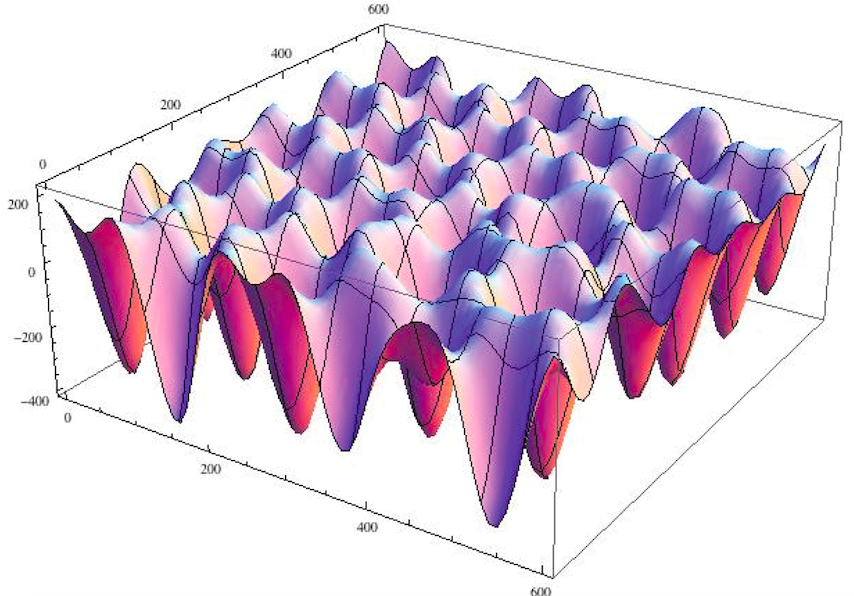}
\end{subfigure}
\caption{(Left) Small intestine morphology, showing developed
villi, arising when the coupling between cell proliferation and curvature is weak, and (Right) colon morphology, showing only crypts, resulting from strong growth-curvature coupling. Units
are micrometers. Reproduced from \cite{Hannezo}. }
\label{intecol}
\end{figure}

So far the theoretical arguments can accurately
describe the villi-like undulations of the small intestine. However, the architecture of the colon is different. In the colon there are no villi
but only crypts extending into the stroma. This asymmetry, which is illustrated in Figs.\ref{intcol} and \ref{intecol}, can be explained by allowing for 
spatially nonuniform cell division rates.  {\it In vivo}, cells
are produced from stem cells in crypts, then flow towards the villi, differentiate, and eventually undergo apoptosis at the villi tips. 
To model nonuniform cell division consistently with this biological input, one assumes that the growth rate depends on the local curvature of the monolayer.

Taking, for simplicity, as a reference system a homeostatic and flat monolayer, we denote the vertical displacement of the monolayer by $w \equiv w(x,y,t)$. The incompressibility condition on the in-plane cellular flow, including sources and sinks, reads
\begin{equation}
    \nabla \cdot \textbf{v} = k_d - k_a =  \xi (\gamma - \gamma_0) + \alpha \nabla^2 w 
    \label{incompr}
\end{equation}
and expresses the pressure as well as curvature dependence of the rates. Note that when the coefficient $\xi$ is positive the growth rate decreases with pressure. Also note that for positive $\alpha$ the growth rate is higher in the crypts (minima of $w$). To proceed to quantify the flow we take into account the friction of the cells against the basement membrane. 
In stationary flow the pressure gradient in the monolayer is balanced by the friction force against the basement membrane,
\begin{equation}
    \zeta {\bf v} = - \nabla \gamma,
    \label{fric}
\end{equation}
with $\zeta >0$ a friction coefficient. 
We consider henceforth small friction and keep only terms to first order in $\zeta$. In this case a pressure deviation linear in the undulation solves \eqref{incompr} together with \eqref{fric},
\begin{equation}
    \gamma = \gamma_0 - \zeta\alpha \, w
\end{equation}
The model features the following equation of motion for the vertical displacement of the membrane, as a result of force balance in the vertical direction, and after a suitable rescaling of variables,
\begin{equation}
    \frac{\partial w}{\partial t}  = -K \Delta^2 w - \gamma \nabla^2 w \\
 - f_{el,z} +f_{memb,z}
\end{equation}
Here, the left-hand side covers the friction force, while the right-hand side features, in order, the curvature force of the monolayer, the pressure force exerted by the
cells, the ($z$-components of the) elastic force due to the stroma and the stretching force of the basement membrane. 

\begin{figure}[h!]
\centering
\includegraphics[width=0.35\linewidth]{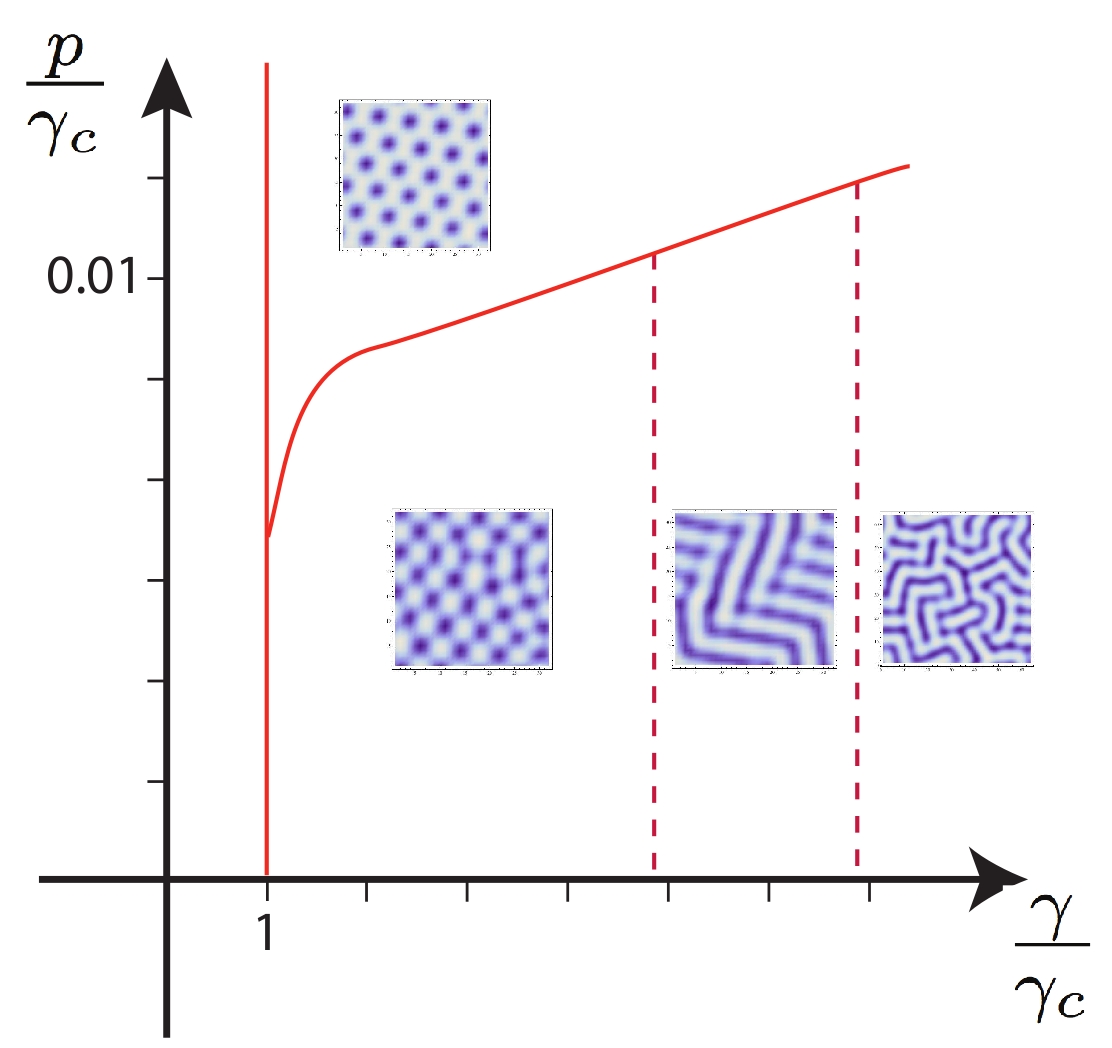}
\caption{Phase diagram showing the various possible villi architectures (in 2$d$ projections): (a) colon-like, (b) small-intestine-like finger-shaped, (c) herringbone, (d) labyrinth. The vertical axis, with $p\equiv \alpha \zeta$, represents the coupling $\alpha$ between cell division rate and curvature, essential to describe the 
villi-to-crypt crossover in colon-like tissue. The horizontal axis is the reduced pressure exerted by the cell monolayer. Inspired by \cite{Hannezo}. }
\label{phadia}
\end{figure}

\begin{figure}[h!]
\centering
\includegraphics[width=0.25\linewidth]{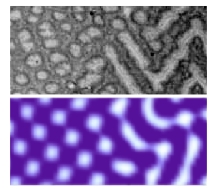}
\caption{Transition between finger-like and herringbone villi, comparing
a transverse section {\it in vivo} (top) and the simulation (bottom). Cell pressure
increases along the horizontal axis from left to right. Inspired by \cite{Hannezo}. }
\label{finghar}
\end{figure}

The calculations based on (refinements of) this model lead to the phase diagram of Fig.\ref{phadia}, which aims at presenting the possible
morphologies of the intestinal tube that are encountered {\it in vivo}.
The three main morphologies are
finger-shaped villi, herringbone pattern or labyrinth
pattern, emerging upon increase of the reduced buckling pressure $\gamma/\gamma_c$. A transition between finger and herringbone is further documented in Fig.\ref{finghar}. Similar patters to those shown here are also seen in skin papillae, for example a hexagonal array of villi, but with stem cells now on top (inverted polarity). Skin papillae form the thin top layer of the dermis (the inner layer of the skin), whose function is to nurture the epidermis (the outer layer of the skin) and help control skin temperature.

Understanding, as we attempted, the profound link
between tissue morphology and the stresses produced by the dividing cells may have practical usefulness in the health sector. Intestinal shape and renewal are to an important extent governed by a balance of physical (mechanical) forces. This balance is perturbed, e.g., in the case of intestinal diseases. It is known that an excess of mechanical pressure can,
in itself, induce colonic cancer. Therefore, advances in (statistical and fluid) mechanics of active and living matter is of paramount interest in biosciences and medicine, notably cancer research.

In this section on tissues as active matter, we have appreciated that much progress can be made by describing tissues as active nematic visco-elastic fluids. Of central significance is the coupling between mechanics and growth. Further, cell division entails a new mechanical relaxation time scale. Elastic instabilities, such as buckling, are key to our understanding of the structure of intestinal tissue. Of outstanding interest are applications of theory and simulations to real and model tissues, particularly in three major arenas: tumor growth, early embryonic development and cell therapies. 

\section{Acknowledgments} 
J.I. thanks the organizers Alberto Rosso, Thomas Speck and Andrea Gambassi of the Summer School FPSPXV for their hospitality in Brunico, Italy. He gratefully acknowledges the hospitality of Nguyen Van Thu at Hanoi Pedagogical University 2, Vietnam, and Kenichiro Koga at Okayama University, Japan, facilitating the writing of these Notes. He received a sabbatical bench fee (K802422N) from the Research Foundation-Flanders (FWO).
{}
\end{document}